\documentclass[review,a4paper]{elsarticle}

\usepackage{lineno,hyperref}
\modulolinenumbers[5]

\journal{Journal of Fluid and Structures}

\usepackage[top=2cm, bottom=2cm, left=2cm, right=2cm]{geometry}
\usepackage{subfigure}
\usepackage{xcolor}
\usepackage{amsmath}
\usepackage{epstopdf, epsfig}

\graphicspath{{Figures/}}

\begin{document}

\begin{frontmatter}

\title{Water entry and exit of 2D and axisymmetric bodies}
\author[add1,add2]{A. Del Buono\corref{mycorrespondingauthor}}
\cortext[mycorrespondingauthor]{Corresponding author}
\ead{alessandro.delbuono@inm.cnr.it}
\author[add2,add1]{G. Bernardini}
\author[add3]{A. Tassin}
\author[add1]{A. Iafrati}

\address[add1]{CNR-INM Via di Vallerano 139, 00128 Rome, Italy}
\address[add2]{University Roma Tre, 00146 Rome, Italy}
\address[add3]{Ifremer, RDT, F-29280 Plouzané, France}

\begin{abstract}
The present paper is dedicated to the development of a numerical model for the water 
impact of two-dimensional (2D) and axisymmetric bodies with imposed motion. The work is 
a first step towards the implementation of a 2D+t procedure to be used for the analysis 
of aircraft ditching. The problem is investigated under the assumptions of an inviscid 
and incompressible fluid, which is modelled by a potential flow model with fully 
non-linear boundary conditions at the free-surface. The unsteady boundary value problem 
with a free-surface is numerically solved through a boundary element method, coupled to 
a simplified finite element method to describe the thinnest part of the jet. The study 
is aimed at describing the entry and exit phases. Specific numerical solutions are 
developed to tackle the exit phase and to improve the stability of the model. Results 
are presented in terms of free-surface shape, pressure distribution and hydrodynamic 
load acting on the impacting body. The model is used to study the water entry and exit 
of a 2D wedge and an axisymmetric cone, for which numerical or experimental results are 
available in the literature. The numerical investigation shows that the proposed model 
accurately simulate both the entry and exit phases. For the exit phase, it is shown 
that the proposed model, being fully non-linear, provides a much better prediction of 
the loads and the wetted area compared to simplified (analytical) approaches. The 
effects of the gravity, usually missing in the approaches available in the literature, are 
also investigated, showing they are rather important, especially, in the exit phase.

\noindent
© 2021. This manuscript version is made available under the  CC-BY-NC-ND 4.0 license
  
http://creativecommons.org/licenses/by-nc-nd/4.0/
\end{abstract}



\end{frontmatter}


\section{Introduction}
Aircraft ditching, despite being a rare event, has to be accounted for in the design 
and certification phases of new aircrafts to guarantee the safety of the occupants. 
Large-scale experimental tests are very expensive and impractical. Therefore, 
aircraft manufacturers need computational tools able to provide a satisfactorily 
accurate description of the hydrodynamics and of the resulting fluid-structure 
interaction during the ditching phase, which govern both the aircraft dynamics and the 
structural response. During the design and certification phases, especially when many 
different configurations have to be analyzed, besides the high fidelity, fully coupled, 
fluid and structural solvers, fast and efficient solvers, although approximate, are 
strongly necessary to reduce the computational effort. A possibile solution is to 
describe the ditching phenomena with a 2D+t procedure, largely used in the past in the 
naval application, e.g. \cite{Iafrati2dt}, and recently applied to aircraft ditching 
application as well \cite{Gropengieber,Dassault}. In such a procedure the 3D problem is 
approximated by a 2D problem in the transverse plane in an earth-fixed frame of 
reference, with the shape of the impacting body changing in time. That is, the problem 
is described by a series of 2D water-impact problems, by neglecting the longitudinal 
derivatives of the hydrodynamic unknowns. Despite the approximations, this approach 
provides reasonably accurate information on the loads acting on the body. By looking at 
the ditching problem from a 2D+t perspective, in the rear part of the advancing 
fuselage the cross section shrinks and the lowest point of the contour rises, thus 
resembling a water exit problem. As such, the 2D solver used in the procedure must be 
able to handle both the entry and exit phases. These phases are characterized by a 
rapid evolution of the wetted surface which expands during the entry phase, when the 
body contour moves downwards, whereas it shrinks during the exit phase, when the body 
contour moves upwards. Moreover, in the combined water entry and exit problem, suction 
loads (hydrodynamic pressures being below the atmospheric pressure) can be observed. 
The suction loads increase as the body decelerates during the entry phase and, if the 
body deceleration is constant, they reach their maximum value at the transition between 
the entry and exit phase \cite{Tassin}. However, CFD results presented in \cite{Piro} 
for a rigid wedge show that the peak can occur before the beginning of the exit phase 
if the jet root starts separating from the body during the entry phase. So, the suction 
load and the flow separation can be closely related, and they can also appear in other 
problems, such as the oblique water entry of wedges \cite{semenov} and curved bodies at 
high horizontal speed \cite{Reinhard}. Recent experimental tests on the combined water 
entry and exit problem of axisymmetric bodies also show the influence of the gravity on 
the exit phase \cite{Breton}. The previuos paper also presents experimental tests on 
pure water exit of a body initially submerged. Pure water exit problems were 
investigated with experiments in \cite{Wuexp} and through a potential flow model in 
\cite{Ni1,Ni2}.\\
The first analytical model for the description of the combined water entry and exit 
events was proposed in \cite{kaplan}. In this paper, the original Wagner model 
\cite{wagner} was used for the entry stage whereas during the exit phase the same model 
was modified by discarding the "slamming term" (depending on the body velocity) and 
considering only the "added mass term" (depending on the body acceleration). In 
\cite{kor2013} the problem is solved via a linearized form of the mixed boundary 
value problem and by enforcing a Kutta-type condition at the contact point during the 
exit stage. The contact point position is predicted by assuming the proportionality 
between the speed of contraction of the wetted surface and the velocity of the 
particles located at the contact points. The latter model was extended in 
\cite{Kor2017a} to include body motions with time-dependent acceleration and water 
exit of a body whose shape changes in time. A semi-analytical model was proposed in 
\cite{Tassin} as a combination of the Modified Logvinovich Model (MLM), for the entry 
phase, and a modified von K\'{a}rm\'{a}n model, for the exit phase. In the second 
one, the position of the contact point is given by the intersection between the body 
and a reference water level corresponding to the maximum vertical position of the 
contact point reached during the entry phase. The pressure is evaluated with the MLM, 
but only the terms depending on the body acceleration are retained during the exit 
stage. Methods \cite{Kor2017a} and \cite{Tassin} were compared in \cite{Breton} with 
experimental results of the water entry/exit of a rigid cone, highlighting the limits 
of analytical models partly due to the lack of gravity effects.\\
In this paper the idea is to use the fully non-linear, two-dimensional incompressible 
potential flow model proposed in \cite{Battistin2003} and \cite{Battistin2004}, and 
already used for the water entry problem with constant velocity. The use of fully 
non-linear boundary conditions at the free-surface enables an accurate prediction of 
the pressure distribution at the spray root, which can be important when modelling the 
fluid-structure interaction. In \cite{Battistin2003} the problem is formulated via a 
boundary-element representation of the velocity potential (BEM model) and the time 
evolution is described by a mixed Eulerian-Lagrangian approach, originally proposed 
in \cite{Longuet-Higgins}. The free-surface evolution is followed using a Lagrangian 
approach by integrating in time the kinematic boundary condition. The model has been 
validated in the case of 2D and axisymmetric body impact with constant velocity 
\cite{Battistin2003}. In the latter paper, the thin jet, generated by the flow 
singularity about the intersection between the free-surface and the body contour, is 
cut off and replaced by a straight panel orthogonal to the solid boundary with a 
suitable boundary condition applied there, as proposed in \cite{Zao1993}. This 
assumption is justified as the pressure inside the thin jet is rather negligible, 
despite the high computational effort needed for an accurate description of the flow in 
such a thin region. In \cite{Zao} the model proposed in \cite{Zao1993} was extended to 
describe the flow in presence of flow separation by enforcing a Kutta condition as soon 
as the jet truncation passes through the separation point: it is assumed that the fluid 
leaves the separation point tangentially and with a finite velocity. The proposed model 
was successfully applied to the case of body contours with geometry singularities, such 
as knuckles. To achieve an improved prediction of the flow separation but limiting the 
high computational effort, \cite{Battistin2004} presents a simplified hybrid BEM-FEM 
model (as an evolution of the BEM solver in \cite{Battistin2003}), which models the 
thinnest part of the jet with control volumes where the velocity potential is 
represented as a harmonic polynomial representation. The model is validated in 
\cite{Battistin2004} for the water entry with constant velocity of a wedge with 
different deadrise angles and in \cite{Iafrati2003}, also for separated flows, solved 
by using the Kutta condition at the separation point.\\
Here, the hybrid BEM-FEM model is extended to deal with the combined water entry and 
exit problem. The numerical simulation of the exit phase is critical, especially if 
subsequent to a water entry phase, and free-surface instabilities might arise, as 
observed in CFD simulations in \cite{Piro} and potential flow simulations in 
\cite{Baarholm}. These instabilities are quite difficult to manage and can cause 
the stop of the numerical simulation. For this reason, two techniques used in the 
original model for the water entry at constant velocity \cite{Battistin2003} are further
extended and exploited here to improve the stability of the numerical solution during 
the exit phase. The first one cuts the jet when its thickness becomes too thin, whereas 
the second one increases the action of the numerical filter used to overcome the 
saw-tooth instability of the BEM solution \cite{Dold}. Furthermore, in the original 
hybrid BEM-FEM approach, the gravity effects were not included, because it was used 
only in the water entry problems at constant velocity, where the gravity effects are 
negligible. Here, in order to enable the model to deal also with the exit phase, the 
gravity effects are included.\\
The proposed model is validated against two different test cases, a rigid 2D wedge 
and a rigid cone, with imposed motion combining water entry and exit phases. 
For the wedge case, the present results are compared with CFD \cite{Maki} and 
semi-analytical \cite{Tassin} computations, in terms of non-dimensional vertical 
force, whereas, for the cone case, the vertical force and the contact line time 
histories are compared with experimental and analytical results presented in 
\cite{Breton}. Results, in terms of free-surface evolution and pressure distribution, 
are also presented.

\section{Hybrid BEM-FEM formulation}
\begin{figure}[h]
	\centering
	\includegraphics[scale=0.4]{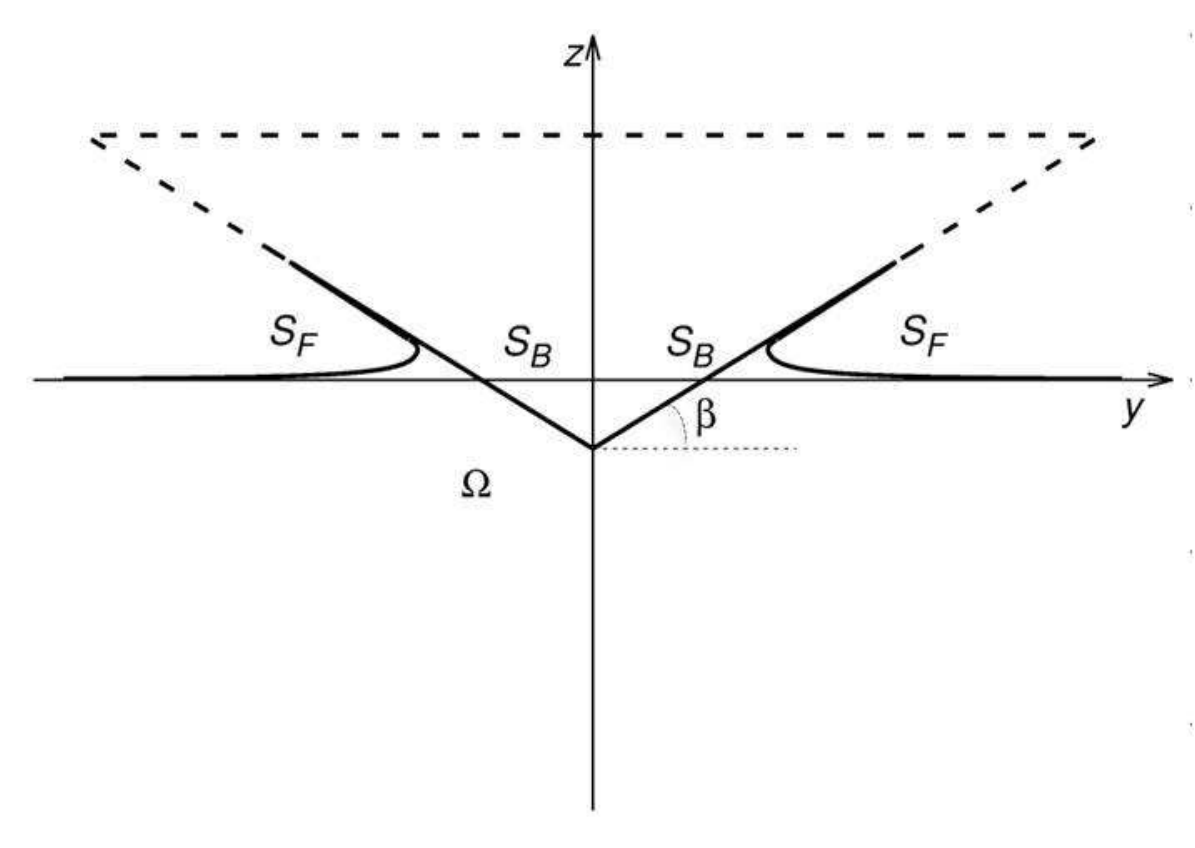}
	\caption{Sketch of the problem}\label{Problem_sketch}
\end{figure}

The vertical water impact problem of 2D and axisymmetric bodies is here investigated.
The study is conducted in an earth-fixed frame of refercence, with $y$ the horizontal 
axis oriented from left to right, $y=0$ being at the symmetry axis, and $z$ the 
vertical axis oriented upwards, with $z=0$ located at the still water level. Owing to 
the symmetry of the problem, only the right half of the plane ($y,z$) is considered 
(see Fig. \ref{Problem_sketch}). 
The fluid domain $\Omega$ is therefore bounded by the free surface, $S_F$, on the top 
and by the body surface, $S_B$, on the left, and it is assumed infinitely deep. The 
impacting body is characterized by the deadrise angle $\beta$, which is the angle 
between the body surface and the horizontal axis. Under the hypotheses of an 
incompressible fluid and irrotational flow, the hydrodynamic problem can be formulated 
in terms of the velocity potential, $\varphi$, which satisfies the Laplace equation 
in the fluid domain $\Omega$, the Neumann boundary condition on the body surface, and a 
Dirichelet boundary condition on the free surface $S_{F}$. The former comes from the 
impermeability of the body contour; the latter is written in terms of a dynamic 
boundary condition, which simply states that the pressure is constant over the 
free-surface. The free-surface position is determined as a part of the problem by 
integrating in time the kinematic boundary condition, (particles located at the 
free-surface remain at the free-surface). The boundary value problem reads as follows
\begin{equation}\label{laplace}
\begin{aligned}
{\nabla^{2}\varphi} = 0 \hspace{3.5cm}    \Omega\\
{\frac{\partial \varphi}{\partial n}} = {\mathbf{V}\cdot\mathbf{n}} \hspace{3cm} S_{B}\\
{\frac{D\varphi}{Dt}} = {\frac{\left| \nabla \varphi \right|^{2} }{2}-gz} \hspace{2cm} S_{F}\\
{\frac{D\mathbf{x}}{Dt}} = \mathbf{u} \hspace{3.5cm} S_{F}
\end{aligned}
\end{equation}
where $\mathbf{V}$ is the body vertical velocity (positive downwards), $\mathbf{n}$ 
is the unit vector normal to the boundary of the fluid domain oriented inwards, 
$\frac{D}{Dt}$ is the total derivative with respect to time $t$, $\mathbf{x}$ denotes 
the position of a particle lying at the free-surface, and $\mathbf{u}$ the velocity at 
$\mathbf{x}$. Although the effect of the acceleration of gravity, $g$, are usually 
neglected in water entry problems, especially for the constant water entry problem 
where $gt/U<<1$, they are accounted for in the present study to assess the role of the 
gravity in water entry/exit problems. The surface-tension effects are always neglected. 
According to the boundary value problem (\ref{laplace}), the velocity potential is 
known on the free-surface, whereas its normal derivative is assigned on the body 
contour. The problem is solved through a mixed Eulerian-Lagrangian approach 
\cite{Longuet-Higgins}. In order to solve the boundary value problem (\ref{laplace}), 
the velocity potential is written in terms of a Boundary Integral Representation. The 
velocity potential on the body contour and its normal derivative on the free-surface 
are retrieved by writing the boundary integral representation at the boundary of the 
fluid domain. In this way, the problem takes the form of a boundary integral equation 
of mixed first and second kind \cite{Battistin2003}
\begin{equation}\label{BIE}
\frac{1}{2}\varphi(P)=\int_{S_B\bigcup S_F}{\left( \frac{\partial \varphi 
		(Q)}{\partial n}G(P,Q) -\varphi (Q) \frac{\partial G(P,Q)}{\partial n}\right)dS(Q)} 
\hspace{1cm} P\in \Omega
\end{equation}
where $G$ is the free-space Green’s function of the Laplace operator which in two 
dimensions is 
\begin{equation}\label{G2d}
G(P,Q)=\frac{1}{2\pi}log\left( \left|P-Q \right| \right),
\end{equation} 

whereas in three dimensions is
\begin{equation}\label{G3d}
G(P,Q)=-\frac{1}{4\pi \left|P-Q \right|}.
\end{equation}
The hydrodynamic load acting on the body is obtained by integration of the pressure 
distribution along the wetted part of the body. The pressure is computed by the 
unsteady Bernoulli's equation:
\begin{equation}\label{Bernoulli}
p-p_{inf}=-\rho\left(\frac{\partial \varphi}{\partial t}+\frac{\left| \nabla \varphi \right|^{2} 
}{2}+gz \right) 
\end{equation}
where $\rho$ is the fluid density, equal to 1000 Kg/m\textsuperscript{3}, and 
$p_{inf}$ is assumed null. The velocity potential along the body is deteminated using 
equation (\ref{BIE}), whereas the time derivative of the velocity potential along the 
body, $\frac{\partial\varphi}{\partial t}$, is computed by formulating a boundary value 
problem similar to (\ref{laplace}). The time derivative of the velocity potential is 
also a harmonic function which satisfies the following Dirichlet condition at the 
free-surface:
\begin{equation}
\frac{\partial \varphi}{\partial t}= -\frac{\left| \nabla \varphi \right|^{2}}{2}-gz 
\end{equation}
which comes from the constant pressure condition. A Neumann boundary condition 
applies on the body side, which is derived from the time derivative of the second 
equation in (\ref{laplace})
\begin{equation} \label{bcpre}
\frac{\partial}{\partial n}\left( \frac{\partial \varphi}{\partial t}\right)=\mathbf{n}\cdot \mathbf{a}-\mathbf{n}\cdot \left(\mathbf{V}\cdot \nabla \right)\mathbf{u}
\end{equation}
The latter equation is rigorously valid only when the normal unit vector, 
$\mathbf{n}$, does not vary in time due to body rotations or deformations. In 
equation (\ref{bcpre}), $\mathbf{a}$ is the body acceleration and $\mathbf{u}$ is the 
fluid velocity along the body contour. As shown in \cite{Battistin2003}, the term 
$\mathbf{n}\cdot \left(\mathbf{V}\cdot \nabla \right)\mathbf{u}$ can be written as
\begin{equation}
\mathbf{n}\cdot \left(\mathbf{V}\cdot \nabla \right)\mathbf{u} = V_s\frac{\partial u_n} {\partial s}-V_n\frac{\partial u_s} {\partial s}+k_{sn} \mathbf{V}\cdot \mathbf{u}
\end{equation}
in the two-dimensional case and as
\begin{equation}
\mathbf{n}\cdot \left(\mathbf{V}\cdot \nabla \right)\mathbf{u} = V_s\frac{\partial u_n} {\partial s}-V_n\frac{\partial u_s} {\partial s}+ k_{sn}V_su_s+(k_{sn}+k_{\theta n})V_nu_n- \frac{r'}{r}V_nu_s
\end{equation}
in the three-dimensional case, where $\frac{\partial} {\partial s}$ denotes the spatial 
differentiation with respect to the tangent to the body contour, $k_{sn}$ and $k_{\theta n}$ 
denote the local curvatures of the body contour, $r$ is the radial position and in $r'$, 
prime denotes differentiation along the meridian contour. The terms $\left\lbrace 
u_s,u_n\right\rbrace$ and $\left\lbrace V_s,V_n\right\rbrace $ are the projections of 
the fluid and body velocities along the tangent and the normal to the body contour, 
respectively.\\
The boundary integral equation (\ref{BIE}) is solved via a zero-order panel method 
with a piecewise constant distribution of the variables on each panel. In the 
numerical model, the impacting body is represented as a set of discrete points which 
are interpolated with a cubic spline. At each time, step a linear extrapolation of the 
positions of the free-surface centroids is used to intersect the body contour and to 
locate the contact point and to identify the wetted area $S_B$ and the free-surface 
$S_F$. The first one is discretizated with straight line panels placing the vertices on 
the body spline curve. The same is done for the free-surface, where the panel vertices 
are located along the cubic spline built by interpolating the free-surface centroids 
position. The discretization of the fluid boundary is started from the intersection 
point of the free-surface with the body contour and the panel size increases linearly 
moving away from it. In the discrete form the influence coefficients of a panel on a 
panel midpoint are evaluated analytically in the 2D case, while in the axisymmetric 
case the integration is performed analytically in the azimuthal direction and 
numerically (through a Gauss formula with an even number of points) along the meridian 
section of the panel \cite{Battistin2003}. The use of an even number of points into the 
Gauss formula allows to avoid the singularity of the kernel, and to mimic the 
computation of the integral as a Cauchy Principal value. The solution of the boundary 
value problem provides the velocity potential on the wetted body contour and its normal 
derivative on the free-surface. The latter, along with the tangential velocity obtained 
by a second-order finite differences of the velocity potential along the boundary, 
defines the velocity at the free-surface. This velocity is used into the kinematic and 
dynamic boundary conditions to compute the free-surface shape and the velocity 
potential on it. The integration in time of the kinematic and dynamic conditions is 
done with a second-order Runge-Kutta scheme. The time step is chosen so as to satisfy 
two different conditions. The first one is to ensure that the maximum displacement 
of the centroid is smaller that a fraction of the panel length. The second one, which 
is active mainly in the early stage, is to prevent the panel centroids from penetrating 
the body. The new configuration of the fluid boundary is found by moving the body 
contour and the free-surface centroids in a Lagrangian way and by discretizing the new 
$S_B$ and $S_F$ as mentioned above. It is worth remarking that, although the 
boundary-value problem (\ref{laplace}) is numerically satisfied at the midpoint of the 
panels, which are delimited by the vertices lying on the spline curve, Lagrangian 
markers, which are the centroids, used to track the free-surface motion are always 
located along the spline curve, thus enabling improved mass conservation properties of 
the numerical scheme \cite{Zao1993}.\\
In order to model the occurrence of flow separation, a Kutta condition is enforced at 
the separation point requiring that the flow leaves the body contour tangentially 
\cite{Iafrati2003,Zao}. In particular, a Neumann condition is applied at the first 
panel of the separated region, requiring that the normal velocity equals that of the 
last panel still attached to the body \cite{Iafrati2003}. For body shapes with hard 
chines, such as a finite wedge or a cone, in the case of constant entry velocity the 
separation point can be assigned \textit{a priori} and corresponds to the point of 
geometric singularity. If the body decelerates, such as in the cases studied in the 
following, the separation may be anticipated and the separation point is unknown and 
has to be determined as a part of the solution. The same happens for the water entry of 
convex-shaped bodies, such as a circular cylinder, when, even in the constant entry 
velocity case, the flow separates. For the latter case, a separation approach based on 
the negative pressure was proposed in \cite{Sun06} where the flow is assumed to detach 
from the body surface when the pressure becomes lower than the atmospheric value in a 
large part of the wetted area. However, negative pressure may occur without flow 
separation as shown in \cite{Kor2017a} for a rigid plate. Therein, it is also shown 
that, in the later stage of the deceleration, the pressure is negative over a large 
portion of the wetted area. Experimental evidence of negative pressures occurring 
without flow separation is provided in \cite{Iafrati2019}. For these reasons, it is 
preferred here to assume that the separation point occurs only at the chine of the body.

\subsection{\textbf{Jet Model}}
\begin{figure}[htbp]
	\begin{center}
		\includegraphics[scale=0.4]{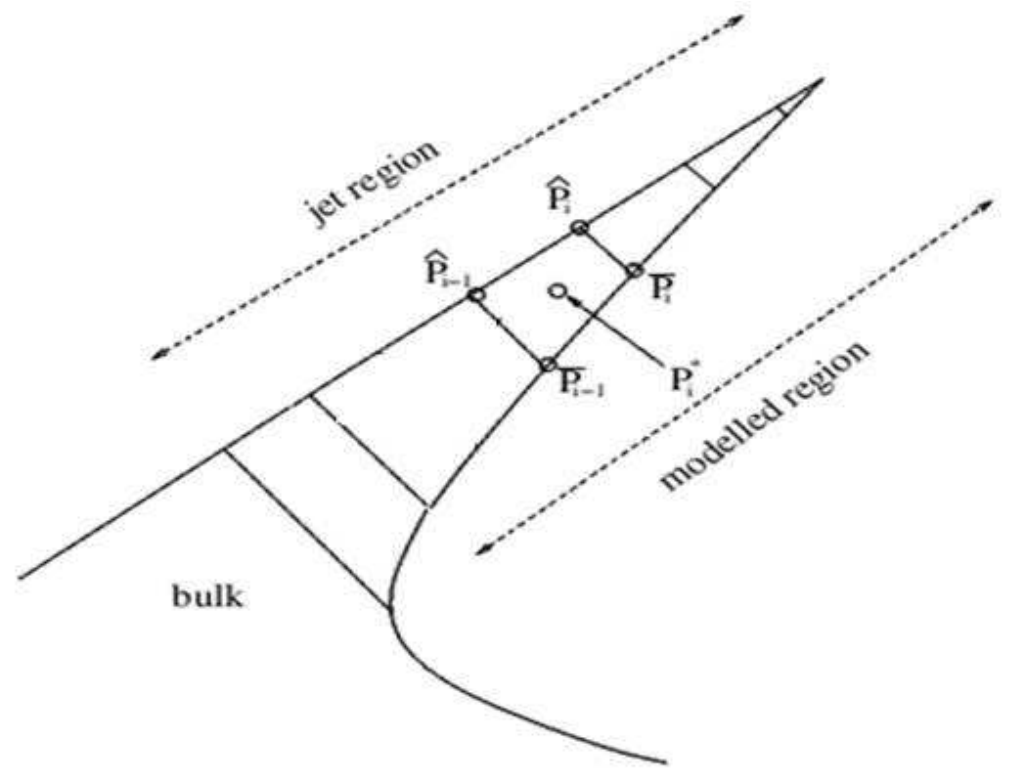}
		\caption{Sketch of the Jet Model}\label{jetmodel}
	\end{center}
\end{figure}

As aforesaid, the flow singularity at the intersection between the free-surface and 
the body \cite{Greenhow} leads to the formation of a thin jet layer along the body 
surface. If solved via BEM, the discretization of such a thin jet would require a panel 
size comparable to the jet thickness, and because of the local high speed of the fluid, 
a quite small time step would be necessary to preserve the stability and accuracy of 
the integration scheme. As a consequence, the computational effort would increase 
substantially. Moreover, due to the quite small angle at the jet tip, there is always a 
region where the boundary element method fails, regardless of the size of the panels. 
In order to overcome such important limitations, a common strategy is to cut the jet 
and replace it with a panel orthogonal to the body contour where appropriate boundary 
conditions are applied \cite{Battistin2003}. Since the pressure inside the thin jet is 
almost uniform, the jet cutting strategy still provides accurate estimates of the 
hydrodynamic loading. However, the cutting procedure may have limitations, as mentioned 
in \cite{Battistin2004} and then, if used, the fluid motion details inside the thin 
jet would be lost, leading for instance to an inaccurate prediction of the separation 
phase. For these reasons, the thin jet is here modelled using the simplified model 
proposed in \cite{Battistin2004,Iafrati2003} and validated in the case of water entry 
with constant velocity of a finite wedge.\\
A portion of the jet region, starting from the tip, is discretized with small control 
volumes (Fig.\ref{jetmodel}). The vertices of each volume $i$ correspond to the panel 
centroids on the body ($\hat{P}_{i-1}, \hat{P}_{i}$) and on the free-surface 
($\bar{P}_{i-1}, \bar{P}_{i}$).
Inside each control volume the velocity potential is written in the form of a harmonic 
polynomial series, up to second order, about the corresponding centroid ($y^*,z^*$)
\begin{equation}\label{phijet}
\varphi_{i}^j(y,z)=A_i+B_i(y-y^*)+C_i(z-z^*)+\frac{1}{2}D_i\left[(y-y^*)^2 -(z-z^*)^2\right]+E_i(y-y^*)(z-z^*) 
\end{equation}
Equation (\ref{phijet}) introduces five additional unknowns, $A_{i},..,E_{i}$, for 
each control volume which require five additional equations. Four of them are derived 
by enforcing the boundary conditions at the body and free surface sides
\begin{equation}\label{jetBC}
\varphi_{i,n}^j(\hat{P}_{i-1})=\varphi_{,n}(\hat{P}_{i-1}), \hspace{0.3cm} \varphi_{i,n}^j(\hat{P}_{i})=\varphi_{,n}(\hat{P}_{i}), \hspace{0.3cm}
\varphi_{i}^j(\bar{P}_{i-1})=\varphi(\bar{P}_{i-1}), \hspace{0.3cm} \varphi_{i}^j(\bar{P}_{i})=\varphi(\bar{P}_{i})
\end{equation}
where $\varphi_{i,n}^j$ is the normal derivative of the velocity potential provided 
by the harmonic expression. The fifth equation is derived by enforcing the continuity 
of the normal  derivative of the velocity potential at adjacent elements
\begin{equation}\label{jetContinuty}
\varphi_{i,n}^j(\hat{P}_{i-1})=\varphi_{(i-1),n}^j(\hat{P}_{i-1})
\end{equation}
The set of equations (\ref{jetBC}) and (\ref{jetContinuty}) are coupled with the 
discretized form of the boundary integral equation (\ref{BIE}) used in the remaining 
part of the fluid domain (hereinafter referred to as the “bulk” of the fluid). The jet 
model is activated only when the thin jet has already developed. Starting from the 
matching point with the modelled region, the bulk region is discretizated as mentioned 
earlier. A similar approach is adopted to discretize the modelled portion of the jet. 
The panel size is increased linearly, with a growth factor generally higher than that 
used for the free surface in the bulk of the domain. The hybrid BEM-FEM approach is 
also used for the evaluation of the time derivative of the velocity potential. Further 
details of the jet model are provided in \cite{Battistin2004,Iafrati2003}.
%
\subsection{\textbf{Water exit phase}}\label{water_exit}
%
A major difficulty in the simulation of the exit phase comes from the development of 
numerical instabilities in the free-surface profile shortly after the start of the exit 
phase. Such instabilities develop in time and eventually lead to a crash of the 
simulation. In order to preserve the stability of the numerical solution, two different 
strategies, which involve the use of techniques already used in the case of water entry 
at constant velocity \cite{Battistin2003}, have been developed, which can either operate 
independently or in combination. 
Both of strategies are not activated at the beginning of the exit phase, when the body 
velocity is zero, but when the exit phase has already started, thus allowing a smoother
transition between the entry and exit phases. In the present case, the strategies are 
activated when $V/V_0 \le -0.10$.\\ 
\noindent
The first strategy consists in cutting the thin jet during the exit phase. This is 
justified as the jet length keeps increasing whereas the jet root shrinks. As a 
consequence, the jet becomes thinner and thinner and its description becomes more and 
more difficult.
On the other hand, as in the entry phase, the pressure in the jet region is essentially 
zero, and thus cutting the jet has no effects on the hydrodynamic loading. 
Furthermore, while in the entry phase the description of the jet is needed to achieve a
more accurate description of the flow separation, particularly from smoothly curved
contours, in the exit phase the separation, if any, has already occurred and thus there 
is no need for an accurate description of the thin jet.
\begin{figure}[h]
	\centering
	\includegraphics[scale=0.4]{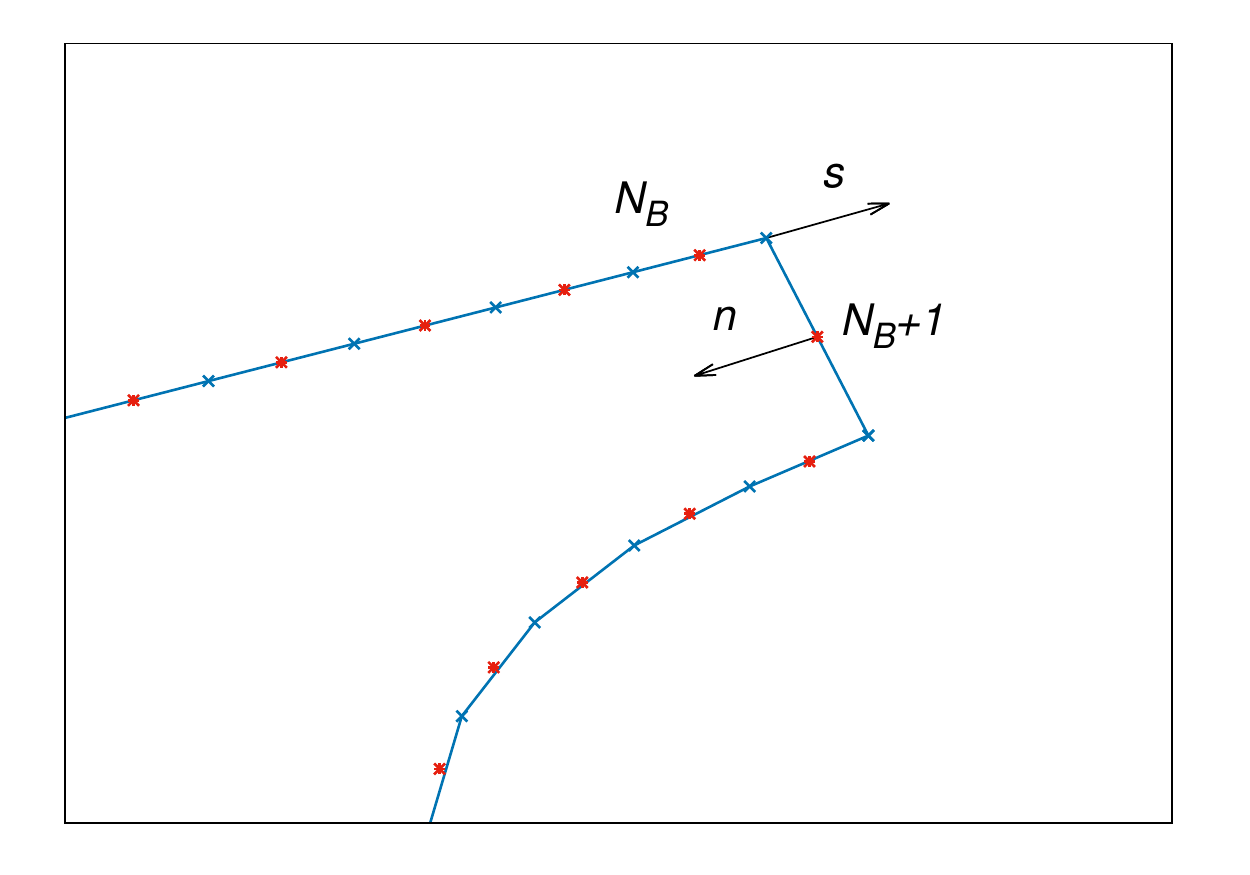}
	\caption{Cut procedure}\label{cut}
\end{figure}
When the jet is cut off, a new panel ($N_B+1$), orthogonal to the local body surface, 
is introduced (Fig.\ref{cut}): as the thinnest part of the jet is cut off, there is no 
need to use the hybrid BEM-FEM approach and the pure BEM is used everywhere in the 
fluid domain. There are two unknowns on the truncation panel: the velocity potential is 
derived from the solution of the boundary value problem whereas an additional equation 
is introduced for the normal derivative of the velocity potential at the panel $N_B+1$, 
which is imposed to be equal to the tangential derivative of the velocity potential 
along the body contour at the panel $N_B$,
\begin{equation}\label{cuteq}
\frac{\partial \varphi}{\partial n}(N_{B+1})=-\frac{\partial \varphi}{\partial s}(N_{B})
\end{equation}
The negative sign accounts for the different orientations of the unit vectors 
$\mathbf{s}$ and $\mathbf{n}$.\\
The second strategy is to increase, during the exit phase, the action of the numerical 
filter used to overcome the saw-tooth instabilities. The smoothing formula used in the 
model was proposed in \cite{Dold} and it is applied to the function $f(\xi)$ by 
subtracting the value of $\delta_M(\xi)$: 
\begin{equation}\label{filter}
f(\xi)=f(\xi)- \delta_M(\xi)
\end{equation}
The value of $\delta_M(\xi)$ depends on the order $m$ of the filter, which involves the 
$M=2m+1$ consecutive panels $(\xi -m,.., \xi,..,\xi +m)$, as follows:
\begin{equation} 
\begin{pmatrix}
2^2\delta_3\\
2^4\delta_5\\
2^6\delta_7\\
2^8\delta_9\\
2^{10}\delta_{11}\\
2^{12}\delta_{13}\\
2^{14}\delta_{15}\\
\end{pmatrix}
= 
\begin{pmatrix}
2 & -1\\
6 & -4 & 1\\
20 & -15 & 6 & -1\\
70 & -56 & 28 & -8 & 1\\
252 & -210 & 120 & -45 & 10 & -1\\
924 & -792 & 495 & -220 & 66 & -12 & 1\\
3432 & -3003 & 2002 & -1001 & 364 & -91 & 14 & -1\\
\end{pmatrix}
\begin{pmatrix}
f(\xi)\\
f(\xi +1)+f(\xi -1)\\
f(\xi +2)+f(\xi -2)\\
f(\xi +3)+f(\xi -3)\\
f(\xi +4)+f(\xi -4)\\
f(\xi +5)+f(\xi -5)\\
f(\xi +6)+f(\xi -6)\\
f(\xi +7)+f(\xi -7)\\
\end{pmatrix} 
\end{equation} 
\noindent
Equation \ref{filter} is for a filter up to 7th order. In the proposed model, the 
filter is used at every time step with an order equal to $3$. The term $\xi$ is 
the panel index of the free-surface and the separated part at the body side. The 
filtering is applied to both the free-surface shape and to the velocity potential on 
it. \textcolor{black}{When the second strategy is applied in the exit phase, the hybrid 
BEM-FEM approach is still used everywhere in the fluid domain but the order of the 
filter is increased to $5$. It was observed that a further increase of the filter
order does not change the results substantially.} Note that if the two strategies are 
adopted in combination and the jet is cut off, the pure BEM is used.

\section{Validation and Results}
The hybrid BEM-FEM approach has been thoroughly validated in 
\cite{Battistin2004,Iafrati2003} for the water entry of a 2D wedge at constant 
velocity. Here the approach, which is extended to deal with the body motion with an 
assigned vertical velocity and with the exit phase, is applied to two test cases. The 
first one concerns the water entry of a two-dimensional rigid wedge with a linearly 
varying vertical velocity, turning from an entry to an exit problem. The second one 
refers to an axisymmetric rigid cone with an imposed sinusoidal vertical motion. The 
two applications are aimed at verifying and validating the capabilities of the model to 
correctly describe the entry and exit phases in terms of free-surface evolution, 
pressure distribution along the body and hydrodynamic loads. As stated above, flow 
separation might occur in water entry problems with non-constant velocity. However, 
the definition of the separation point is a very challenging problem, and the criteria 
based on the negative pressure (see, e.g. \cite{Sun06}), seems too strong. A kinematic 
criterion is under development at present, but without a careful validation, it is 
preferred to assume that the flow does not separate from the body surface but it can 
separate from the chine if the tip of the jet reaches it. In the following the results 
obtained from the original model are denoted \emph{old}, whereas \emph{cut} and 
\emph{filter} are used to indicate the use of the jet cutting or of the increased 
filtering of the free-surface, respectively, as strategy to improve the stability of 
the solution in the exit phase.

\subsection{\textbf{2D Wedge}}
\begin{figure}[thb]
	\centering
	\subfigure[$V/V_0=-0.40$]{%
		\includegraphics[scale=0.55]{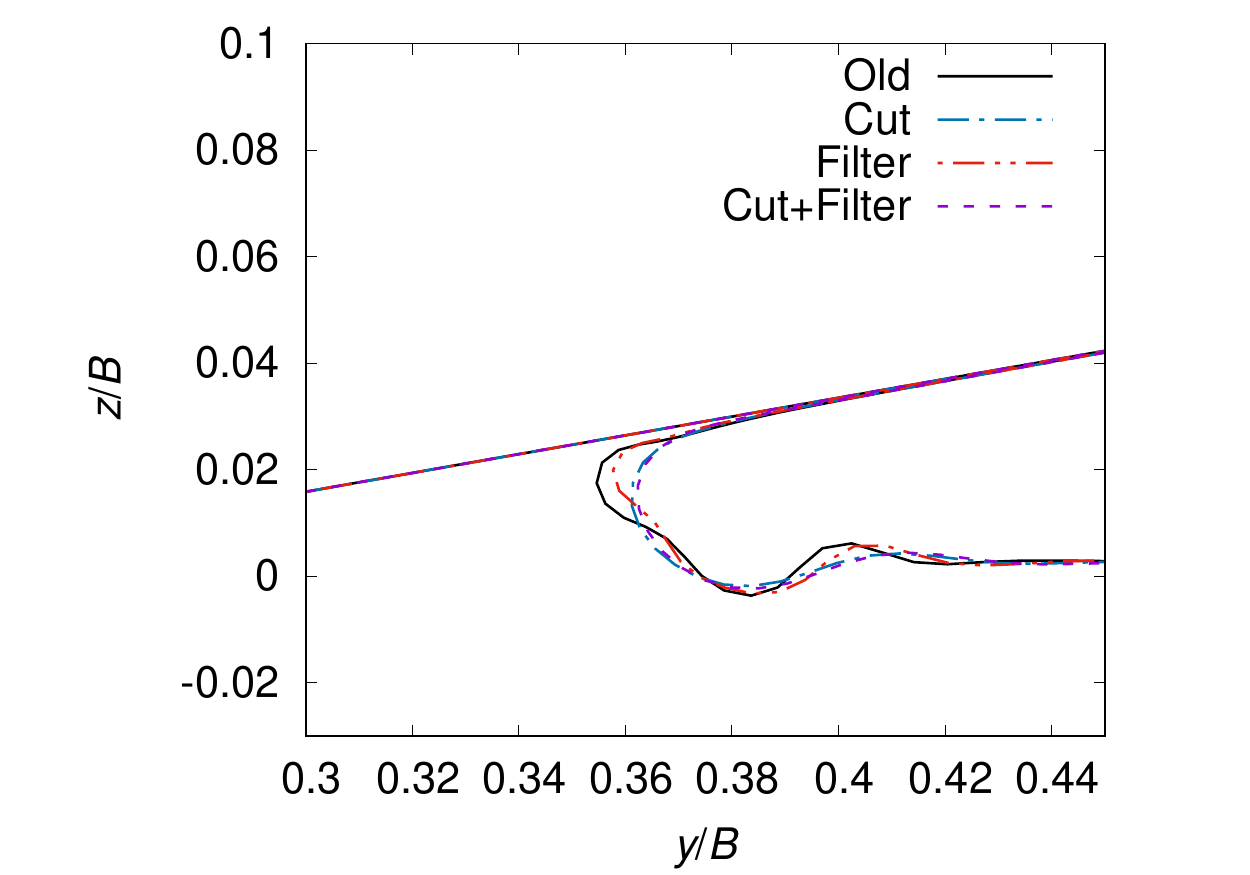}}
	\quad
	\subfigure[$V/V_0=-0.50$]{%
		\includegraphics[scale=0.55]{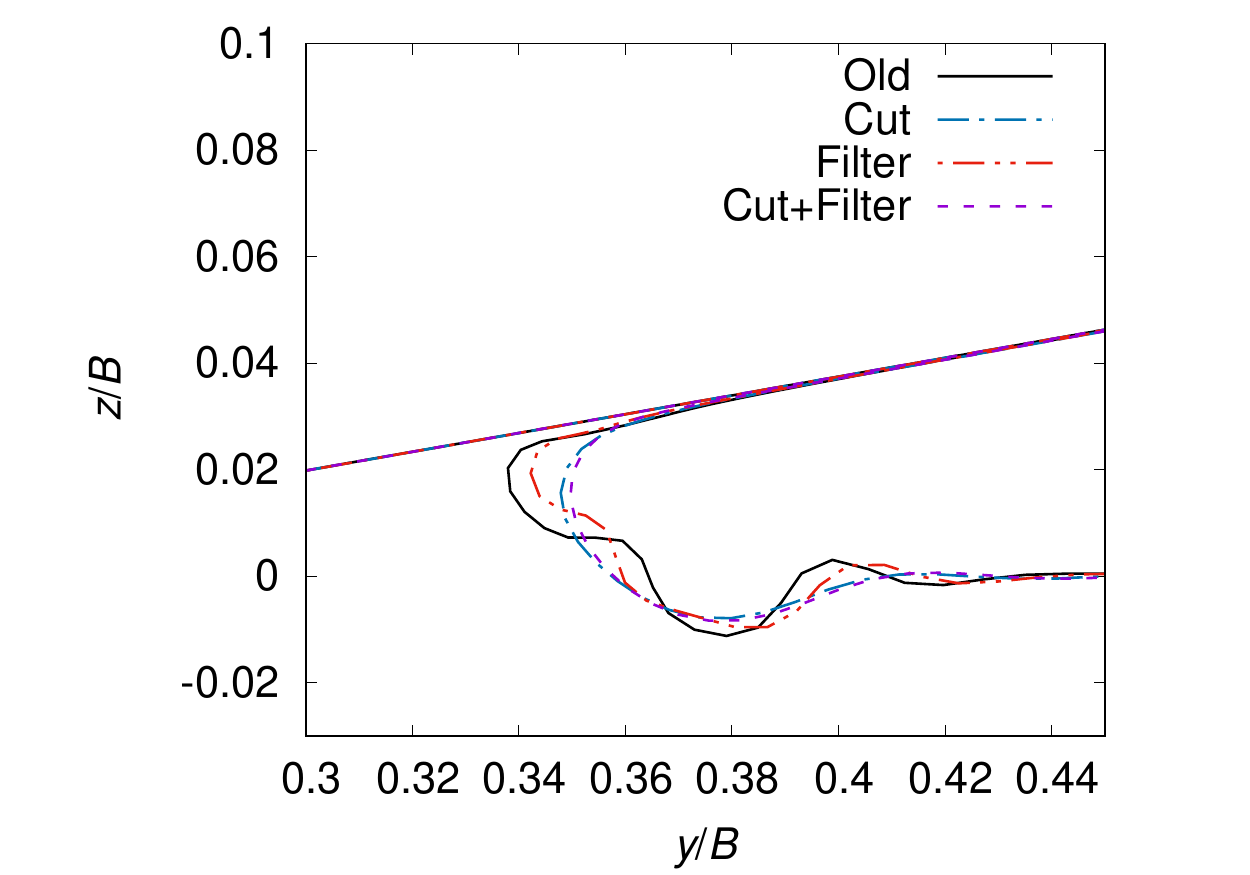}}
	\quad
	\subfigure[$V/V_0=-0.60$]{%
		\includegraphics[scale=0.55]{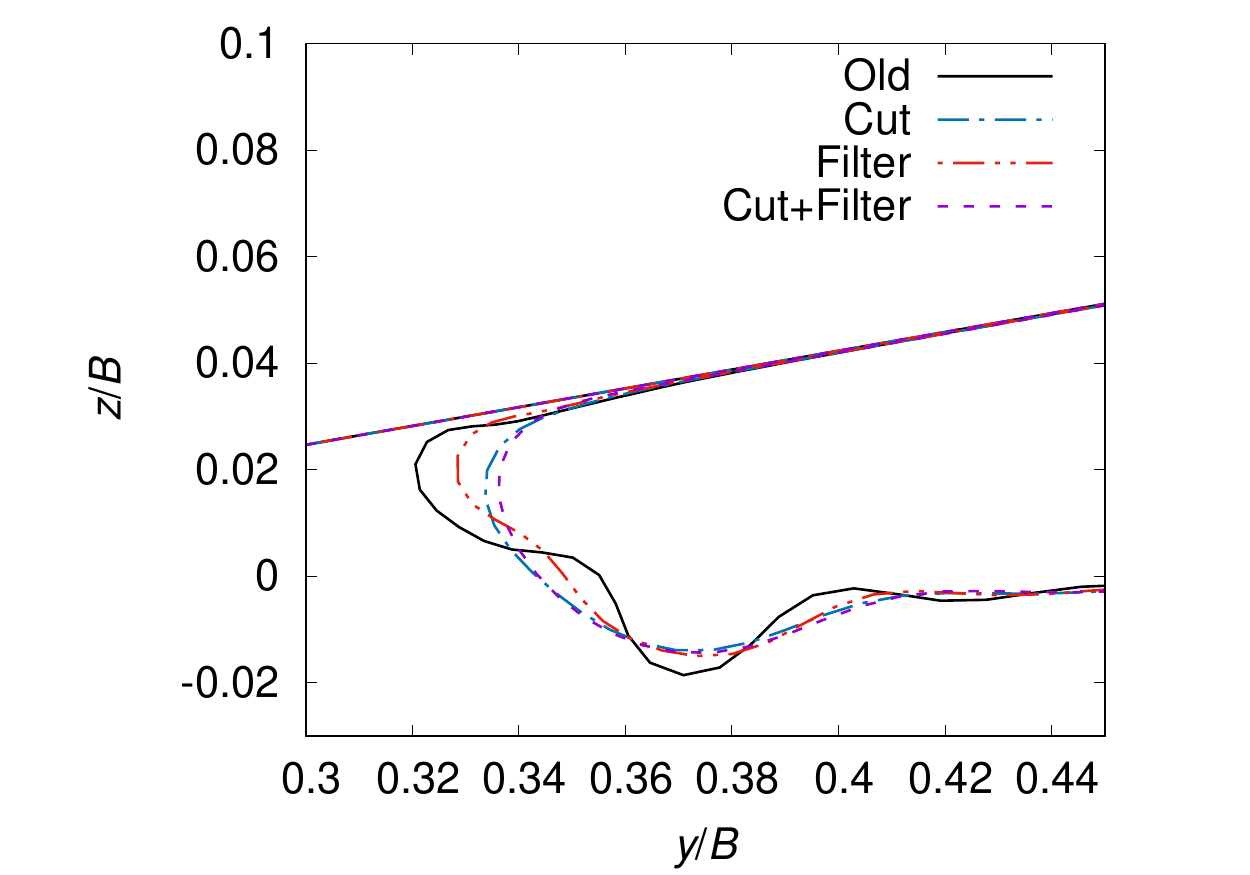}}
	\caption{\textcolor{black}{Wedge: close-up view of the free-surface shape in the jet root region at three different time steps during the exit phase.}}\label{W_exit}
\end{figure}

The flow about a two-dimensional wedge-shaped rigid body, with a deadrise angle 
$\beta=10^\circ$ and assigned vertical motion is here investigated. The velocity 
imposed to the impacting body is linear in time such that $V(t)=V_0-a t$, where 
$V_{0}=4m/s$ is the initial, downwards, entry velocity. The acceleration, $a$, is set 
such that when the body stops, at non dimensional time $t^*=V_0/a$, the maximum 
penetration depth, $d_{max}$, is such that $d_{max}/(L\sin\beta)=0.5$, where $L$ is the 
length of the wedge along the diagonal. Such a condition ensures that the jet root does 
not reach the chine of the finite wedge \cite{Maki}. This test case was investigated in 
\cite{Maki} via CFD and in \cite{Tassin} via a semi-analytical approach. Gravity 
effects are not included here, to be consistent with the previous studies. 
\noindent
\textcolor{black}{As anticipated, in the mixed entry/exit problem, numerical stability 
issues develop shortly after the beginning of the exit phase. 
As shown in Fig. \ref{W_exit}, when using the original model (\emph{old}), some
perturbations, of numerical origin, appear on the free-surface profiles. Such 
perturbations grow as the time elapses eventually leading the simulation to stop.
By introducing the new water exit models, as described in section 
\ref{water_exit}, a significant improvement of the solution is achieved with both
strategies enabling a smoothing of the instabilities from the very beginning. 
The combined use of the two strategies increases the robusteness of the model and 
reduces the computational effort, mainly because of the cut of the thin jet}. 
The use of the new models allows to extend the simulation even beyond the instant when 
the body is above the still water level, i.e. $t^*>2$.
\begin{figure}[thbp]
	\centering
	\subfigure[$V/V_0=0.80$]{
		\includegraphics[scale=0.5]{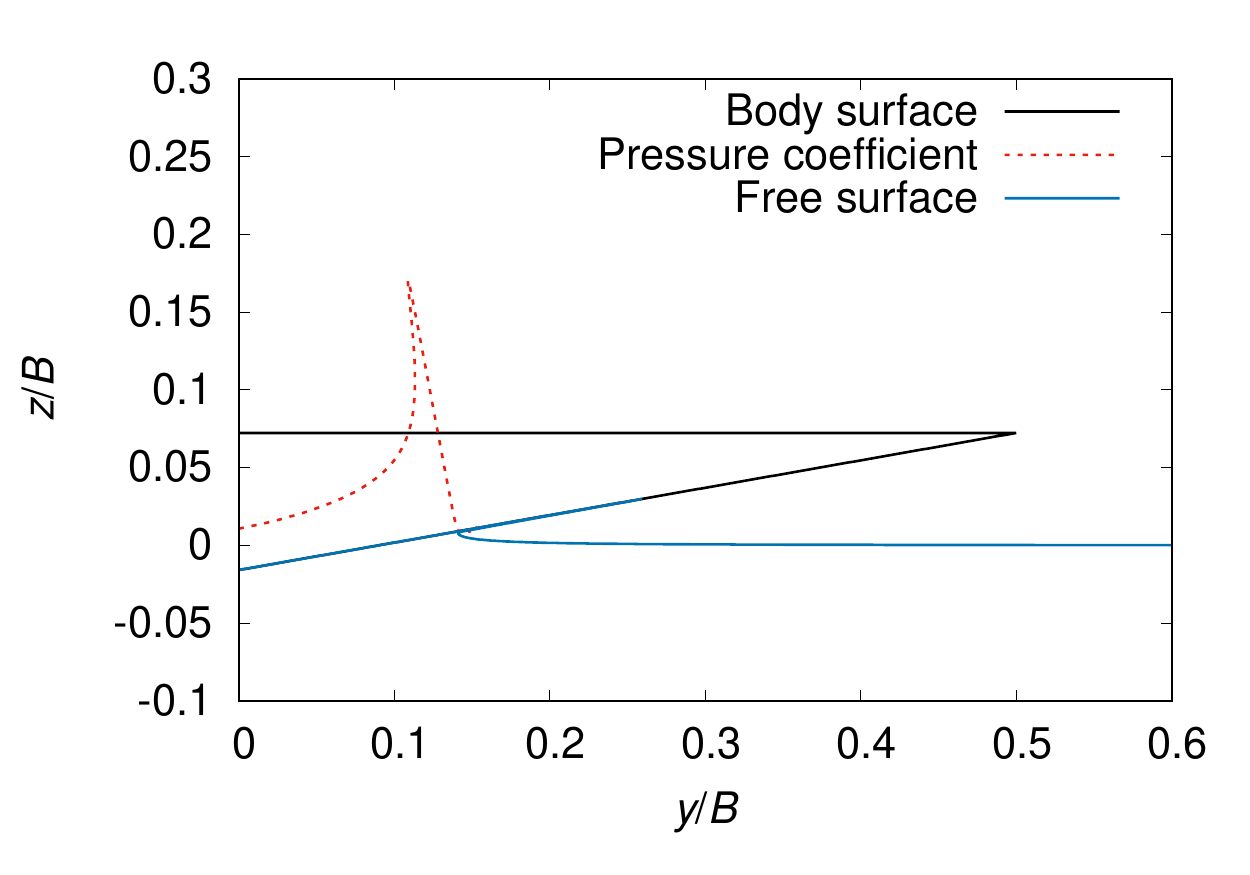} \hspace{2cm}
		\includegraphics[scale=0.5]{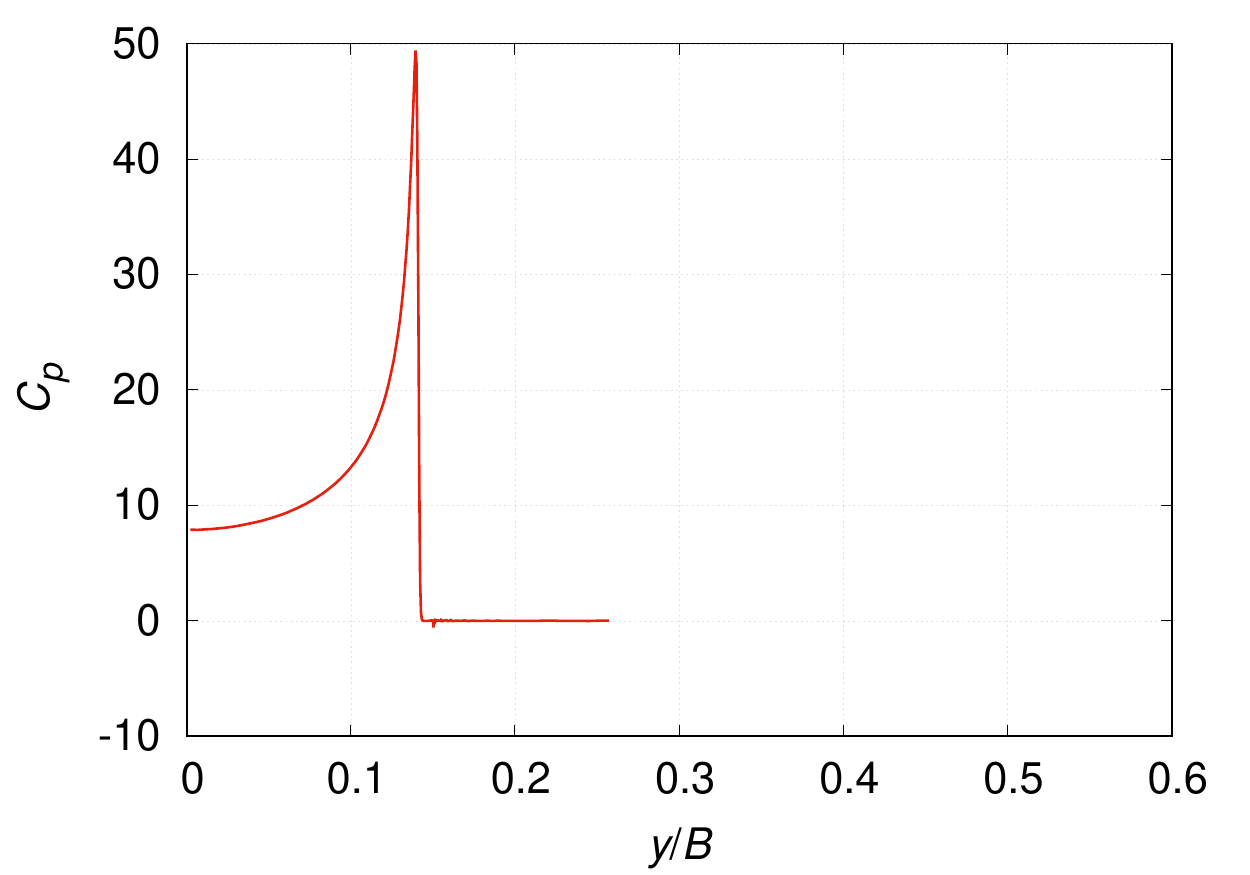}}\\
	\subfigure[$V/V_0=0.50$]{
		\includegraphics[scale=0.5]{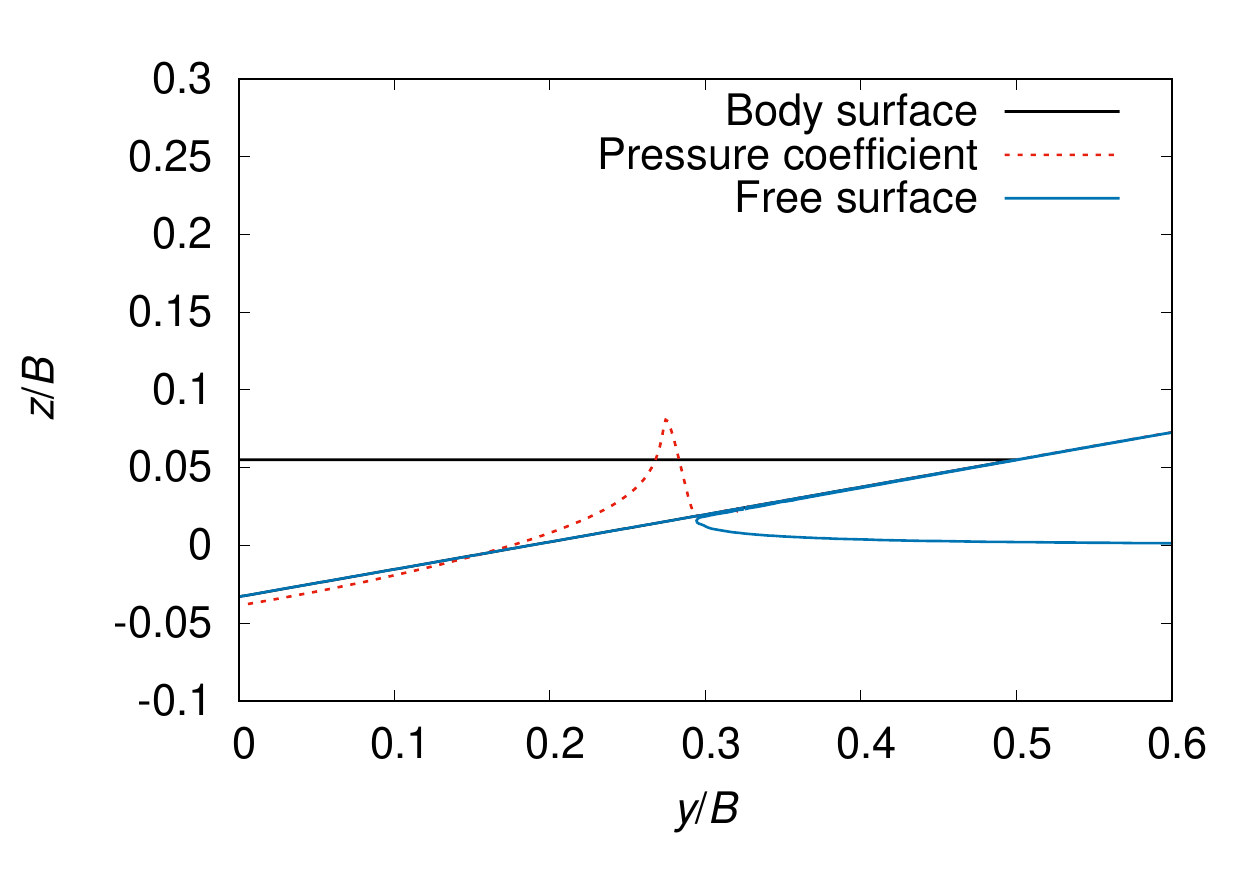} \hspace{2cm}
		\includegraphics[scale=0.5]{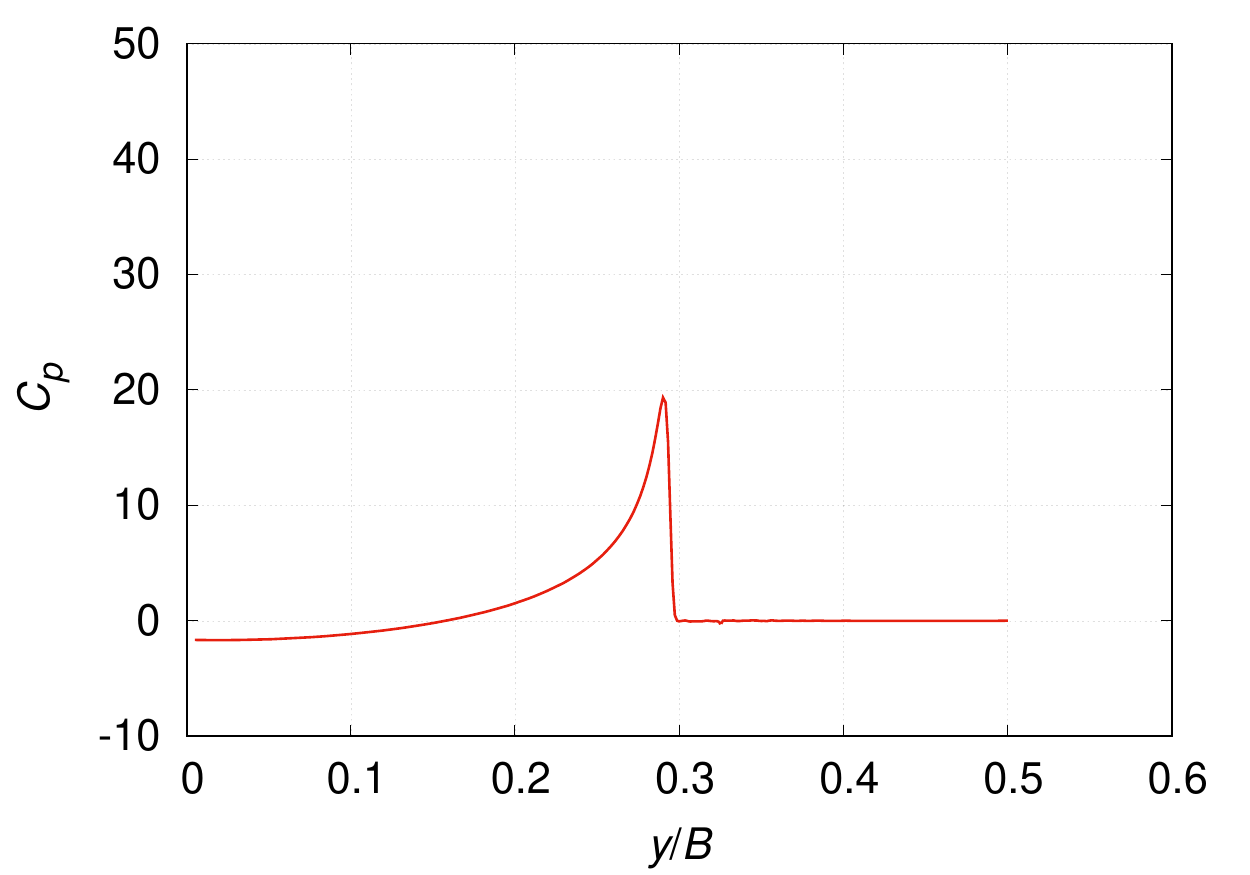}}\\
	\subfigure[$V/V_0=0.20$]{%
		\includegraphics[scale=0.5]{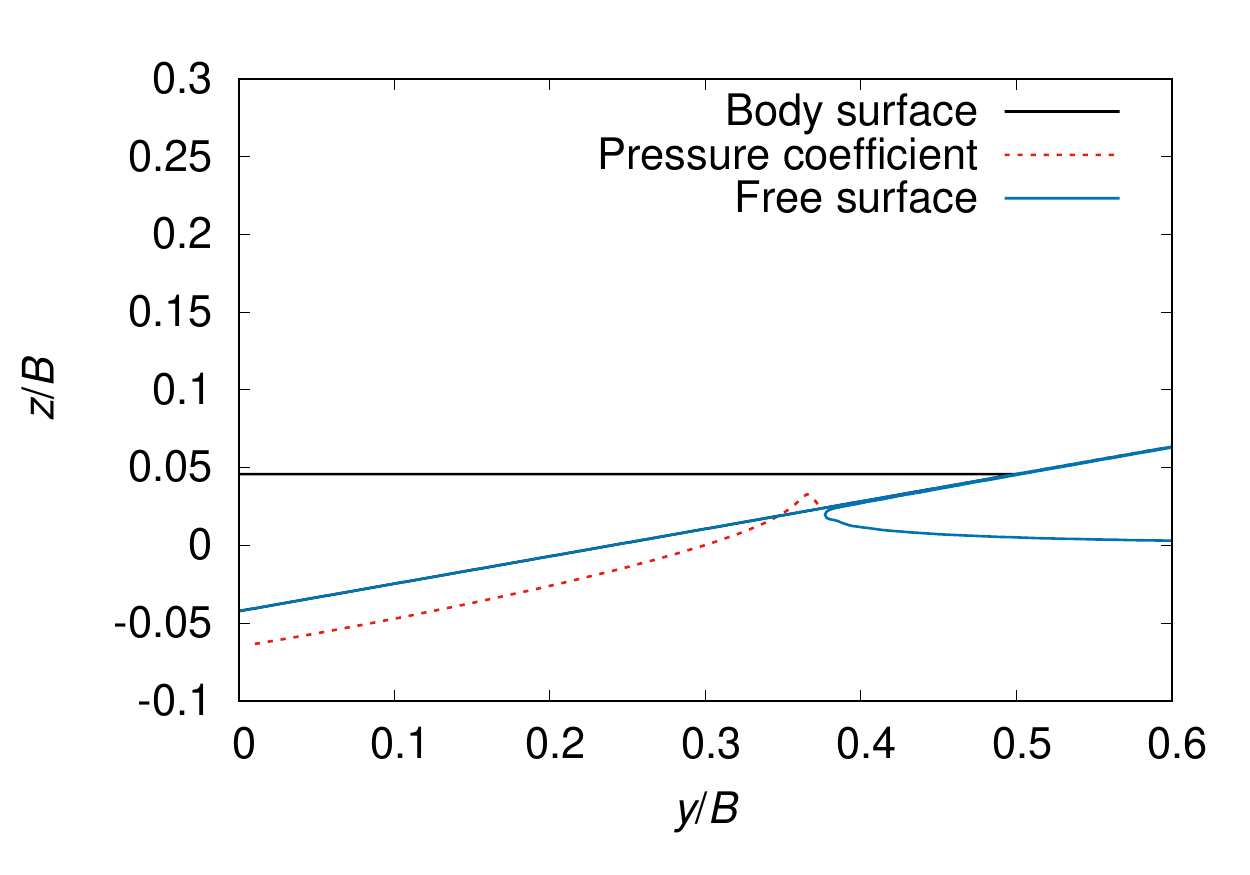} \hspace{2cm}
		\includegraphics[scale=0.5]{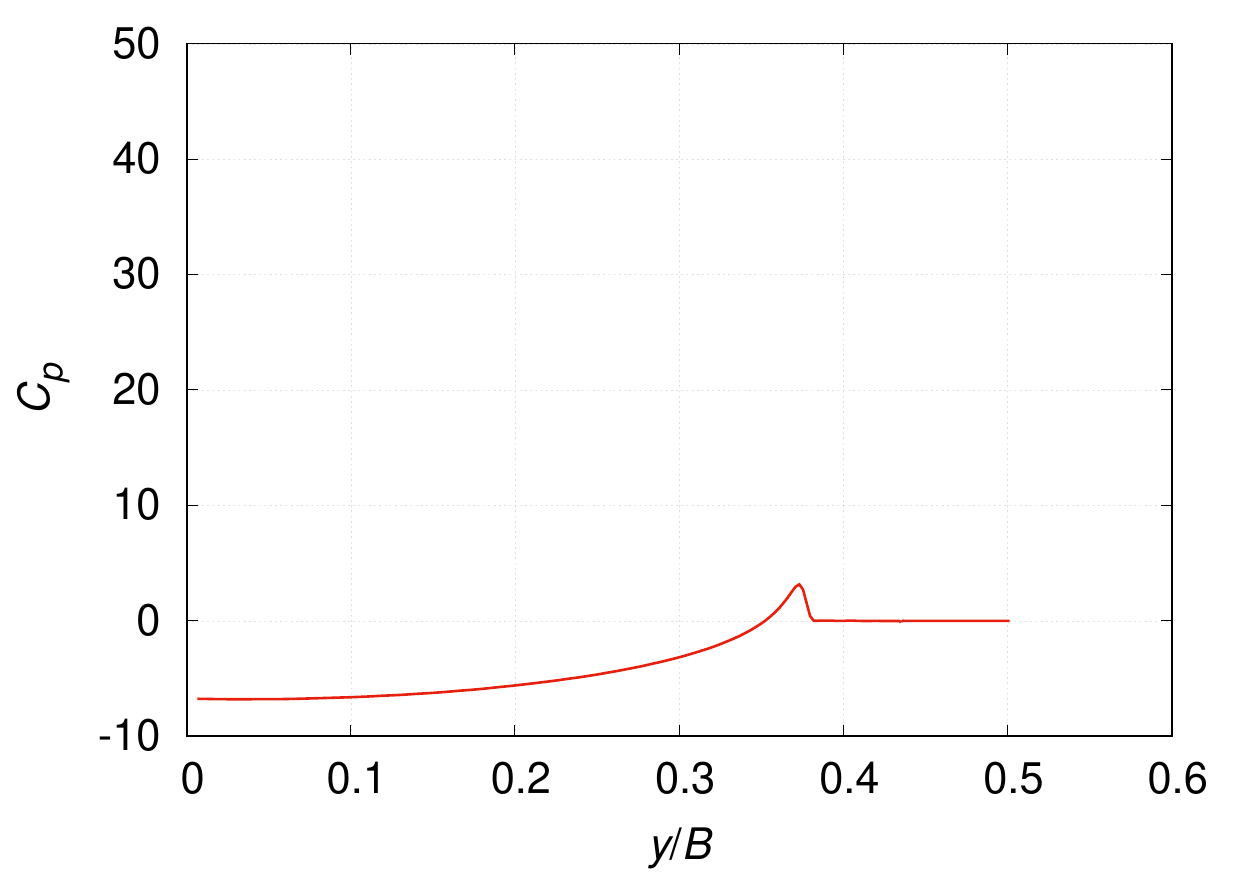}}\\
\subfigure[$V/V_0=0$]{%
	\includegraphics[scale=0.5]{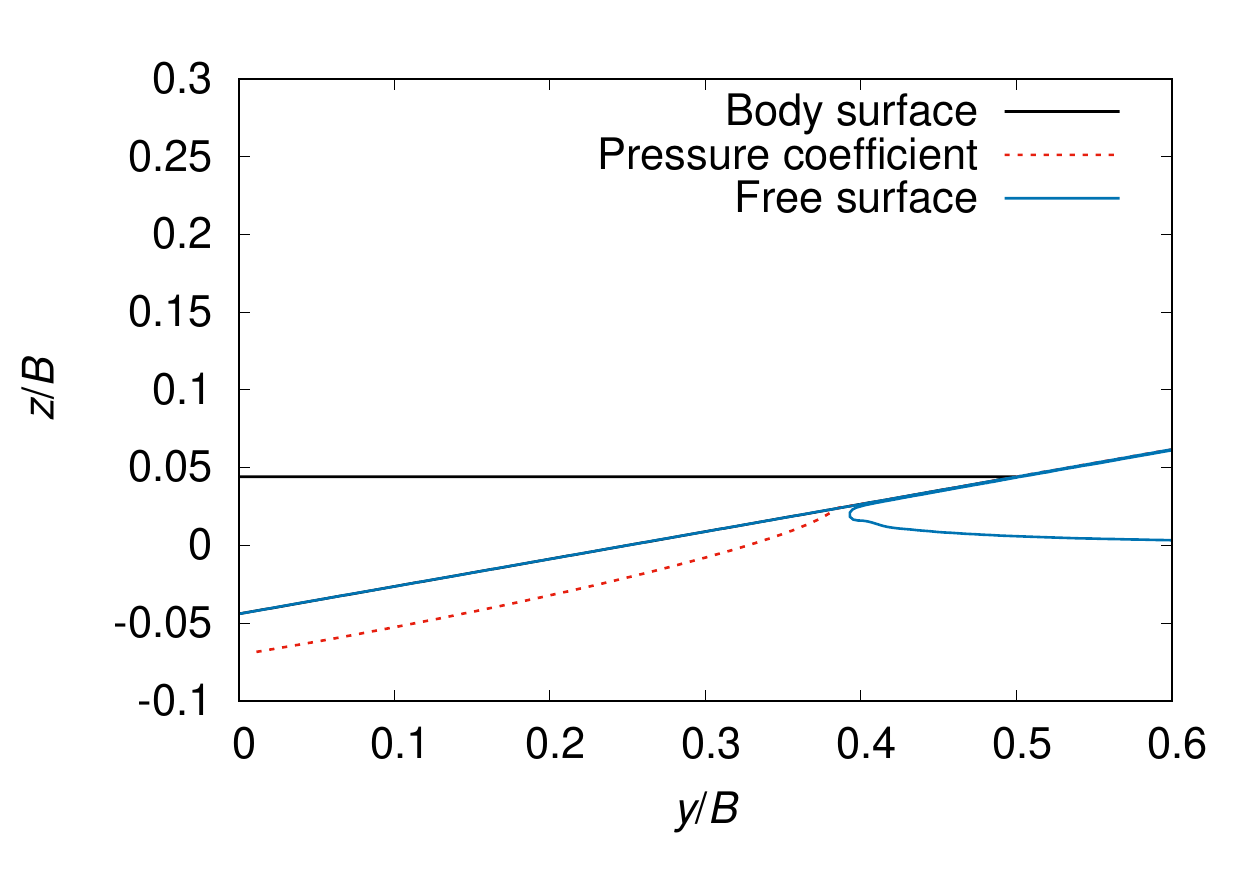} \hspace{2cm}
	\includegraphics[scale=0.5]{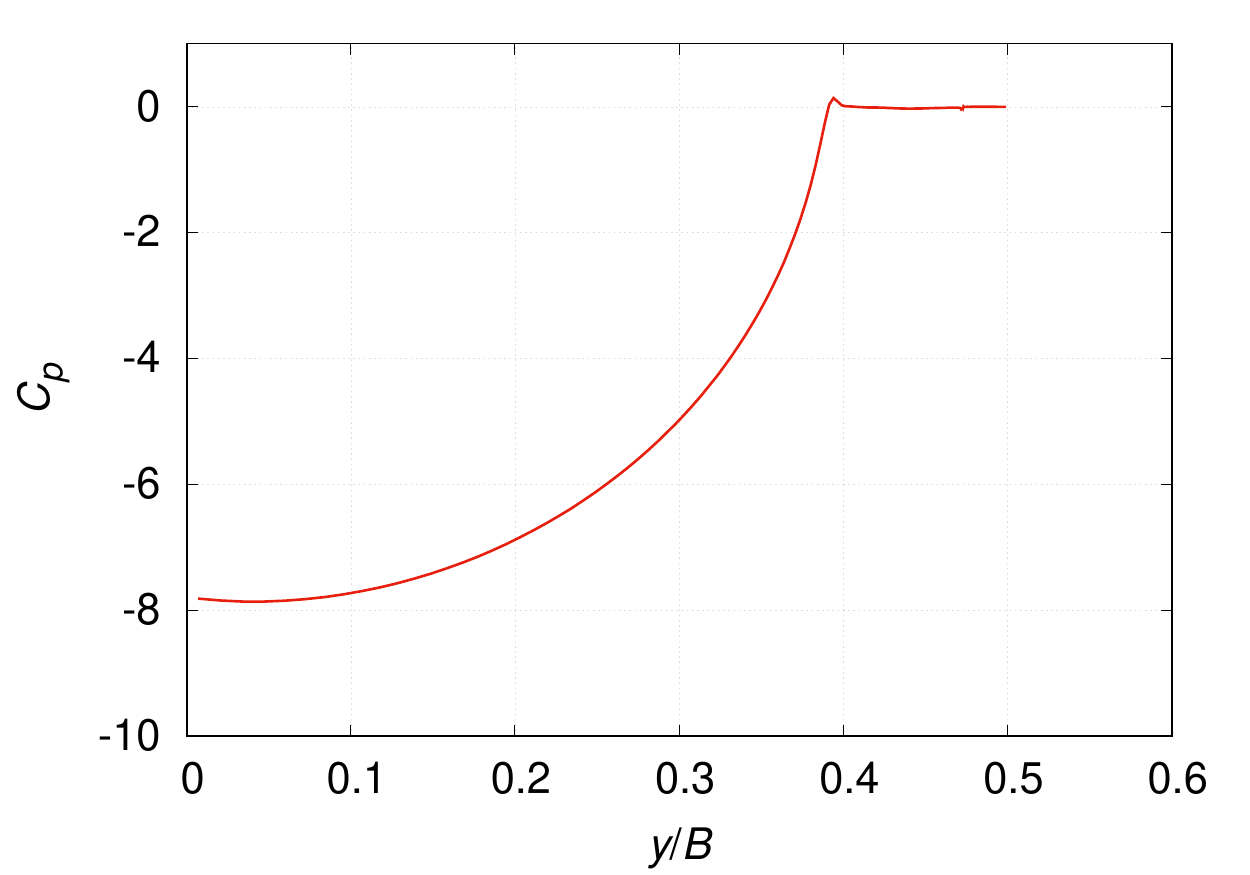}}\\
\end{figure}
\begin{figure}
\centering	
\subfigure[$V/V_0=-0.20$]{%
	\includegraphics[scale=0.5]{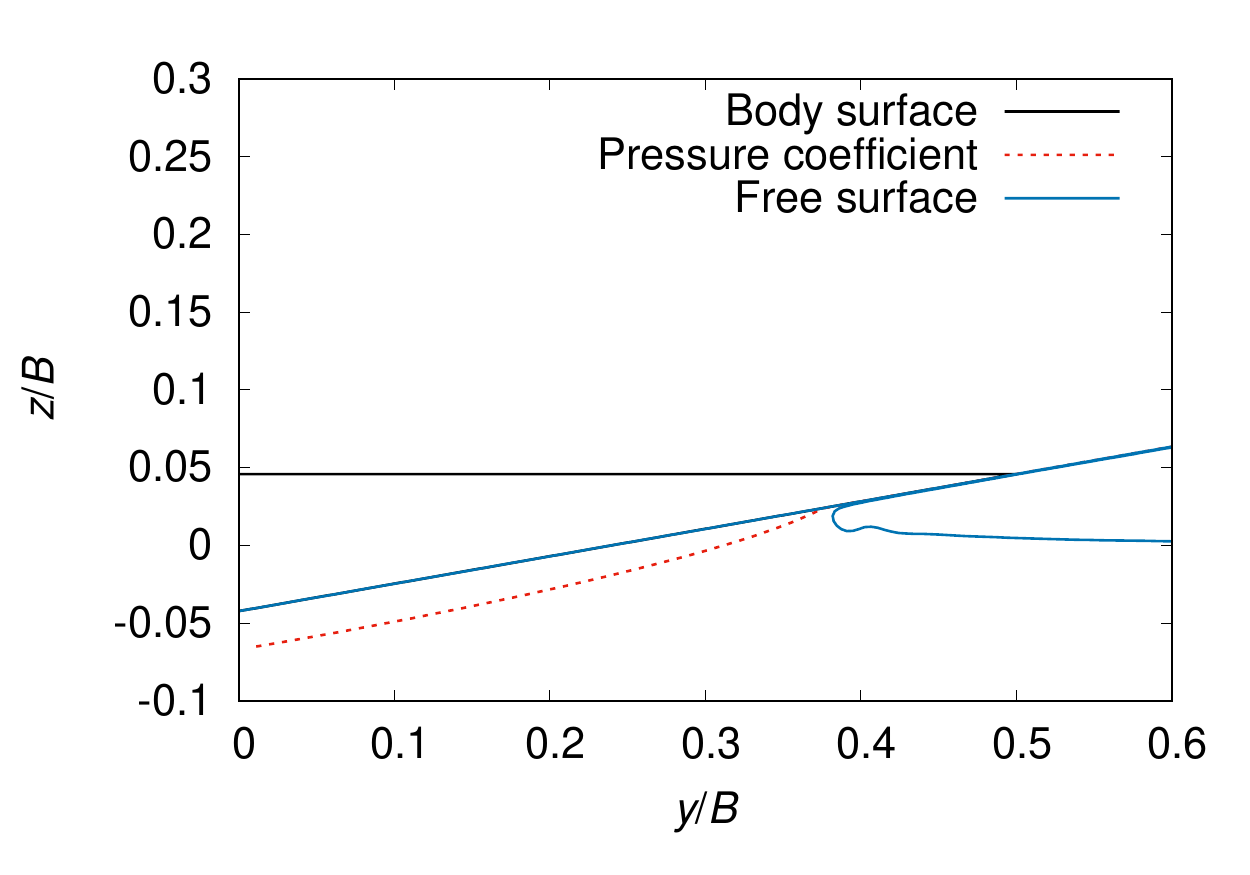} \hspace{2cm}
	\includegraphics[scale=0.5]{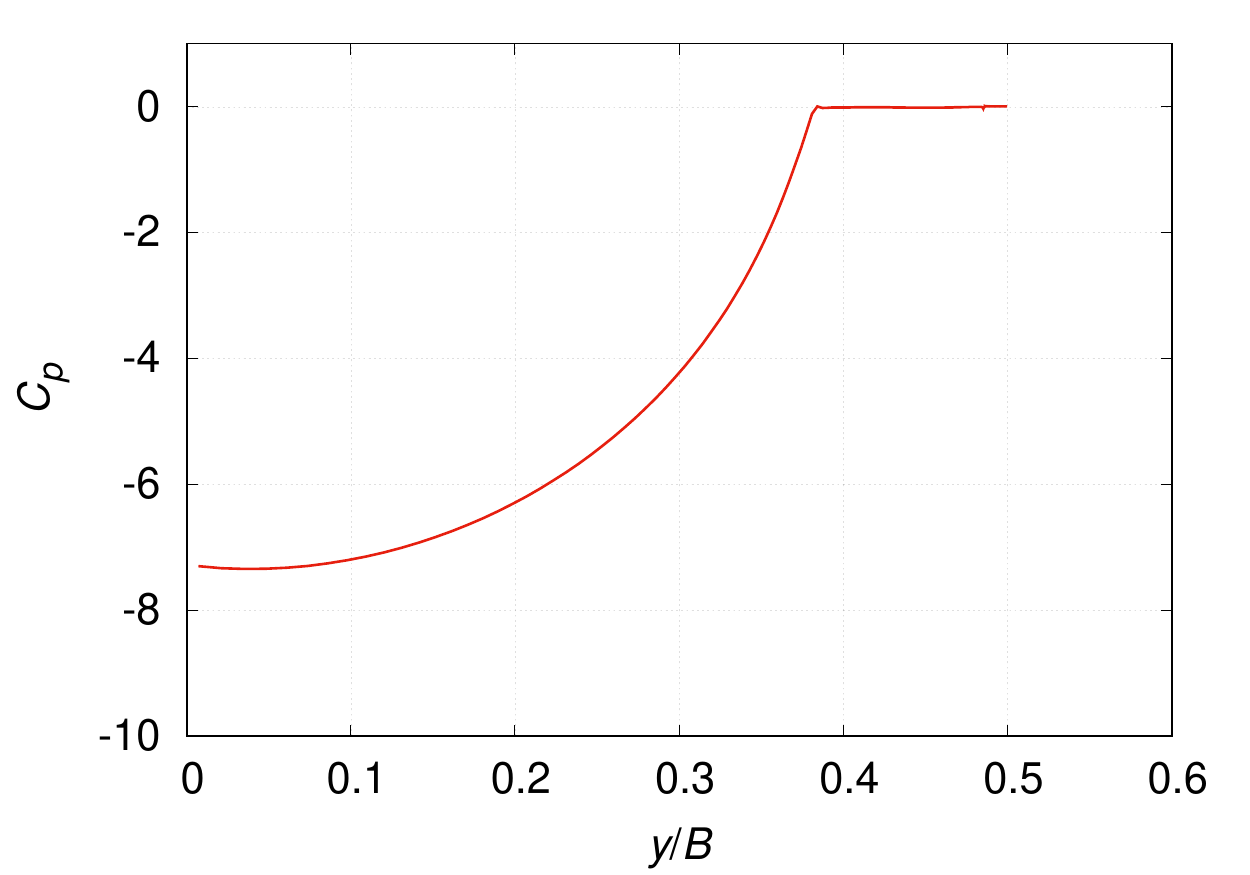}}\\
\subfigure[$V/V_0=-0.70$]{%
	\includegraphics[scale=0.5]{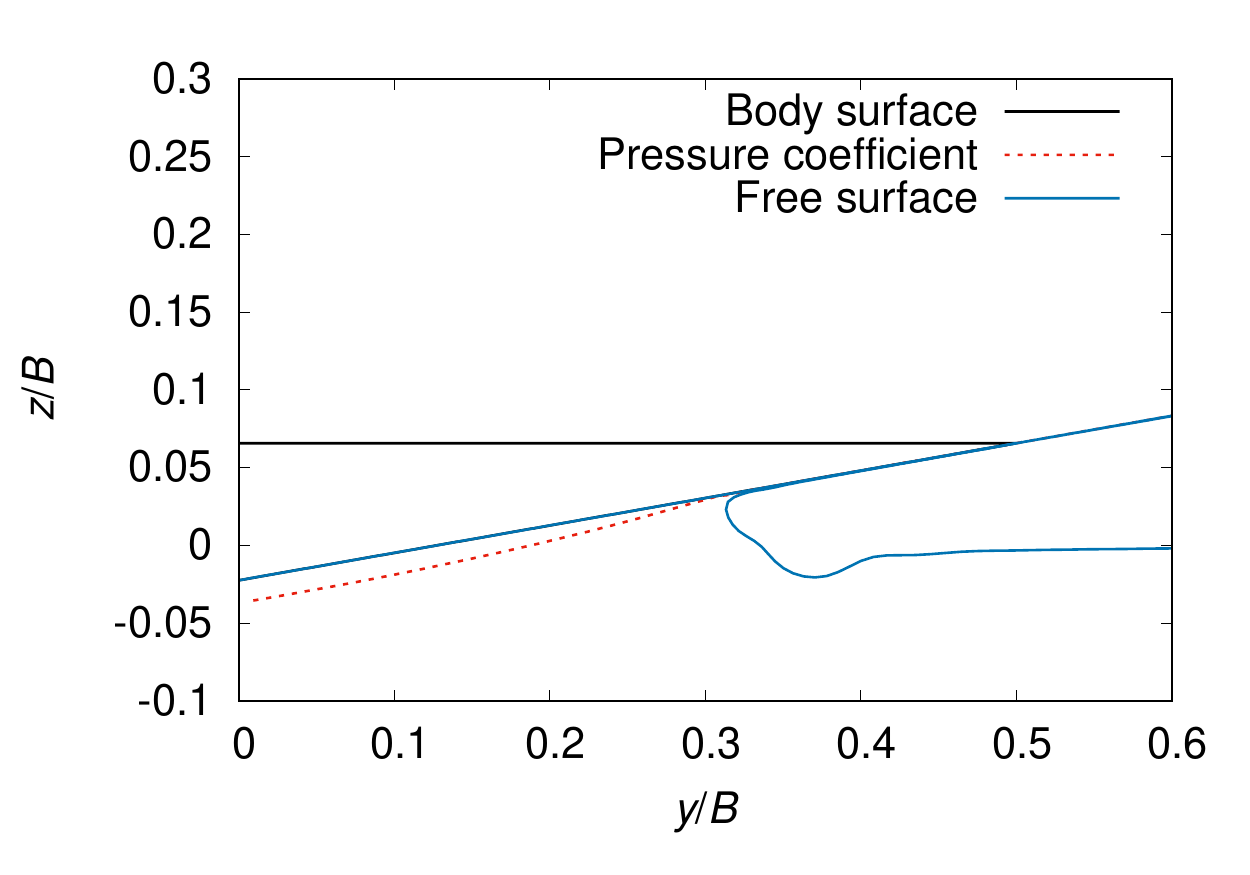} \hspace{2cm}
	\includegraphics[scale=0.5]{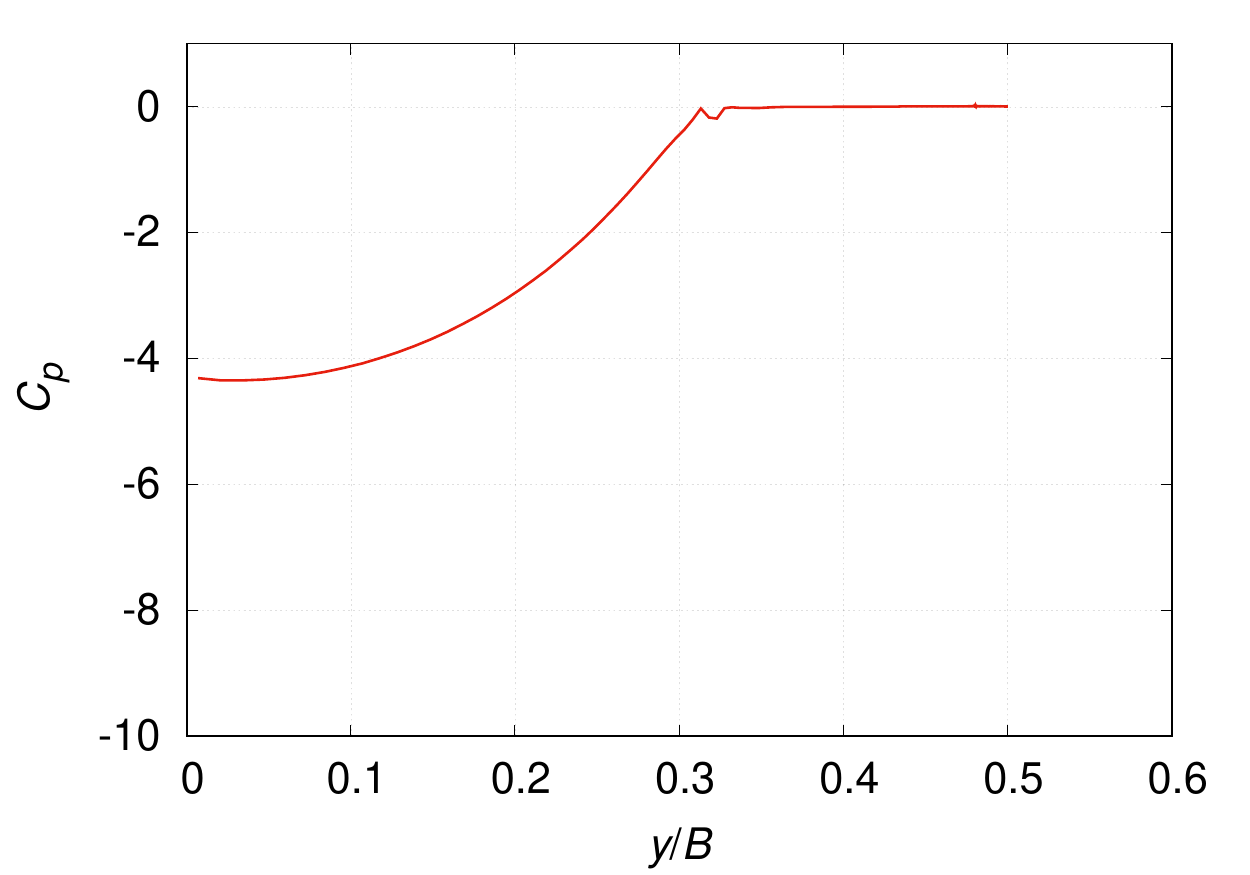}}\\
\subfigure[$V/V_0=-1.30$]{%
	\includegraphics[scale=0.5]{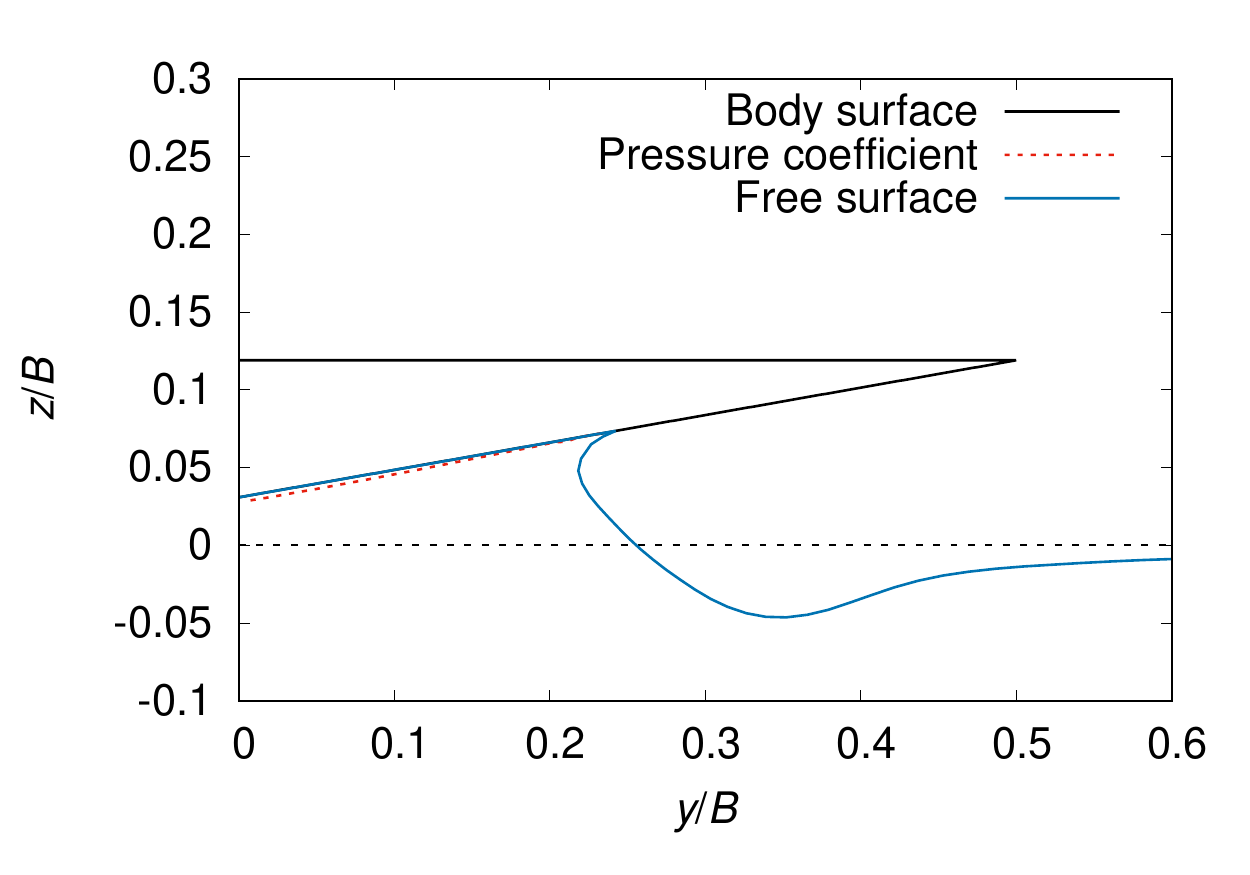} \hspace{2cm}
	\includegraphics[scale=0.5]{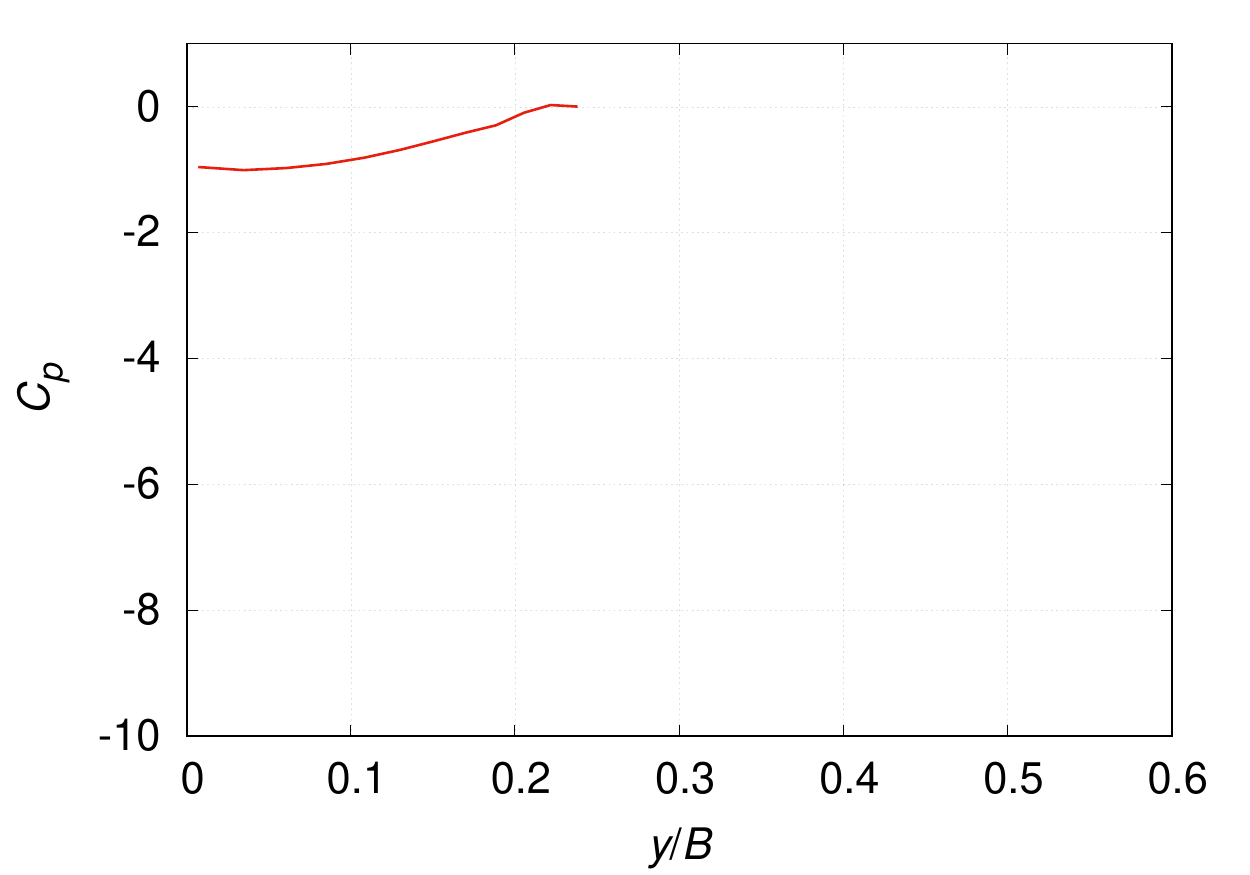}}		
	\caption{\textcolor{black}{Wedge: free-surface evolution and pressure coefficient distribution}}\label{W_fs_cp}
\end{figure}

\begin{figure}
	\centering
		\includegraphics[scale=0.6]{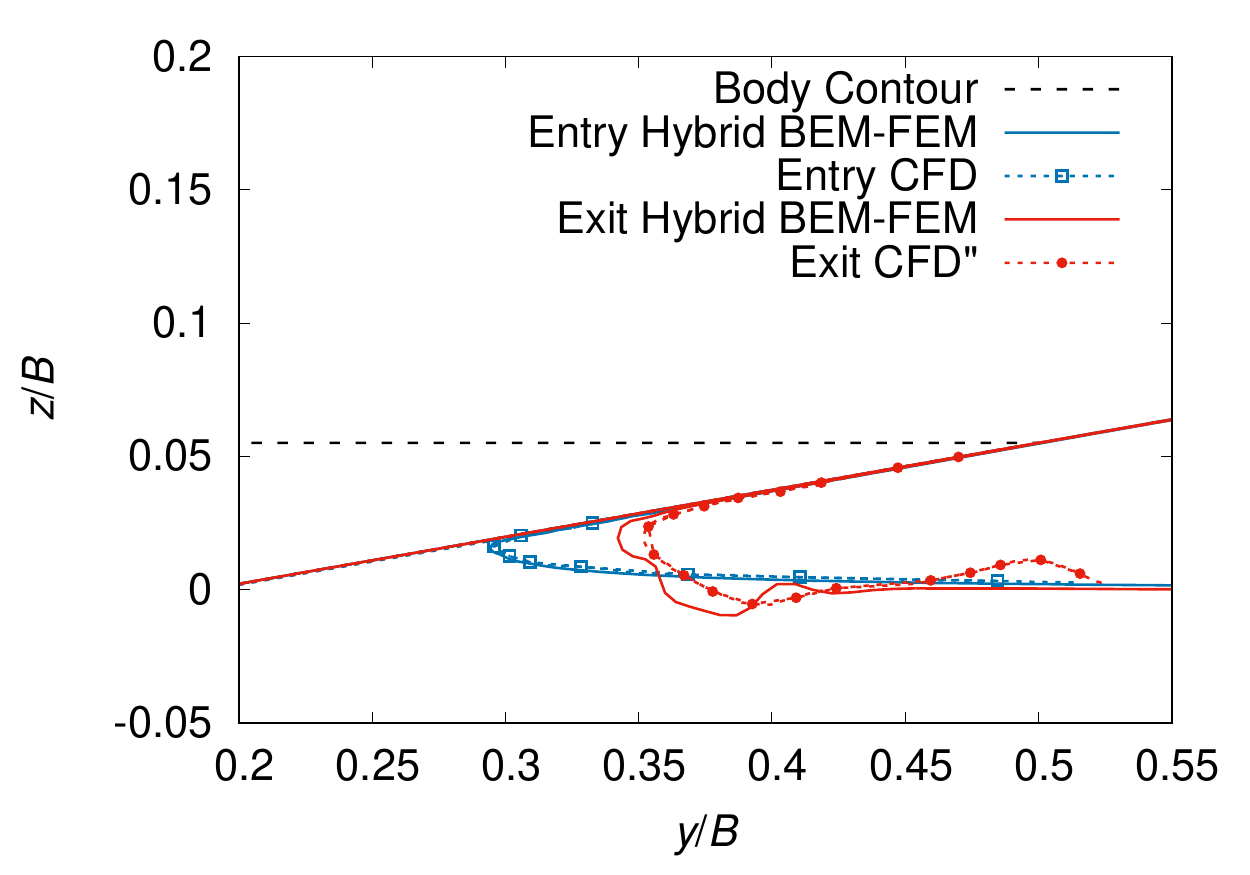}	
		\caption{Wedge: free-surface profiles obtained in the entry and exit phases at the same body position and absolute velocity. Results obtained by CFD in \cite{Maki} are also drawn for comparison.}\label{W_en_ex}
\end{figure}

\begin{figure}[h]
	\centering
	\subfigure[]{%
		\includegraphics[scale=0.5]{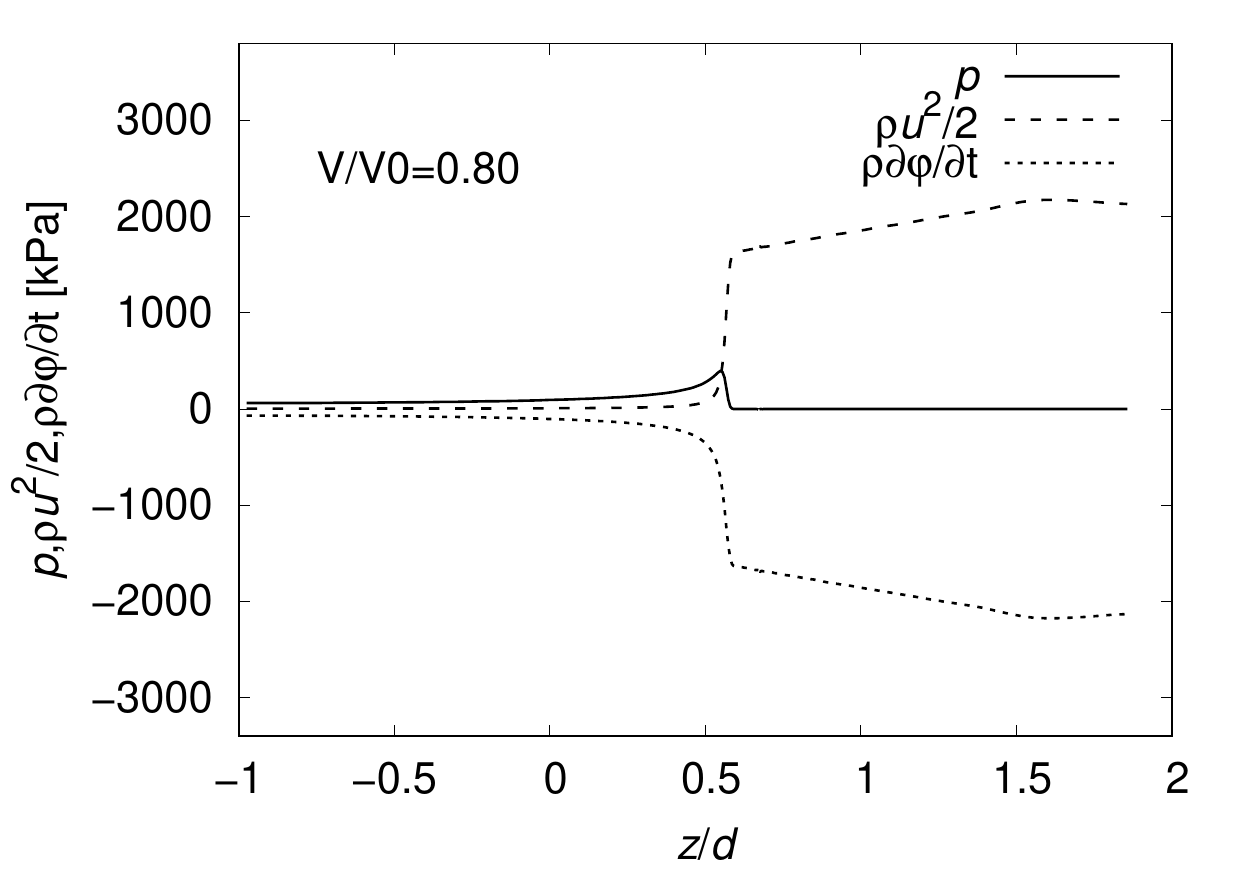}}
	\quad
	\subfigure[]{
		\includegraphics[scale=0.5]{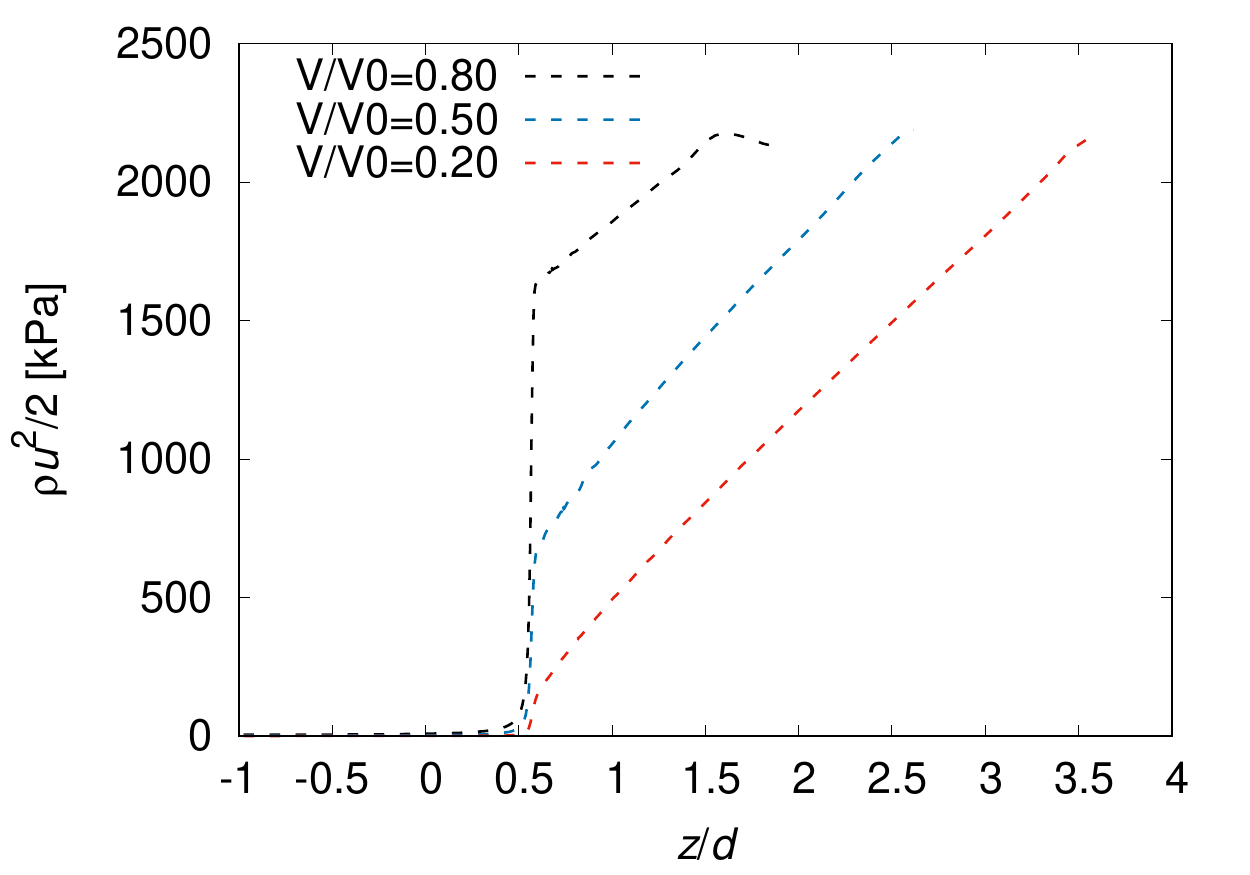}}
	\centering
	\subfigure[]{%
		\includegraphics[scale=0.5]{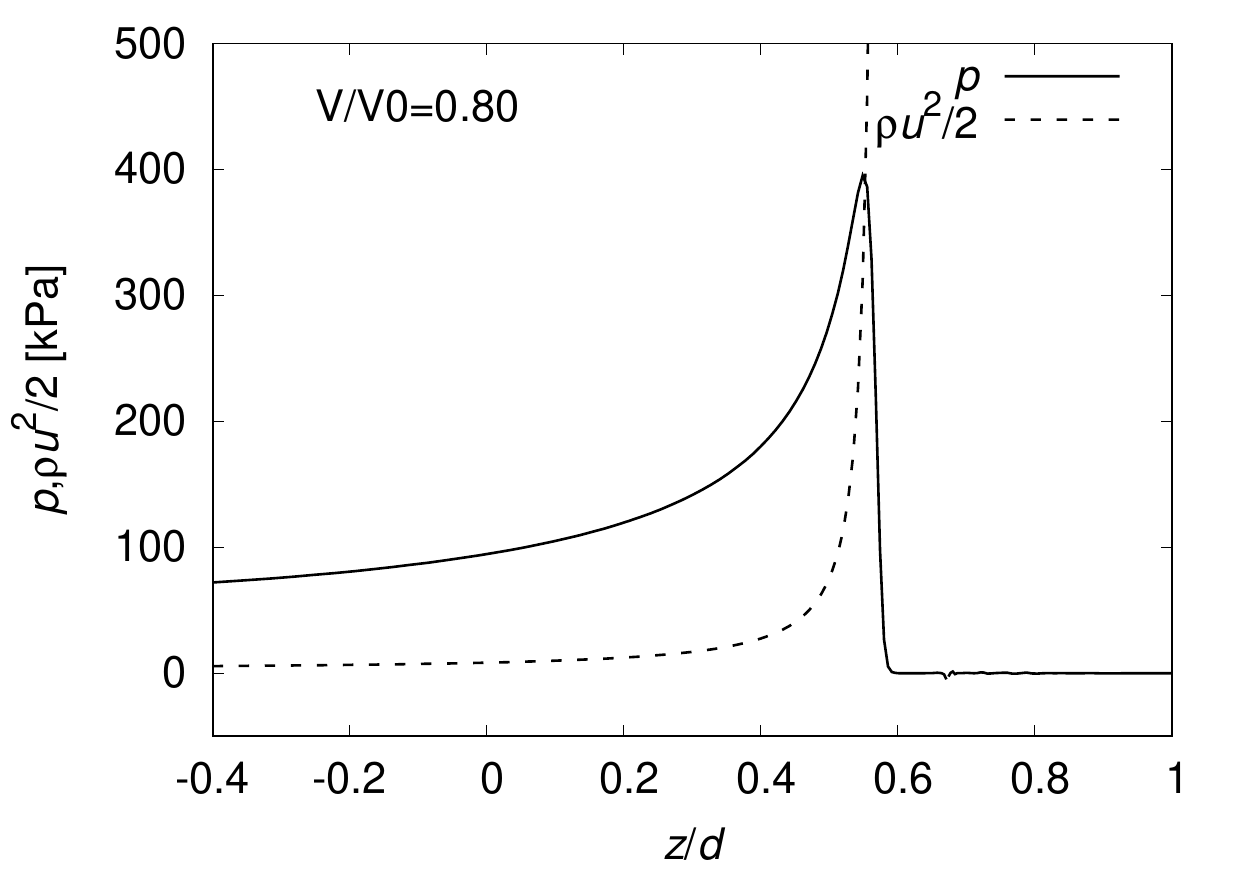}}
	\quad
	\subfigure[]{
		\includegraphics[scale=0.5]{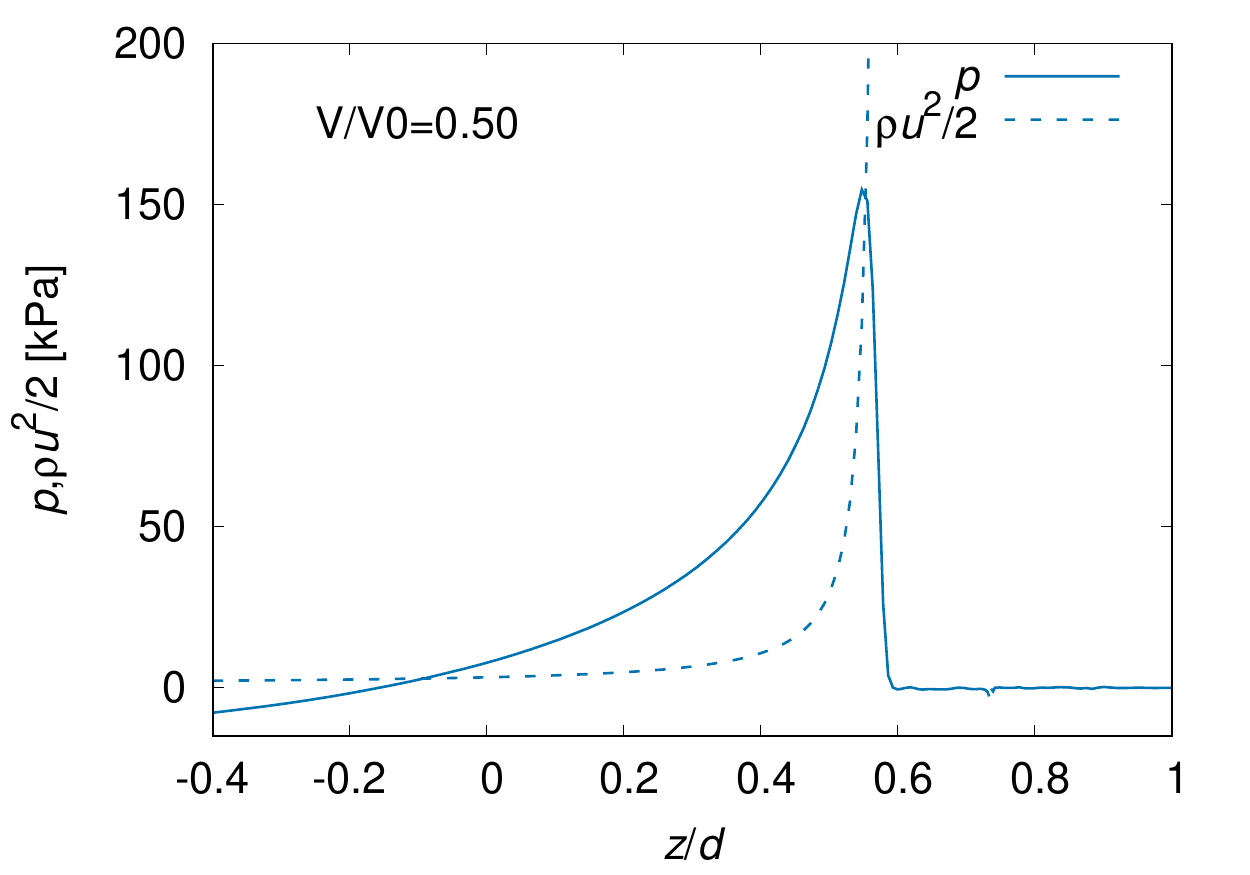}}\\
	\subfigure[]{%
		\includegraphics[scale=0.5]{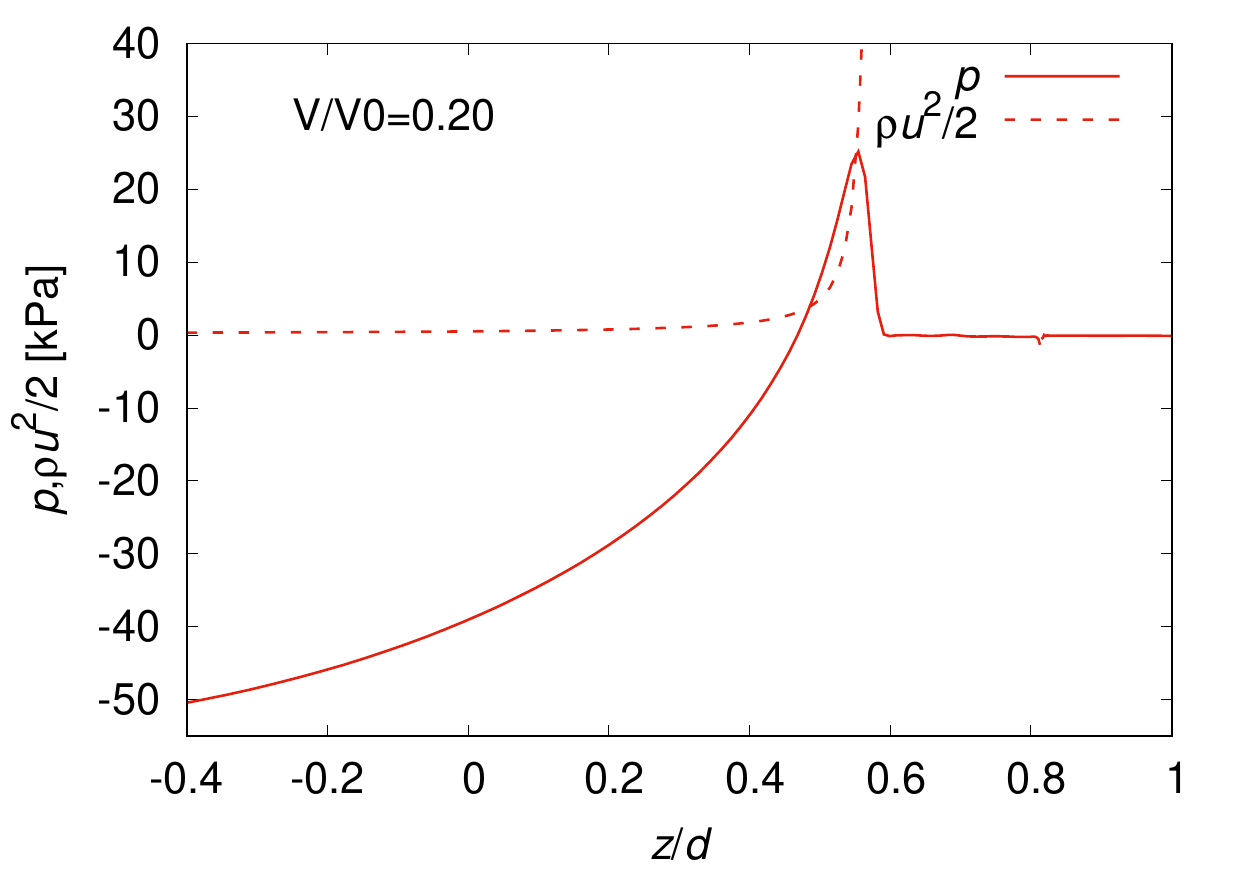}}
	\quad
	\subfigure[]{%
		\includegraphics[scale=0.5]{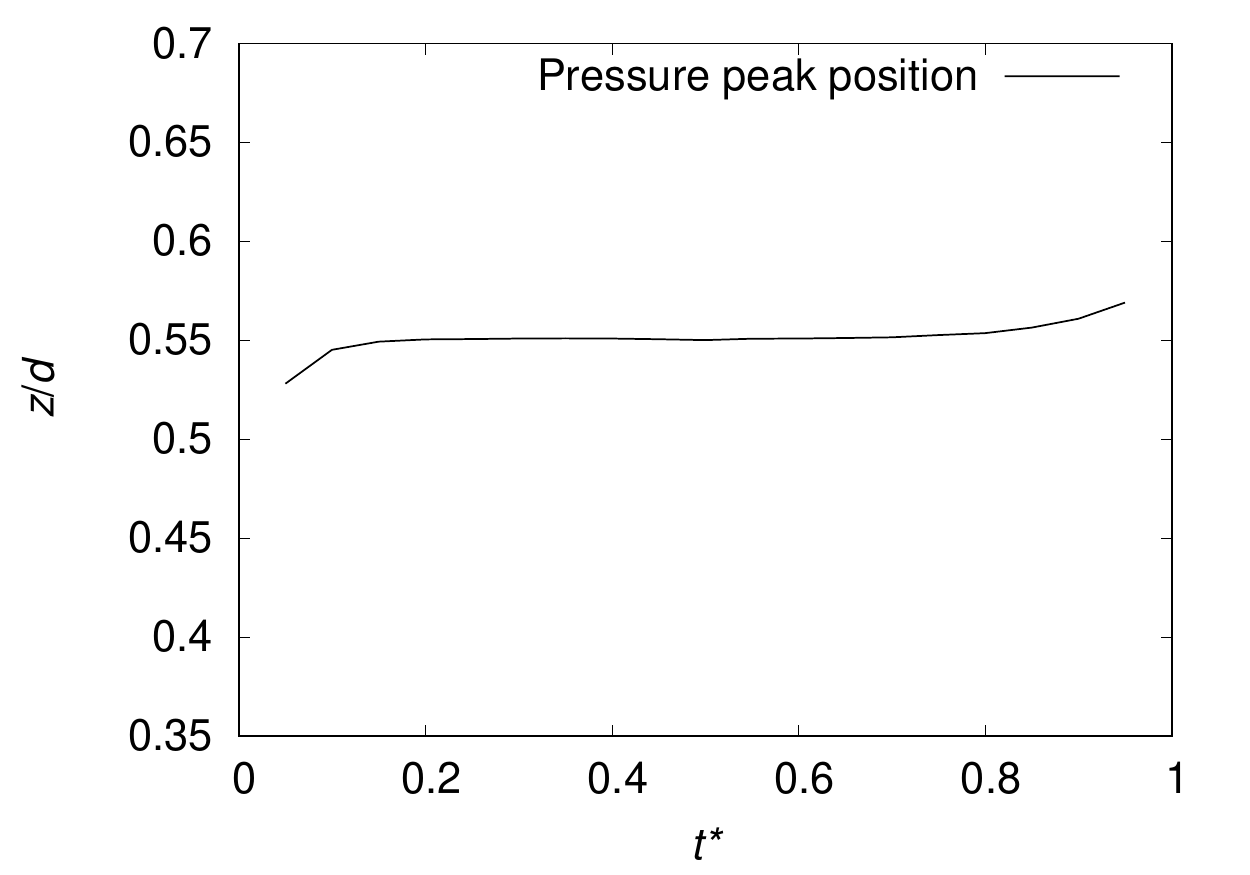}}
\caption{Wedge: pressure details. a) Comparison between the pressure, the non-linear term and the unsteady term of the Bernoulli's equation at $V/V_0=0.80$. b) Non-linear term of the Bernoulli's equation at three different time steps. c-e) Close-up view of the pressure (solid line) and the non-linear term of the Bernoulli's equation (dashed line) at $V/V_0: c=0.80, d=0.50, e=0.20$. f) Time history of the pressure peak position scaled with the current depth.}\label{W_pres}	

\end{figure}

\begin{figure}
	\centering
	\subfigure[]{%
		\includegraphics[scale=0.5]{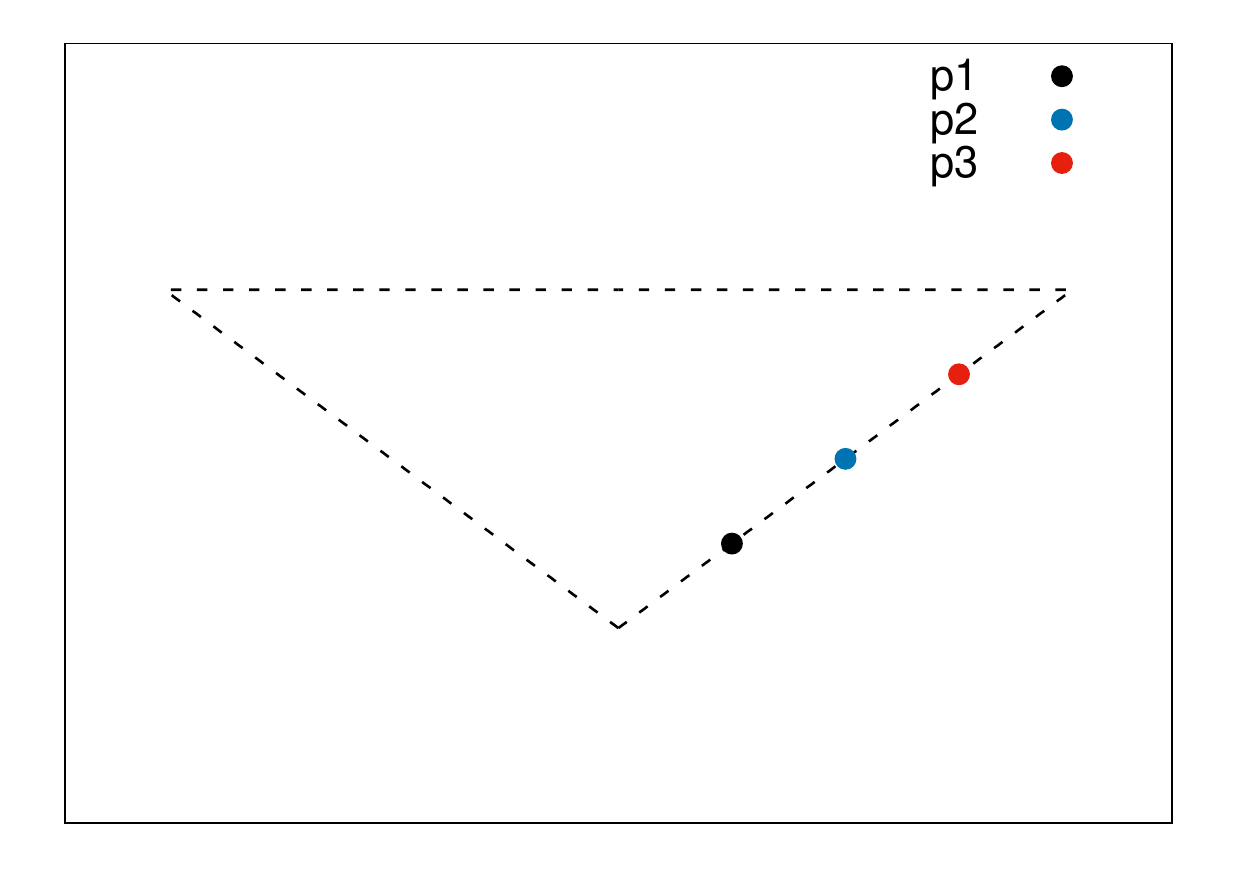}}
	\qquad
	\subfigure[]{%
		\includegraphics[scale=0.5]{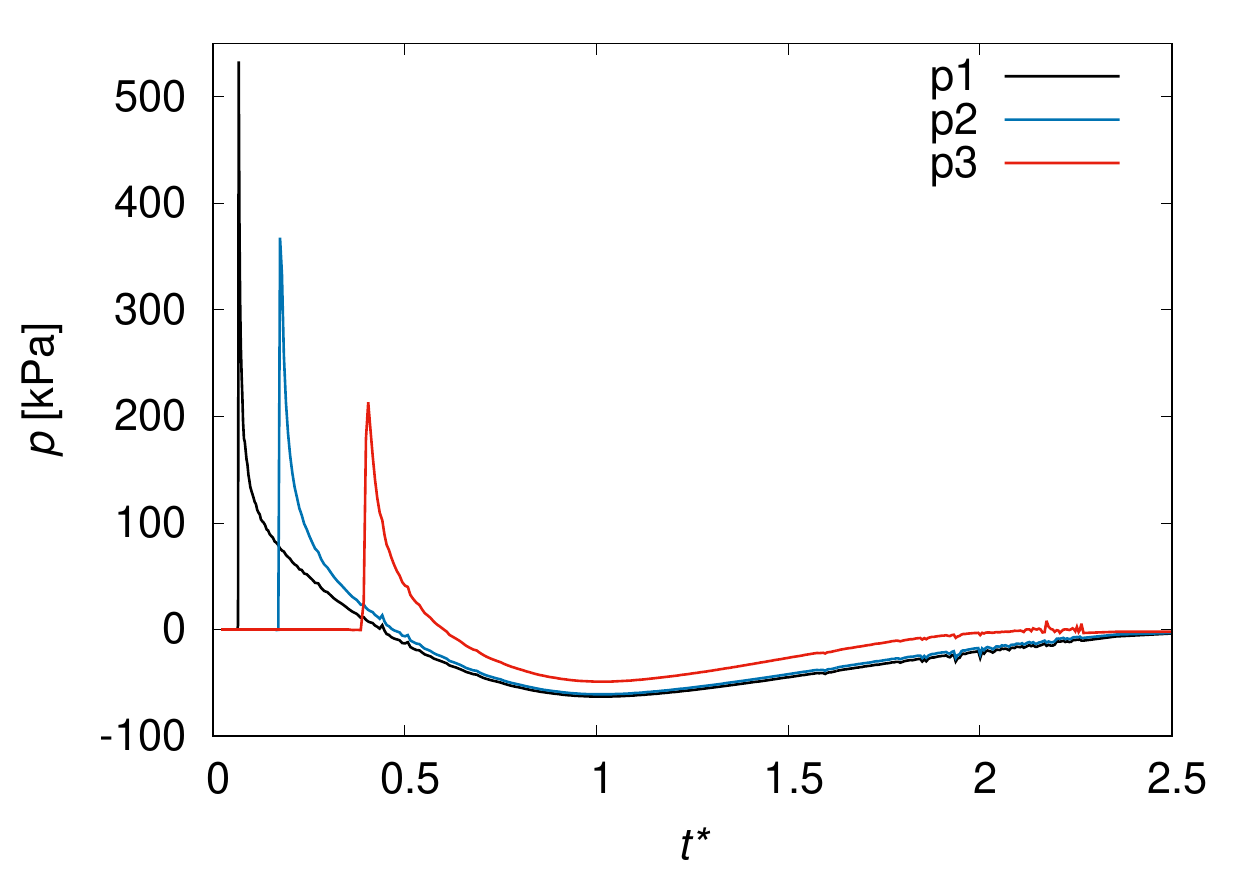}}
	\caption{Wedge: a) Pressure sensors position. b) Pressure time history for each sensor}\label{W_time}
\end{figure}

\begin{figure}
	\centering
	\includegraphics[scale=0.7]{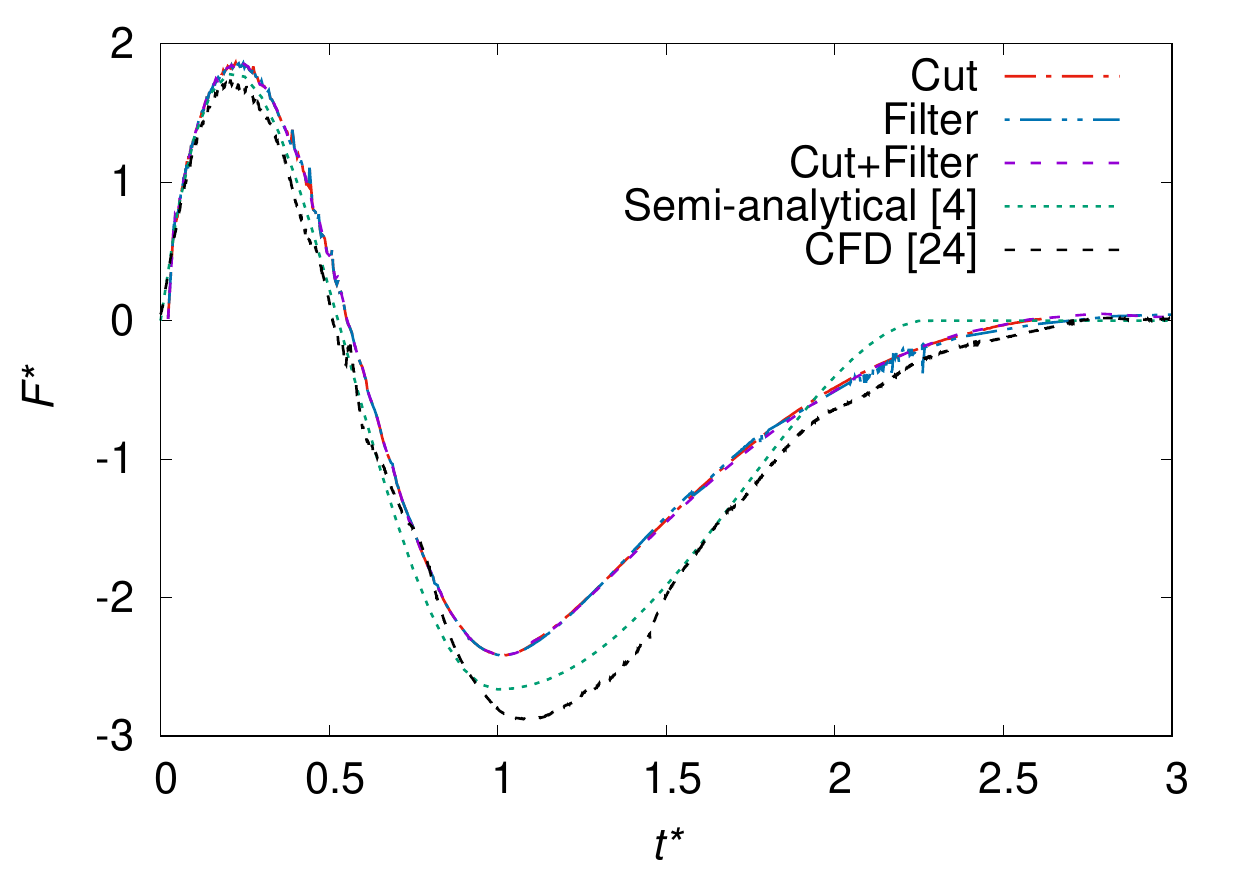}
	\caption{Wedge: non dimensional vertical hydrodynamic force time history.}\label{W_force}
\end{figure}

\noindent
Figure \ref{W_fs_cp} shows the free-surface evolution and the pressure coefficient 
distribution at different time steps, which is expressed in terms of the ratio between 
the current velocity $V$ and the initial velocity $V_0$. During the entry phase, when 
the body is going downwards but the velocity is decreasing, the flow rises along the 
body contour and a thin jet develops and propagates along the body surface. The 
pressure distribution resembles that in the water entry at constant velocity, presented 
in\cite{Battistin2004,Zao1993}, but the pressure peak, occurring behind the jet root, 
progressively diminishes. The same happens to a lesser extent also to the whole 
pressure distribution, which becomes negative over a large portion of the wetted 
surface. In the exit phase, when the body is moving upwards and the velocity grows in 
amplitude, the flow moves in the opposite direction and the pressure remains negative 
but diminishes in amplitude. At $t^*=2$ the body is completely above the still water 
level and the exit velocity is equal in amplitude to the initial velocity. However, the 
fluid follows the body and a portion of the body contour is still wet. The exit phase 
continues for a while until the fluid leaves the body completely. For both phases, the 
pressure inside the thin jet is essentially negligible. Figure \ref{W_en_ex} shows the 
free-surface shape, in proximity of the jet region, obtained in the entry and exit 
phases at the same body position and for the same velocity absolute value. It can be 
seen that in the entry phase the wetted length increases faster than it shrinks in the 
exit phase and this behavior is consistent with what was observed in the CFD results 
\cite{Maki}.
It is worth noting that the results displayed in Figs. \ref{W_fs_cp} and \ref{W_en_ex} 
are obtained by using the filter strategy, but essentially similar results are obtained 
when using the jet cut off procedure. Figures \ref{W_pres} and \ref{W_time} provides 
further details on the pressure behavior, in particular in the entry phase. 
Figure \ref{W_pres}a shows the pressure distribution together with the 
unsteady term ($\rho \partial \varphi/\partial t$) and the non-linear term of the 
Bernoulli's equation, ($\rho u^2/2$). Both terms exhibit a sharp growth approaching the 
spray root, and thus the pressure peak, and a sudden change to a much lower growth rate 
inside the jet. When reducing the speed, the sharp growth is limited and then the 
transition to the lower growth rate in the jet appears much milder, as shown in the 
figure \ref{W_pres}b for the non-linear contribution of the pressure. Inside the jet 
the two contributions balance each other, thus resulting in a zero pressure. Figure 
\ref{W_pres}c shows that, close to the peak, the pressure coincides with the non-linear 
term of the Bernoulli's equation, as it happens for the self-similar solution of the 
case with constant velocity \cite{Zao1993}. Although the solution is no longer 
self-similar, and the pressure peak decreases, such behavior remains valid when the 
entry velocity decreases (Fig. \ref{W_pres}d-e). Furthermore, the pressure peak 
position, in terms of ratio between the z-coordinate and the current depth $d$, is 
almost constant over time (Fig.\ref{W_pres}f). 
In order to better investigate this point, the pressure time histories recorded at three 
probes are analyzed (Figs. \ref{W_time}a,b). As the jet root approaches the probe, the 
pressure rises quite sharply and reaches the peak. Due to the decrease of the entry 
velocity, the pressure peak decreases over time. For each probe, the pressure decreases 
and becomes negative, during the entry phase ($t^*<1$). The pressure remains negative 
during the exit phase ($t^*>1$) but decreases in amplitude until approaching zero. 
Figure \ref{W_force} shows the time history of the vertical hydrodynamic force which is 
non-dimensionalized using the initial velocity $V_0$, the width of the wedge $B$ and the 
water density $\rho$, $F^*=F/(\rho V_0^2B)$. The force increases and is positive in the 
first part of the entry stage. It decreases and turns negative afterwards; indeed, the 
positive contribution of the "slamming term", which is proportional to the velocity, 
decreases as the velocity decreases, whereas the effect of the acceleration, which 
introduces an additional added-mass effect, negative contribution to the force, 
increases as the wetted area increases. The force remains negative, approaching zero as 
the body exits from the water. The negative peak in the hydrodynamic loading occurs at 
the transition between the entry and exit phases, about $t^*\simeq 1$, and it is greater 
than the positive one occurring during the entry phase. The use of different strategies 
during the exit phase, also in combination, does not affect the results and a quite 
satisfactory agreement with the CFD \cite{Maki} and semi-analytical 
\cite{Tassin} results available in the literature is achieved. 
Furthermore, for $t^*>2$ the force approaches zero slowly, highlighting the capabilities 
of the numerical model to follow the exit phase until the body has completely emerged, 
as it happens in the CFD simulations, whereas the semi-analytical model approaches zero 
earlier showing its limitations in describing the latest stage of the exit phase when 
the body is above the still water level but the flow is still attached.
\subsection{\textbf{Cone}}
\begin{figure}[h]
	\centering
	\subfigure[$V/V_0=-0.40$]{%
		\includegraphics[scale=0.5]{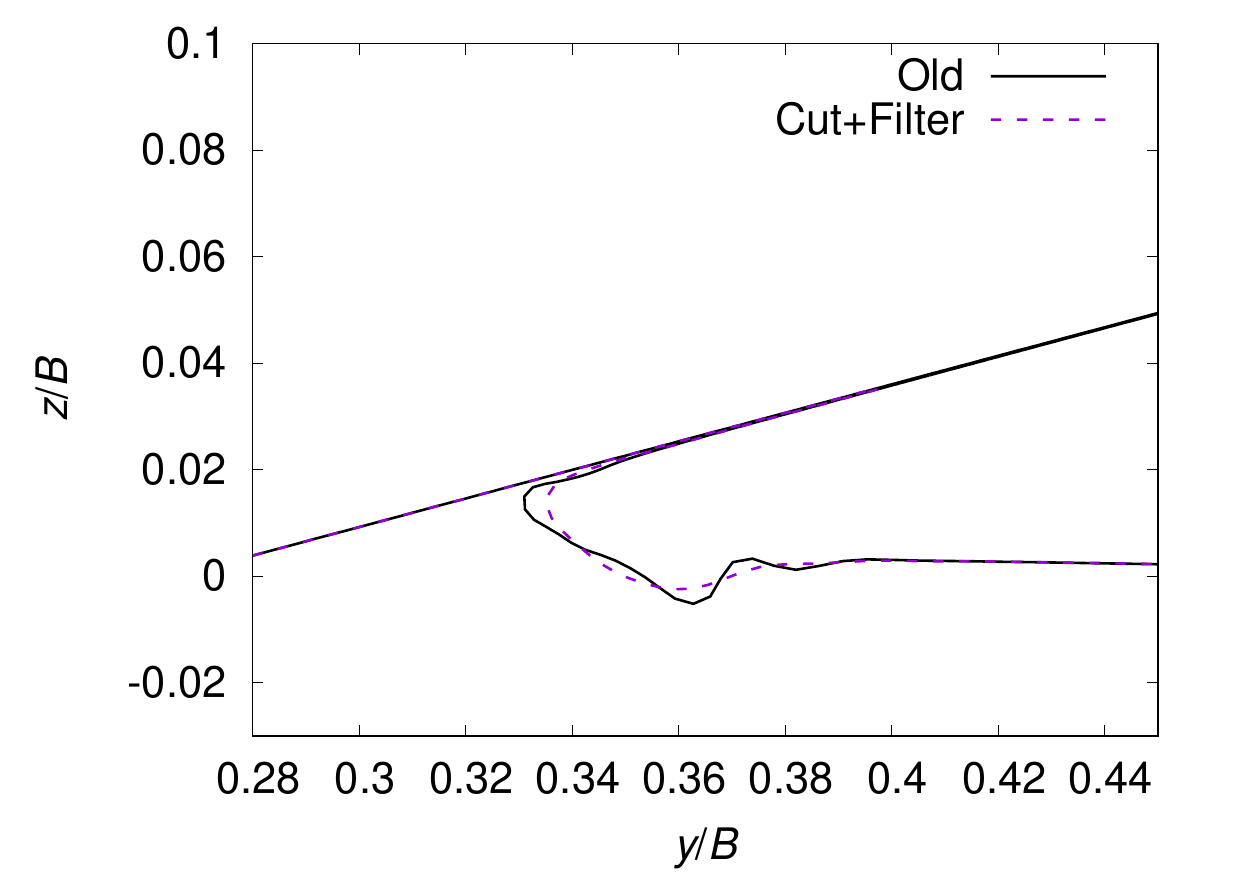}}
	\hfil
	\subfigure[$V/V_0=-0.50$]{%
		\includegraphics[scale=0.5]{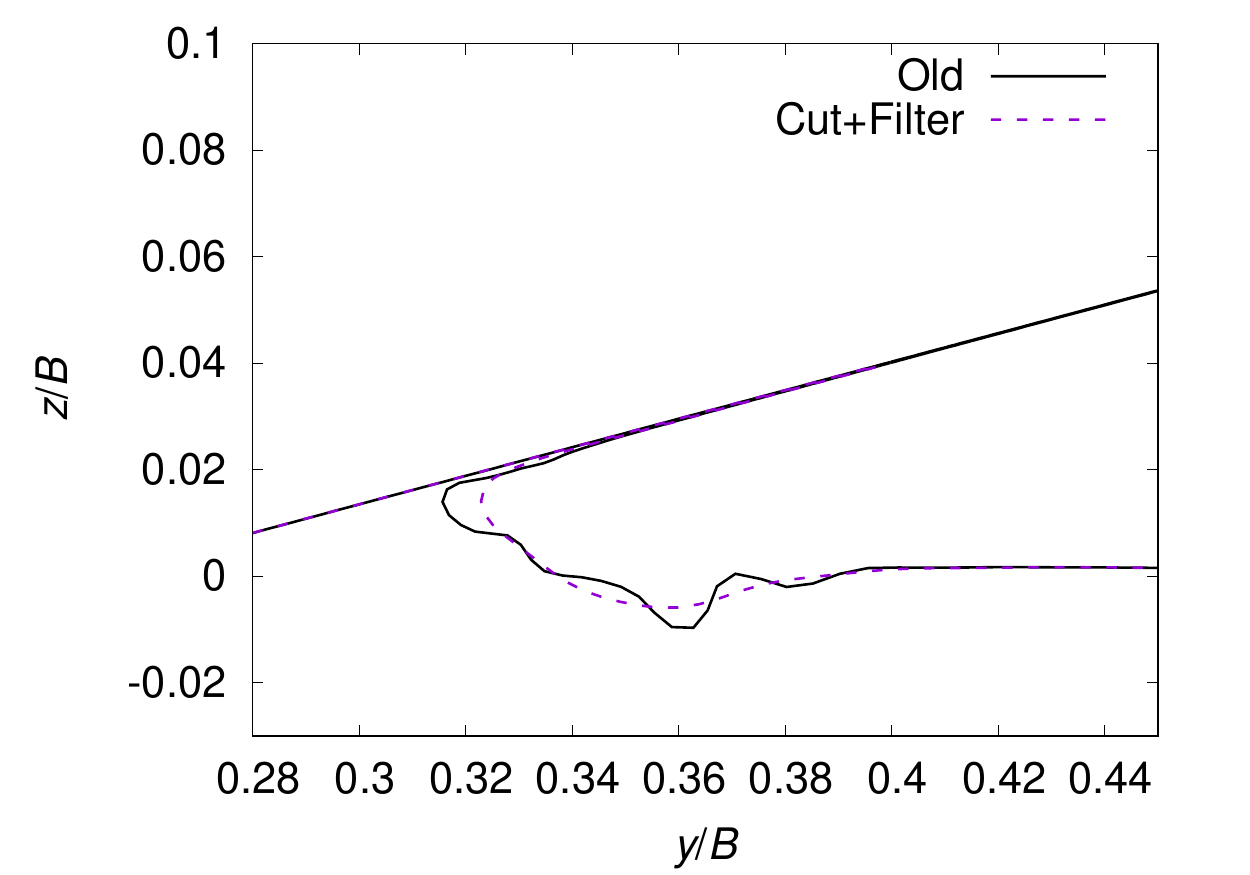}}
	\hfil
	\subfigure[$V/V_0=-0.60$]{%
		\includegraphics[scale=0.5]{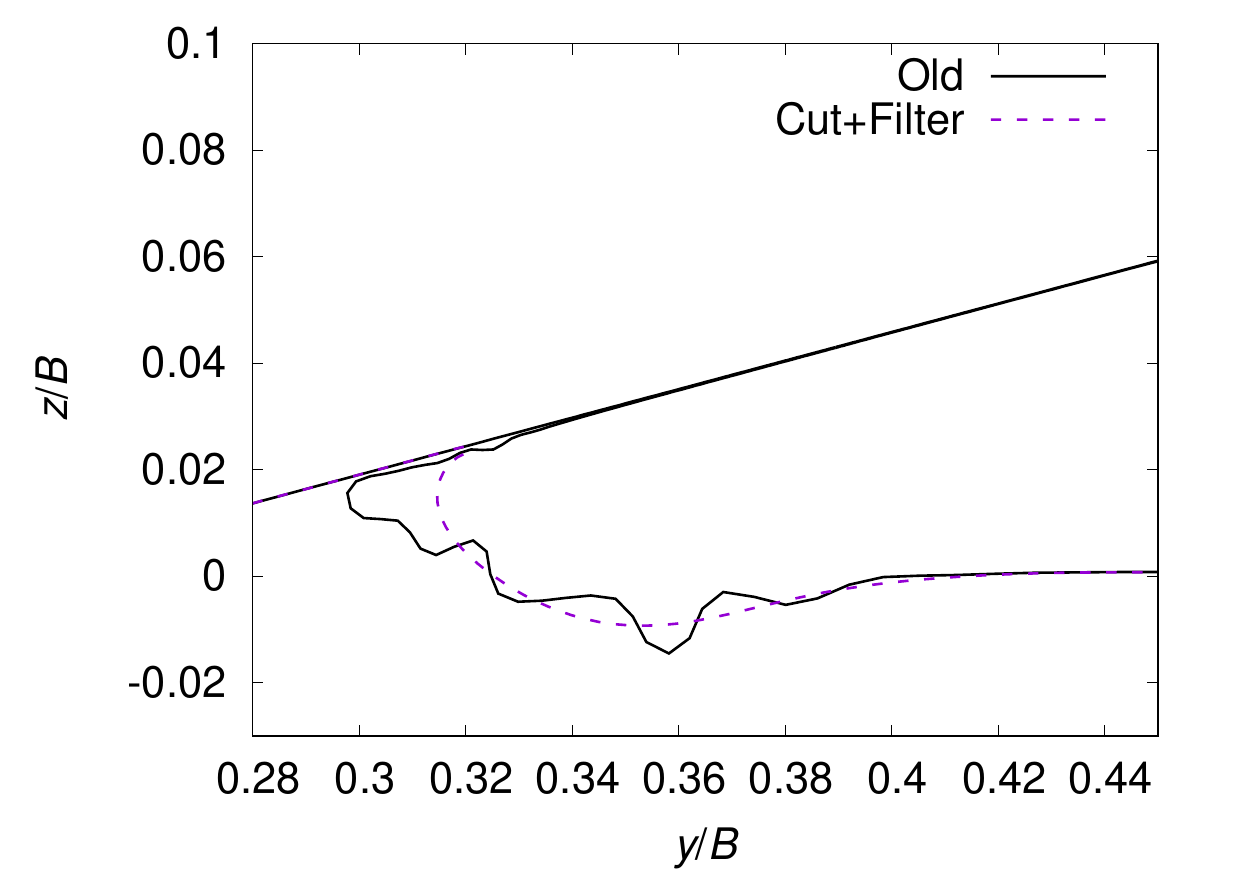}}
	\caption{Cone: close-up view of the free-surface shape in the jet root region at three different time steps during the exit phase.}\label{c_exit}
\end{figure}

\begin{figure}
	\centering	
	\subfigure[$V/V_0=-0.10$]{%
		\includegraphics[scale=0.5]{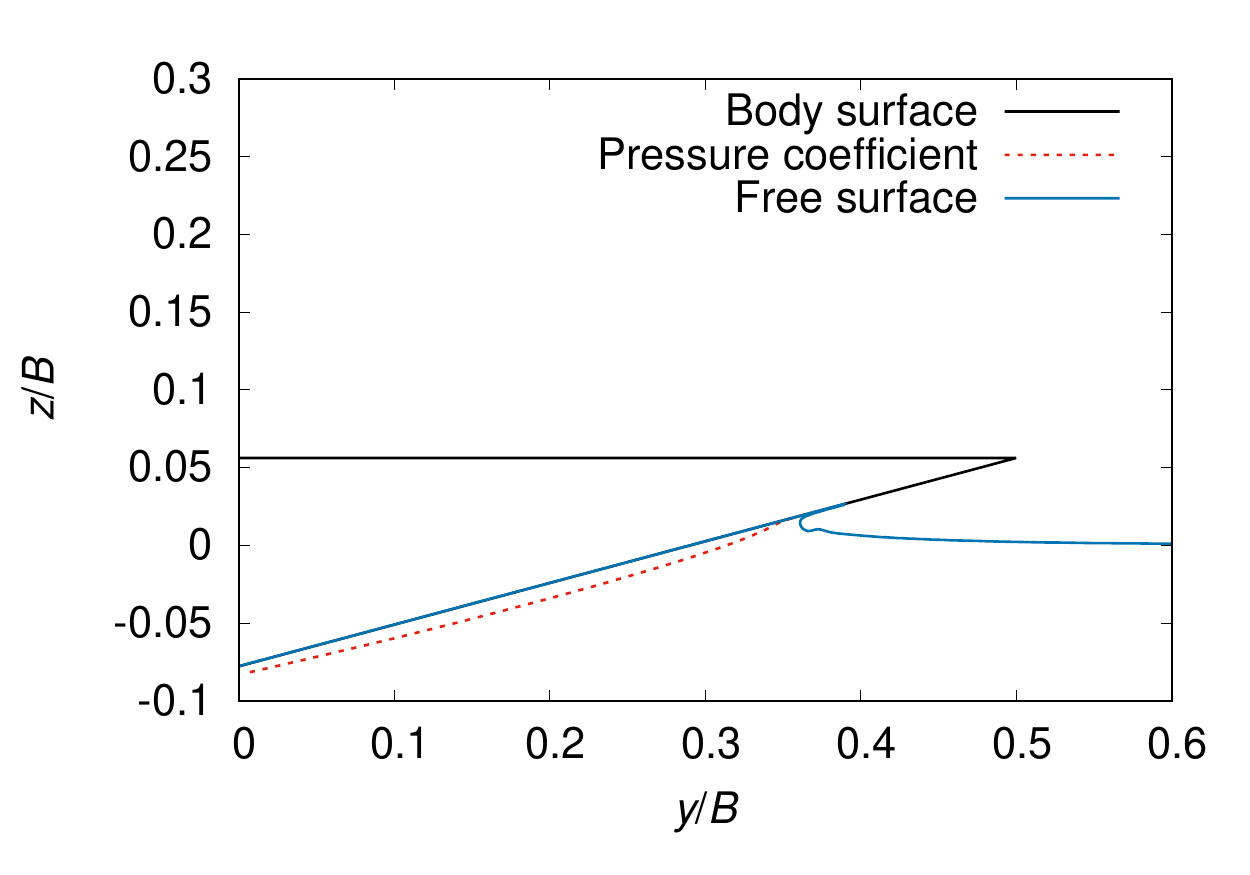} \hspace{2cm}
		\includegraphics[scale=0.5]{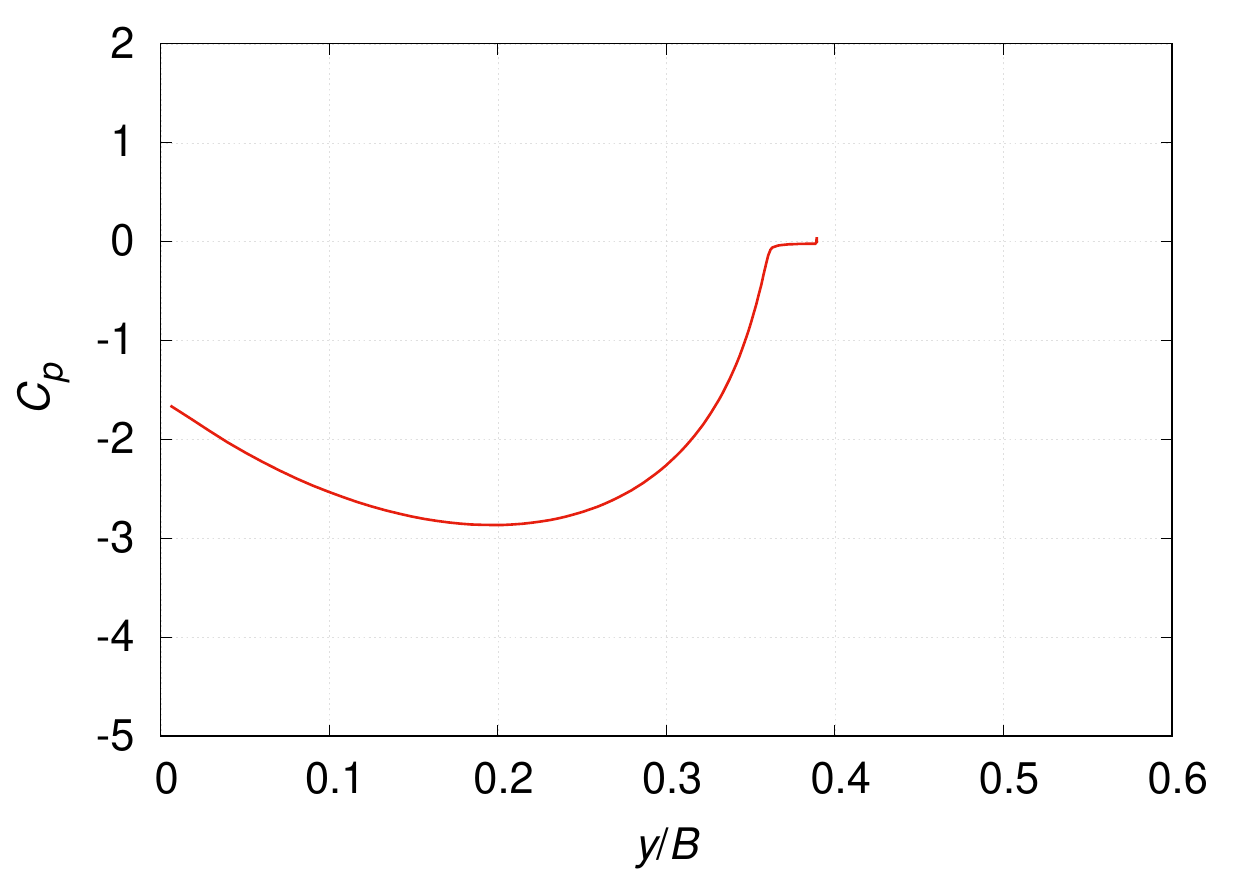}}\\
	\subfigure[$V/V_0=-0.50$]{%
		\includegraphics[scale=0.5]{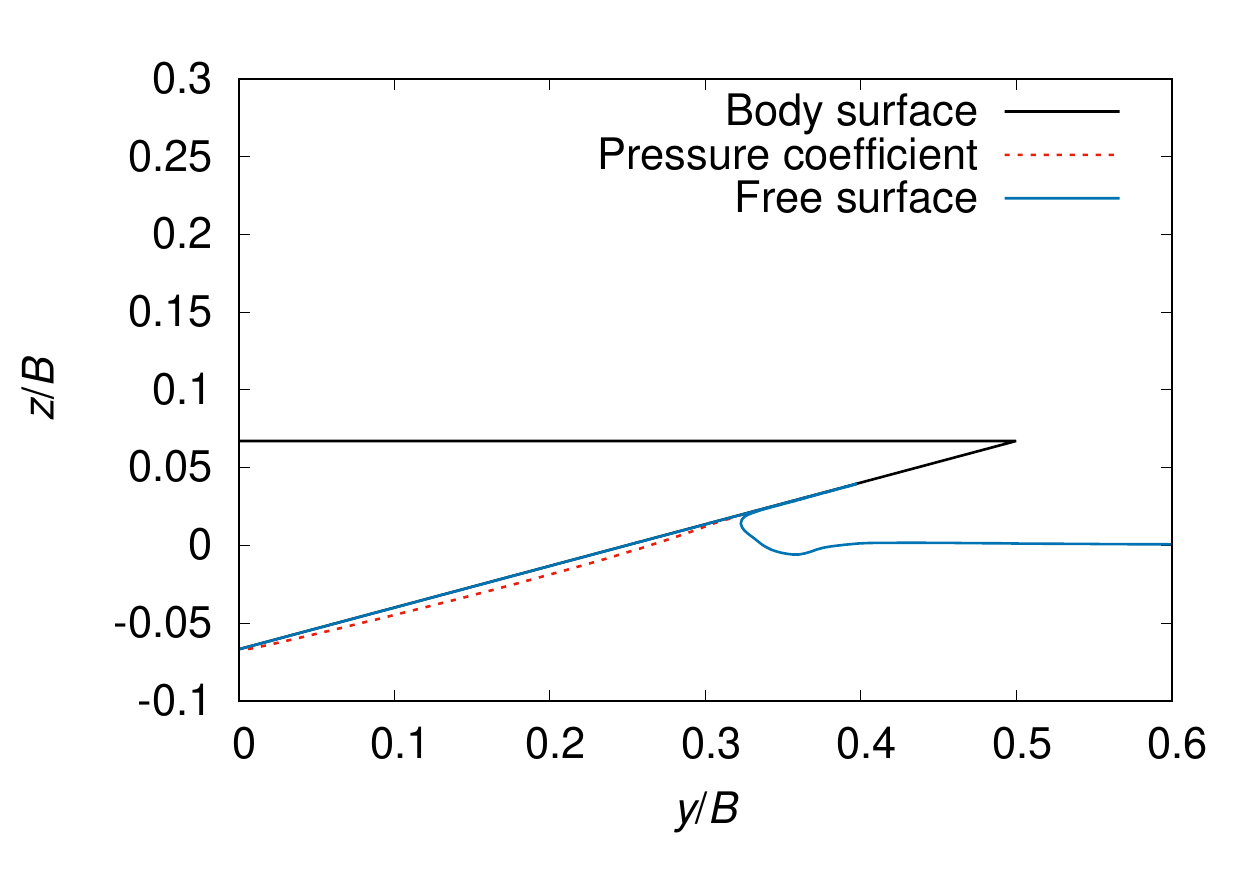} \hspace{2cm}
		\includegraphics[scale=0.5]{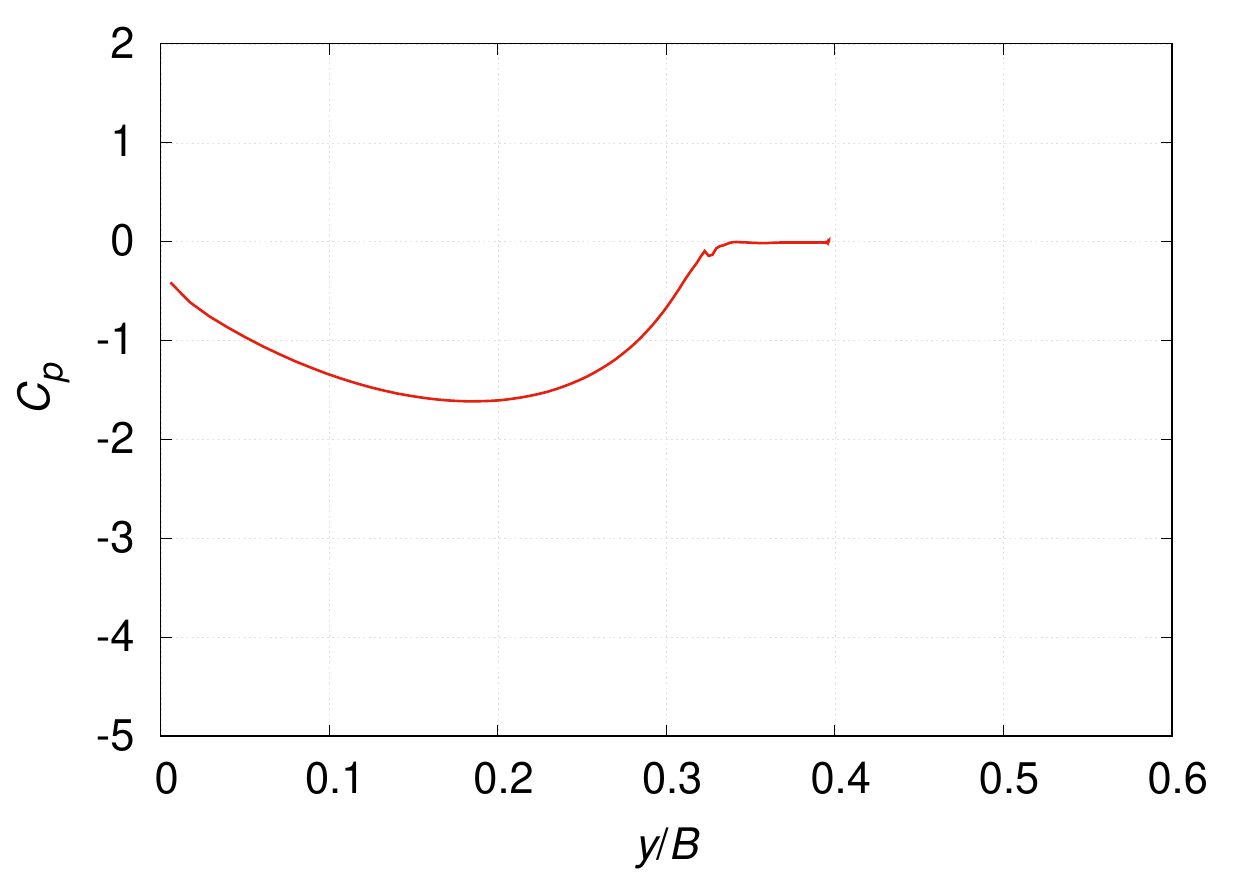}}\\
	\subfigure[$V/V_0=-0.90$]{%
		\includegraphics[scale=0.5]{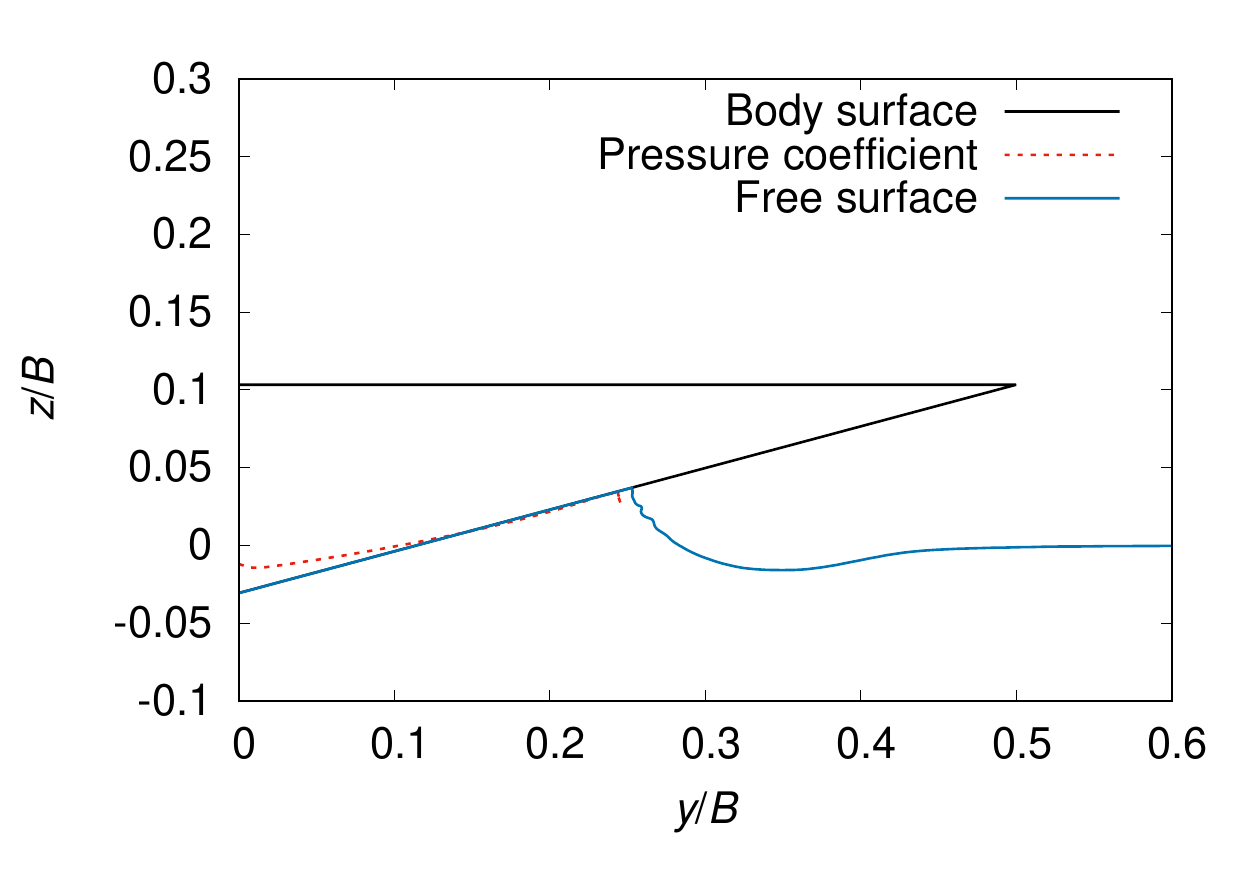} \hspace{2cm}
		\includegraphics[scale=0.5]{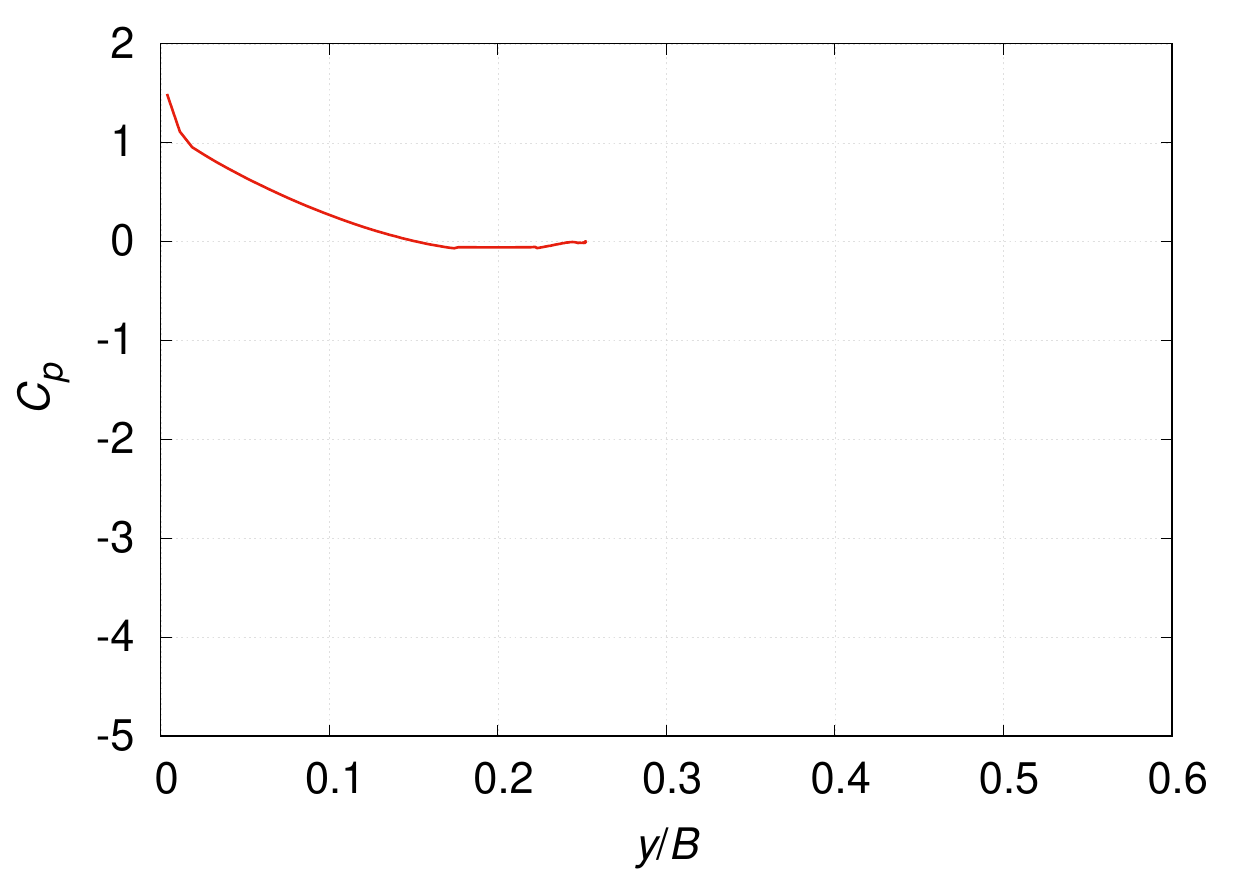}}		
	\caption{\textcolor{black}{Cone: free-surface evolution and pressure coefficient distribution.}}\label{c_fs_cp}
\end{figure}

\begin{figure}
	\centering
	\subfigure[]{%
		\includegraphics[scale=0.5]{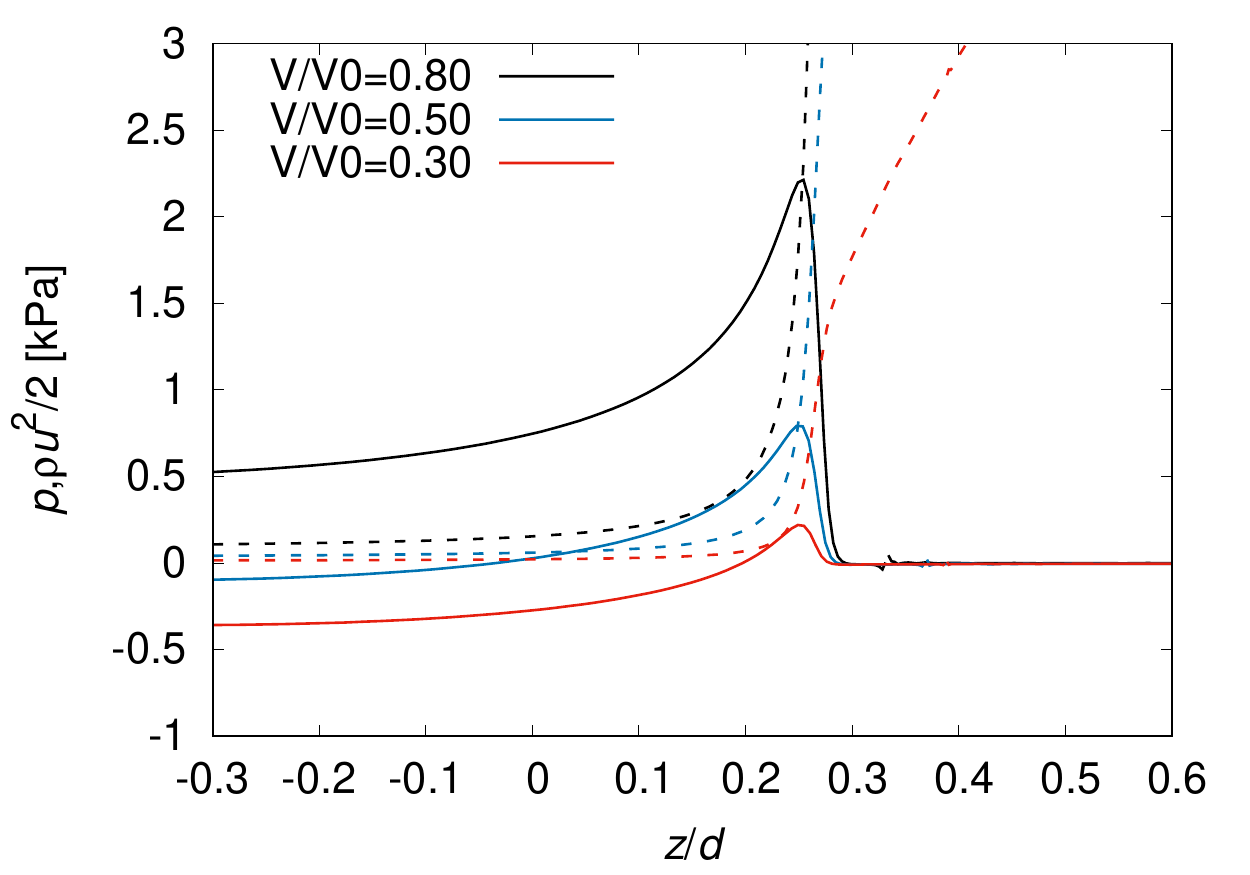}}
	\hspace{2cm}
	\subfigure[]{%
		\includegraphics[scale=0.5]{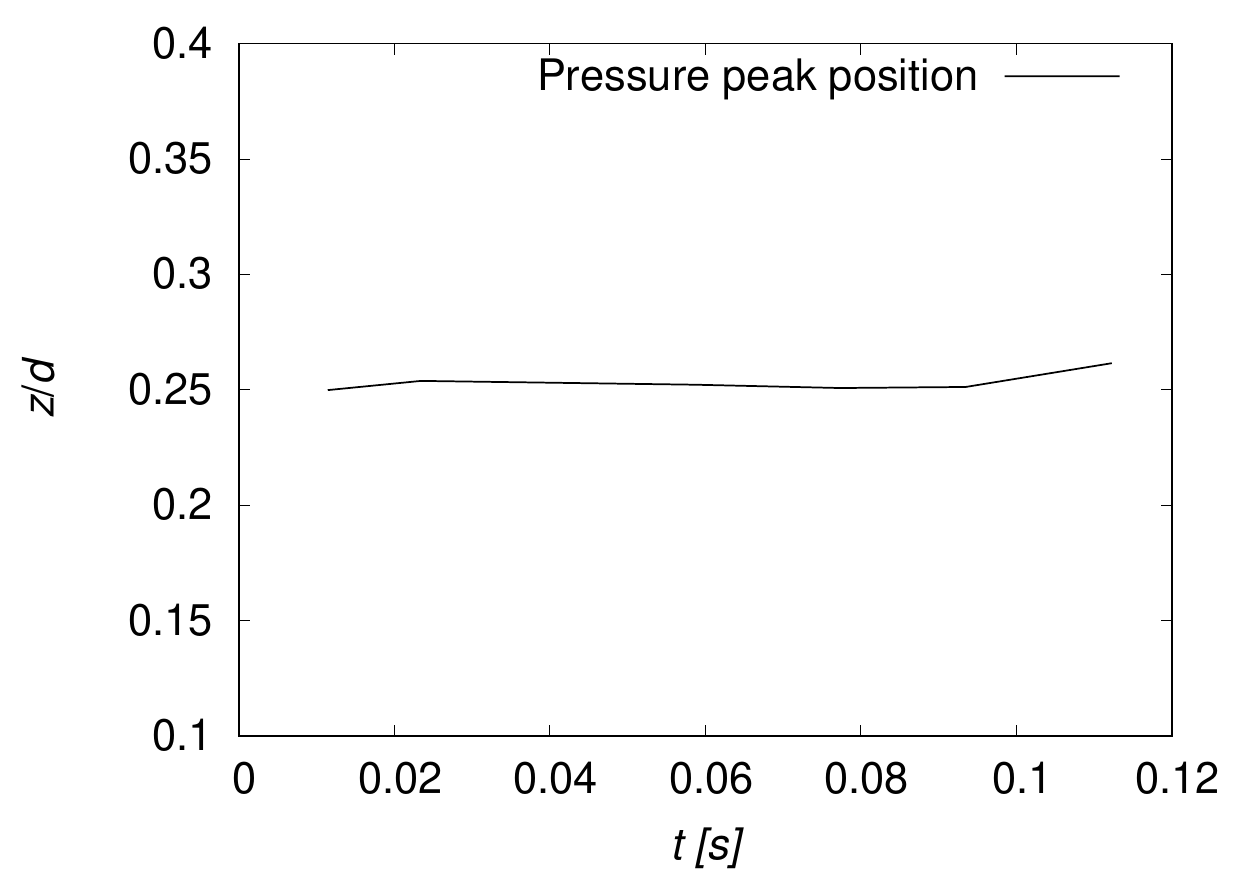}}
	\caption{Cone: pressure details. a) Comparison between the pressure (solid line) and the contribution of non-linear term of the Bernoulli's equation (dashed line) at different time steps. b) Time history of the pressure peak position scaled with the current depth.}\label{c_pres}
\end{figure}

\begin{figure}
	\centering
	\subfigure[]{%
		\includegraphics[scale=0.5]{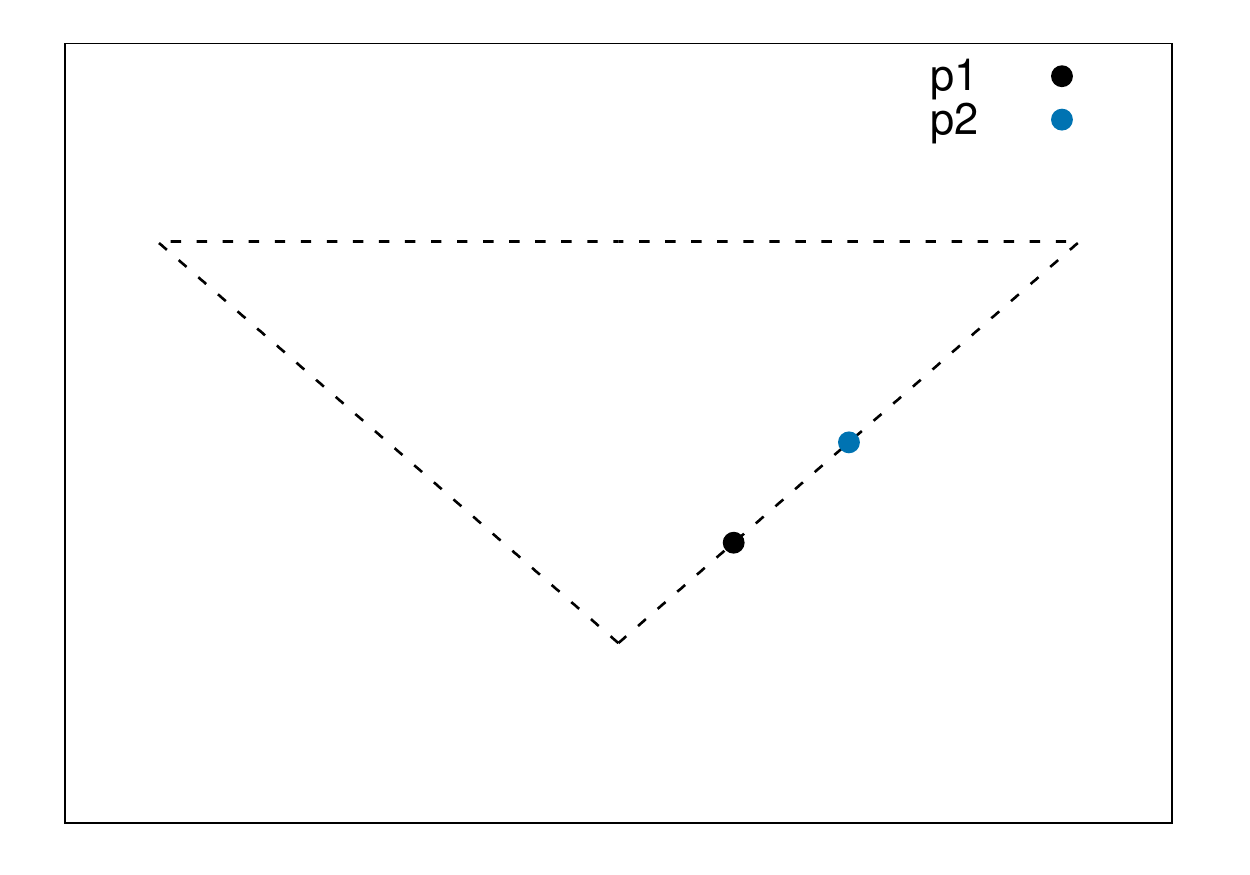}}
	\hspace{2cm}
	\subfigure[]{%
		\includegraphics[scale=0.5]{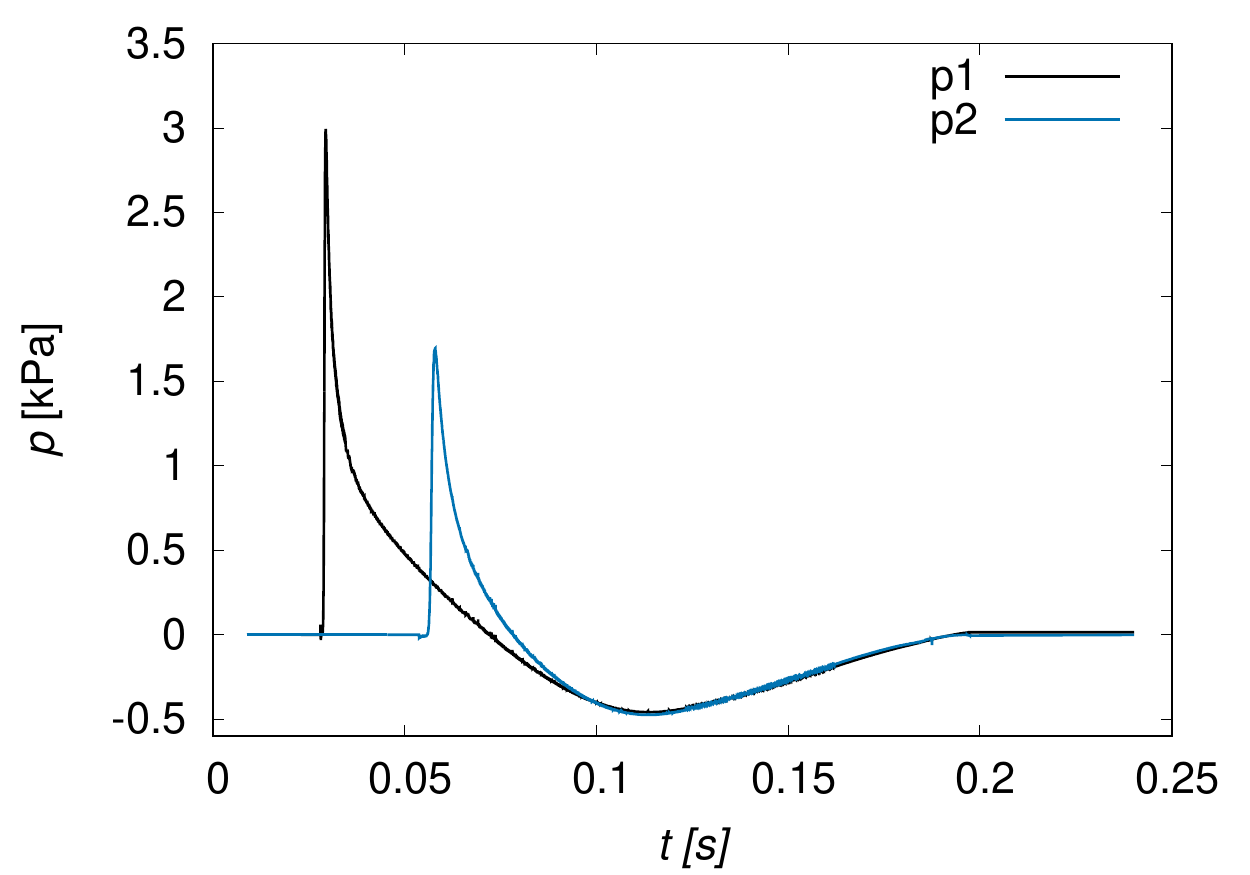}}
	\caption{Cone: a) Pressure sensors position. b) Pressure time history for each sensor.}\label{c_time}
\end{figure}

\begin{figure}
	\centering
	\includegraphics[scale=0.7]{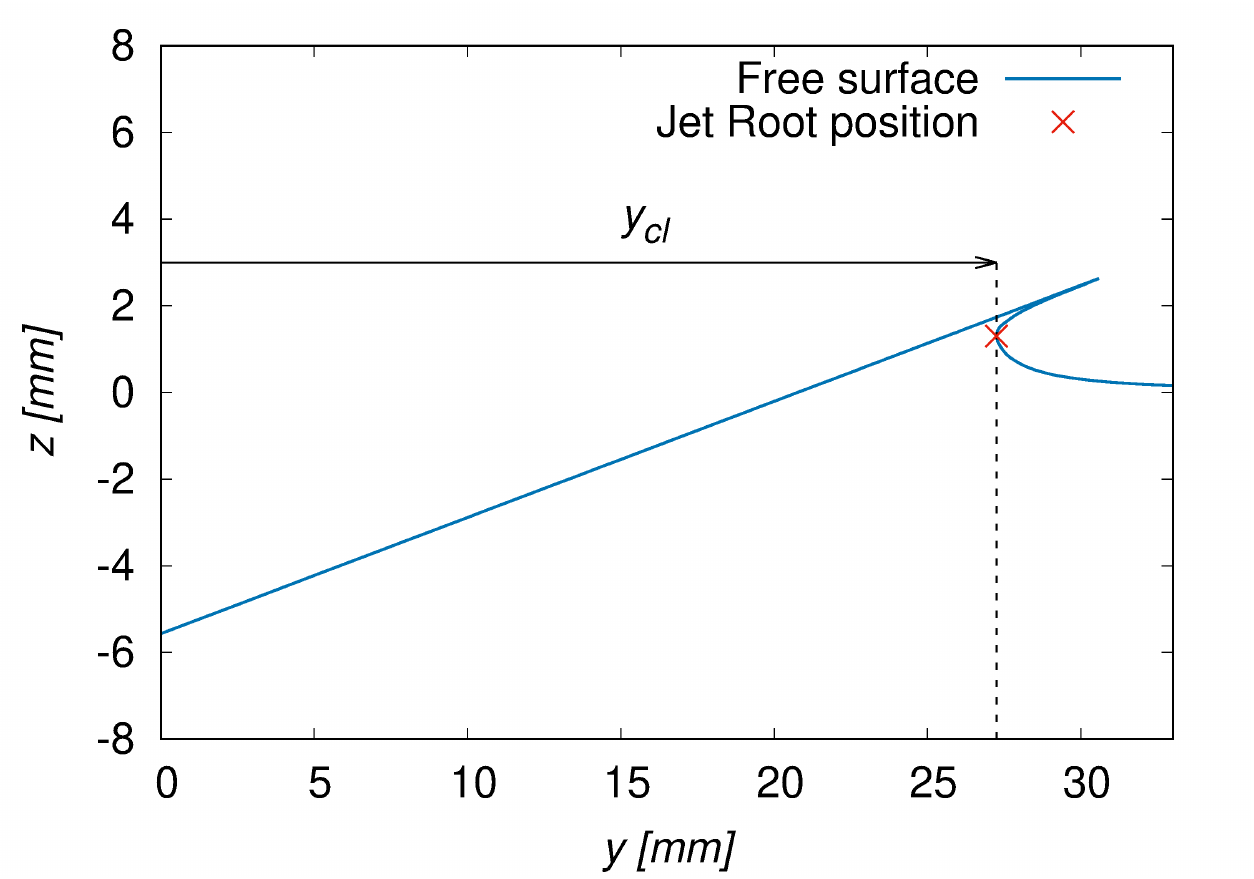}
	\caption{Definition of the contact line}\label{clip}
\end{figure}

\begin{figure}
	\subfigure[]{%
		\includegraphics[scale=0.65]{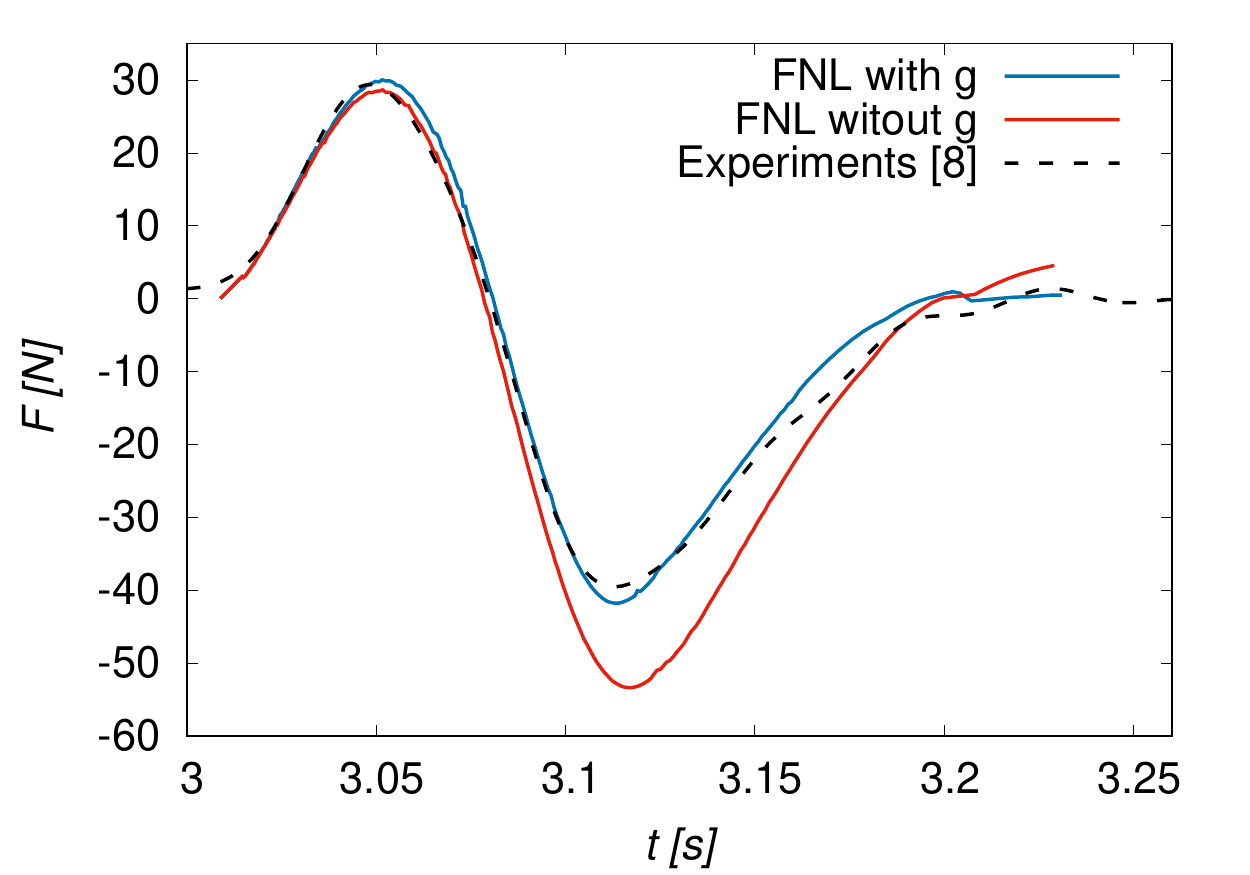}}
	\qquad
	\subfigure[]{%
		\includegraphics[scale=0.65]{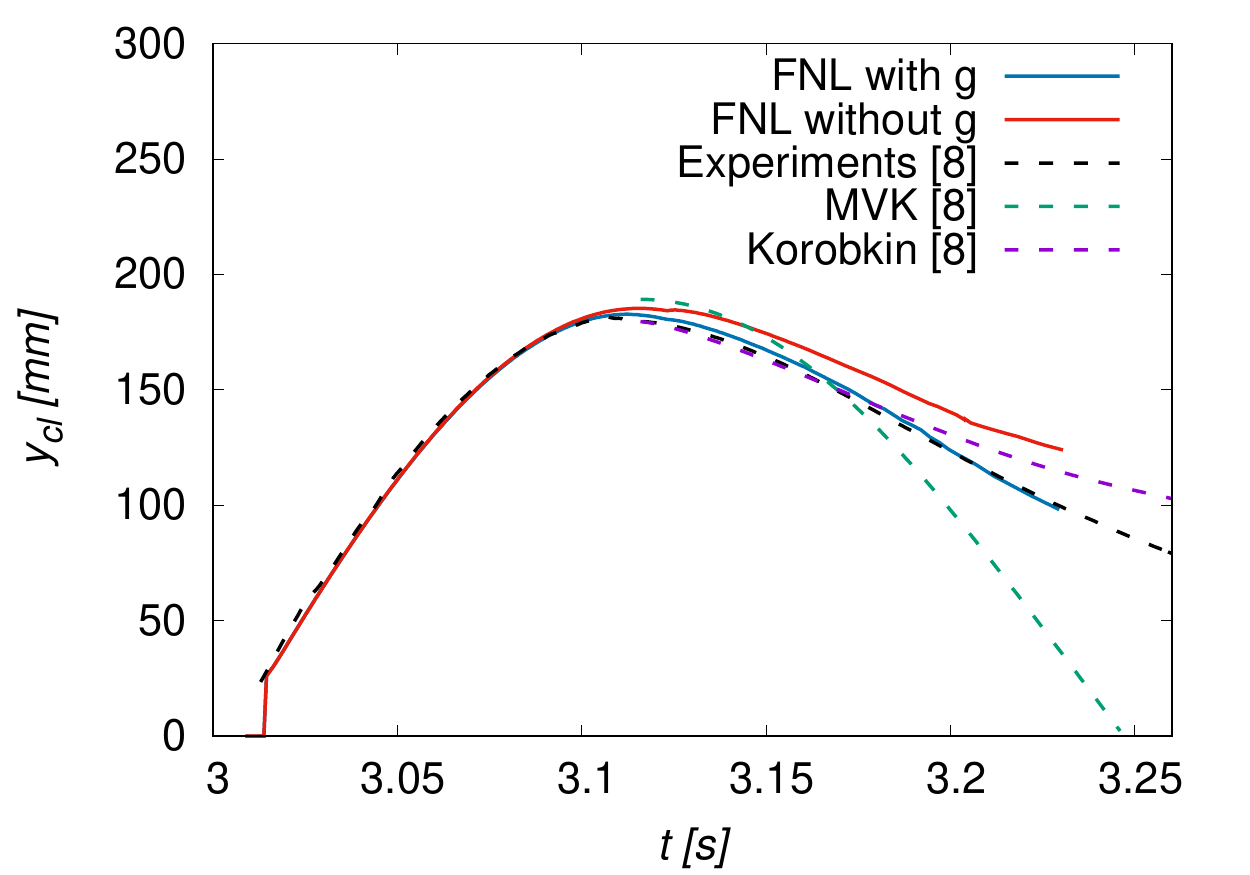}}
	\caption{Cone: evolution of the total force acting on the body (a) and of the contact line (b) as a function of time}\label{C_force}
\end{figure}

The vertical water entry and exit with imposed motion of a rigid cone with a $15^\circ$ 
deadrise angle, $\beta$, is here presented. The test case is based on the experimental 
condition used in \cite{Breton}. In the experiments, the hydrodynamic force acting on 
the impact body was measured and the use of a transparent mockup and a LED 
edge-lighting system with a high-speed video camera placed above the mock-up allowed to 
follow the evolution of the wetted surface during the combined entry and exit event. 
Starting from the time at which the body touches the water, the cone is moved with a 
vertical velocity given by the following equations
\begin{equation*}
\begin{cases}
V(t)=U_{max}\cos(\omega t), \hspace{10mm}     t\leq T/2\\
V(t)=-U_{max},  \hspace{19mm}           t>T/2
\end{cases}
\end{equation*}
where $\omega=U_{max}/H_{max}$ with $U_{max}$ initial, downwards, velocity, $H_{max}$ 
maximum penetration depth and $T=2\pi H_{max}/U_{max}$. The value of $H_{max}$ is 
derived from the Wagner's condition for the wetted surface \cite{Breton}
\begin{equation*}
H_{max} = \frac{c_{max}\pi \tan\beta}{4}
\end{equation*}
where $c_{max}$ is the maximum wetted length. The test case with 
$U_{max}=0.57$m/s and $c_{max}=200$mm has been chosen. The gravity effects, which could 
be significant especially in the exit phase \cite{Breton}, are included here. 
\textcolor{black}{In the previous section, the use of different strategies to improve 
the numerical stability during the exit phase has been shown. Although the two different
approaches provide good results, the combined use of \textit{cut} and \textit{filter} 
strategy is the most preferable. In fact, the jet cut, beside improving the stability, 
allows to reduce the computational effort whereas the \textit{filter} 
strategy can guarantee greater robustness to the solver. 
Based on the above considerations, the \textit{cut+filter} strategies are used here, 
thus managing to stabilize the solution during the exit, as shown in the Figure
\ref{c_exit}.}
\\
Figure \ref{c_fs_cp} displays the results obtained in terms of free-surface evolution 
and pressure distribution. Strong similarities with the solution obtained in the wedge 
case can be observed. During the entry phase the jet rises along the body surface and 
the pressure decreases, becoming negative. During the exit phase, the flow evolves in 
the opposite direction and the negative pressure decreases in amplitude. Note that, due 
to the action of the gravity the pressure returns a little bit positive during the exit 
phase (see Fig. \ref{c_fs_cp} g). Similarly to the wedge case, the pressure peak 
decreases but its position, scaled with the current depth, is almost constant in time, 
and close to the peak the pressure overlaps the value of the non-linear term of the
Bernoulli's equation (Fig. \ref{c_pres}). Moreover, the time history of the pressure 
recordered at different probes is similar (in shape) to what was observed with the 
wedge (Fig. \ref{c_time}). 
\begin{figure}[thbp]
	\centering
	\subfigure[$V/V_0=0.90$]{
		\includegraphics[scale=0.5]{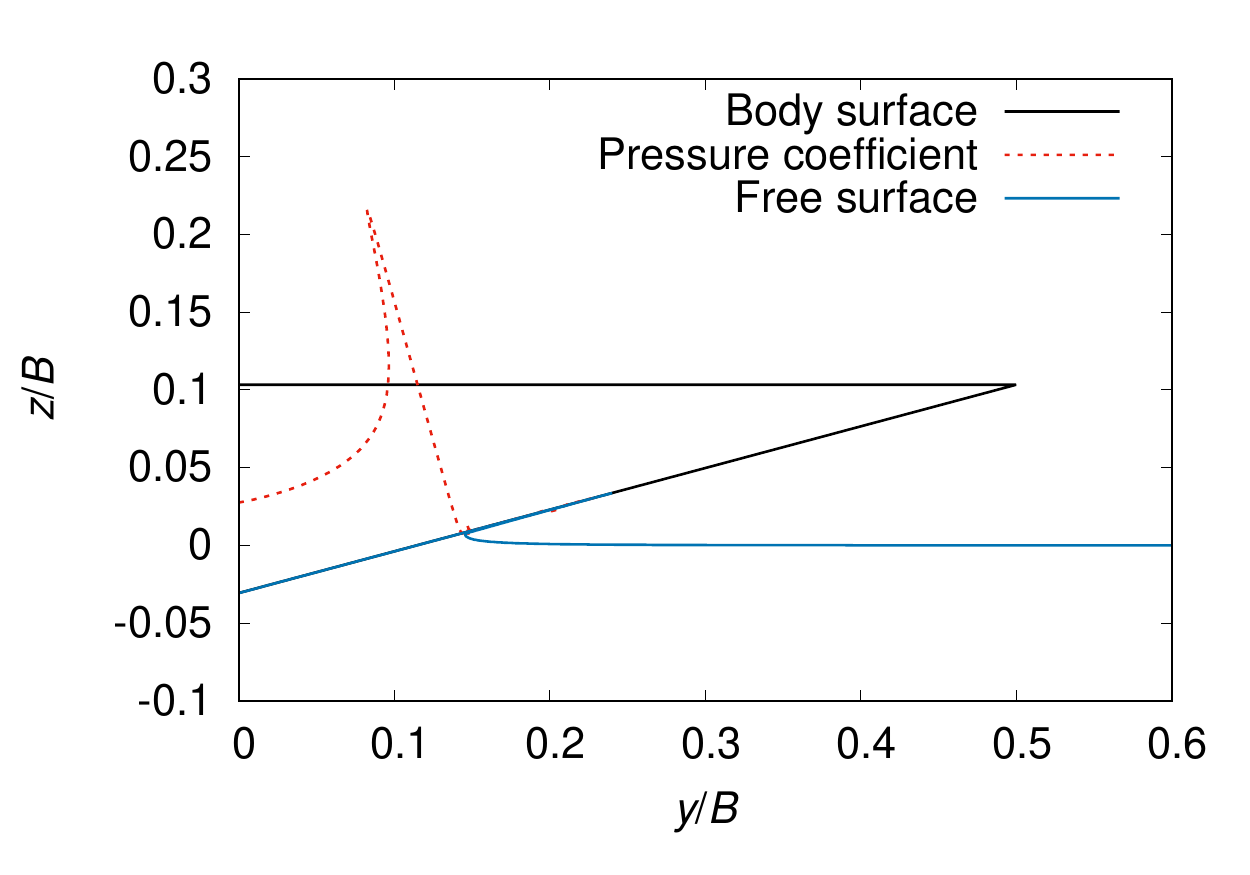} \hspace{2cm}
		\includegraphics[scale=0.5]{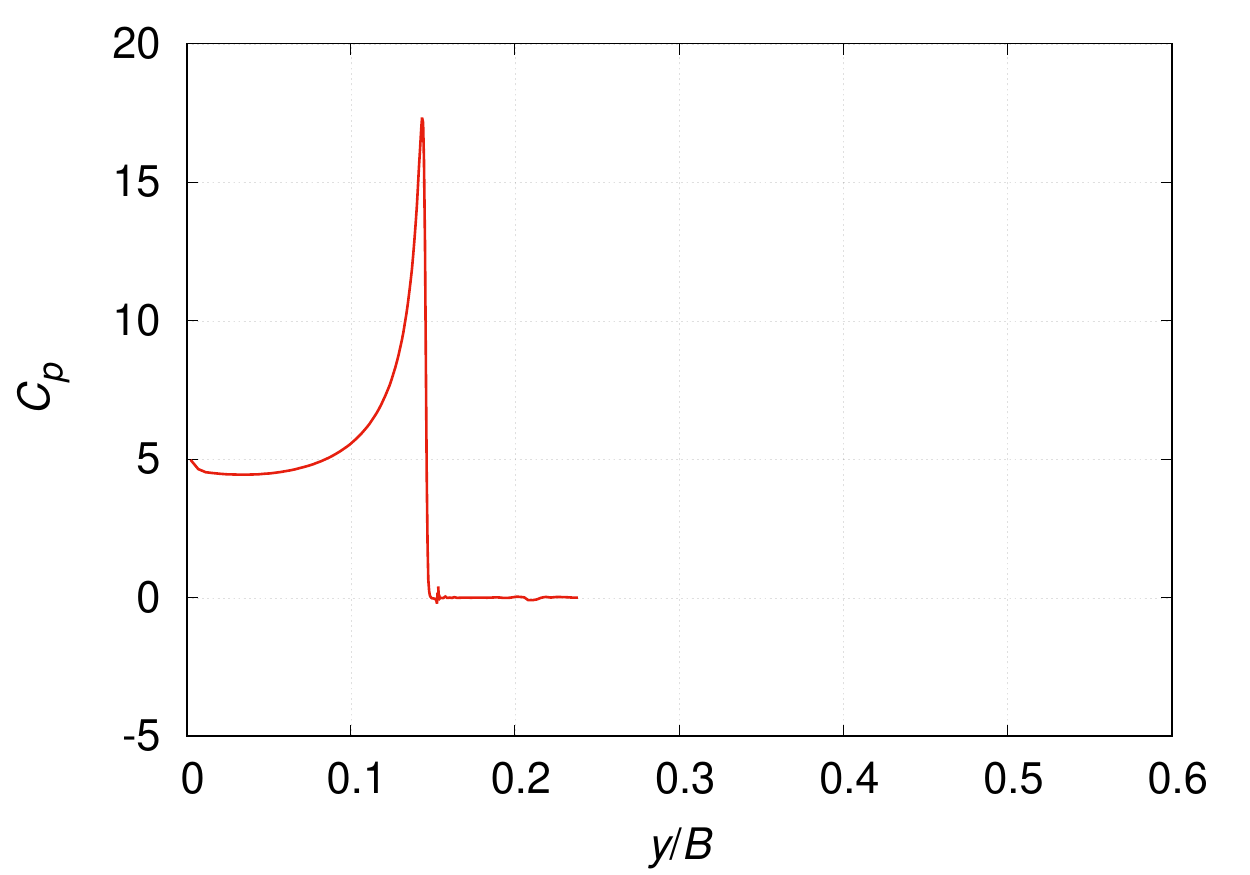}}\\
	\subfigure[$V/V_0=0.50$]{
		\includegraphics[scale=0.5]{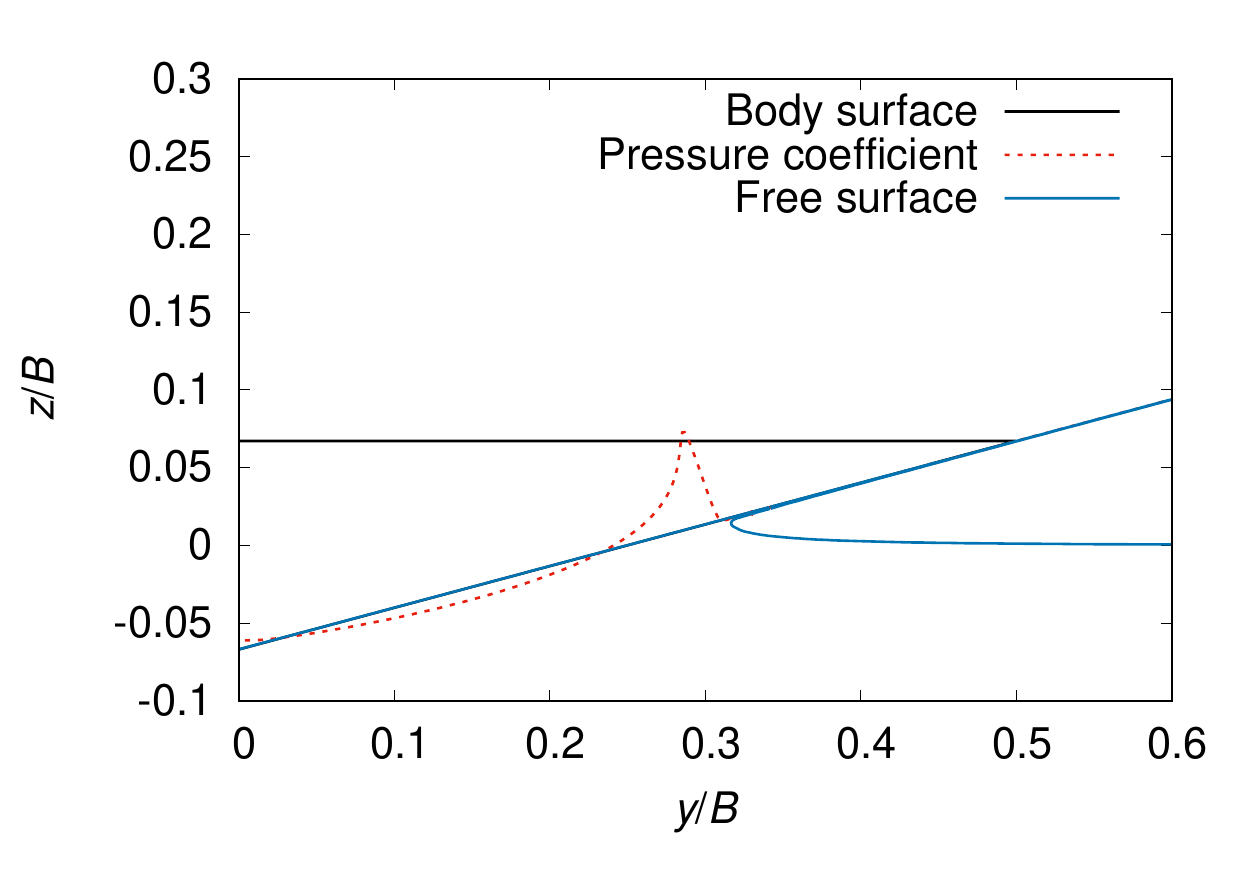} \hspace{2cm}
		\includegraphics[scale=0.5]{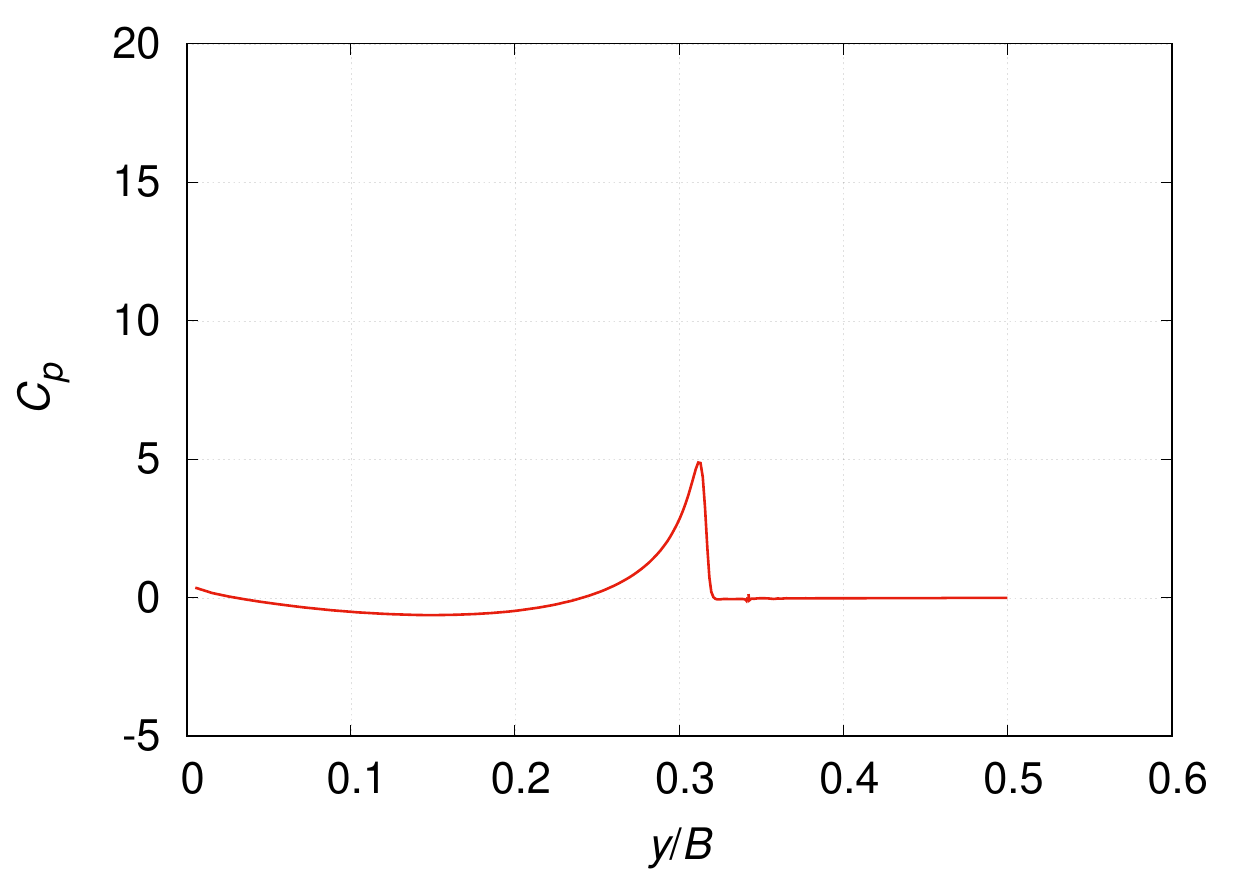}}\\
	\subfigure[$V/V_0=0.10$]{%
		\includegraphics[scale=0.5]{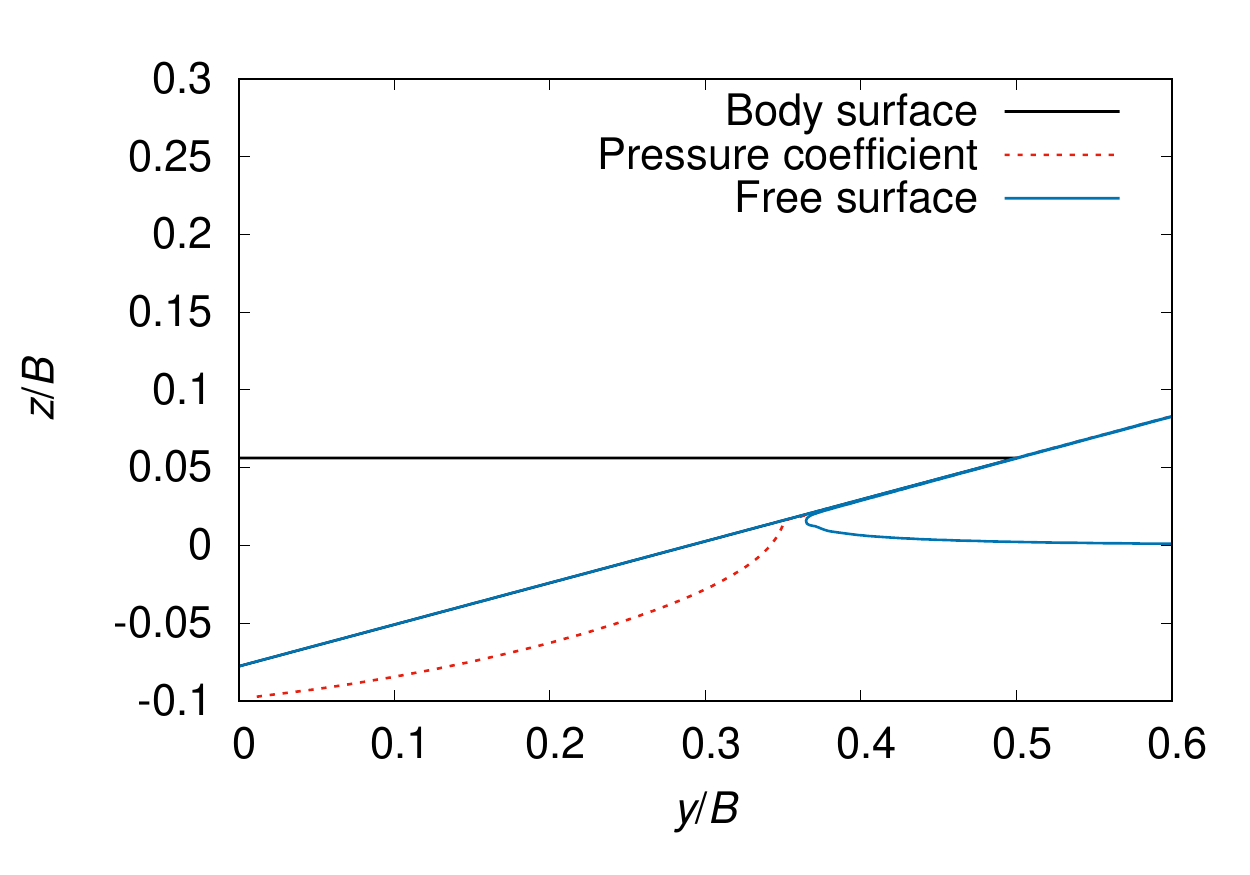} \hspace{2cm}
		\includegraphics[scale=0.5]{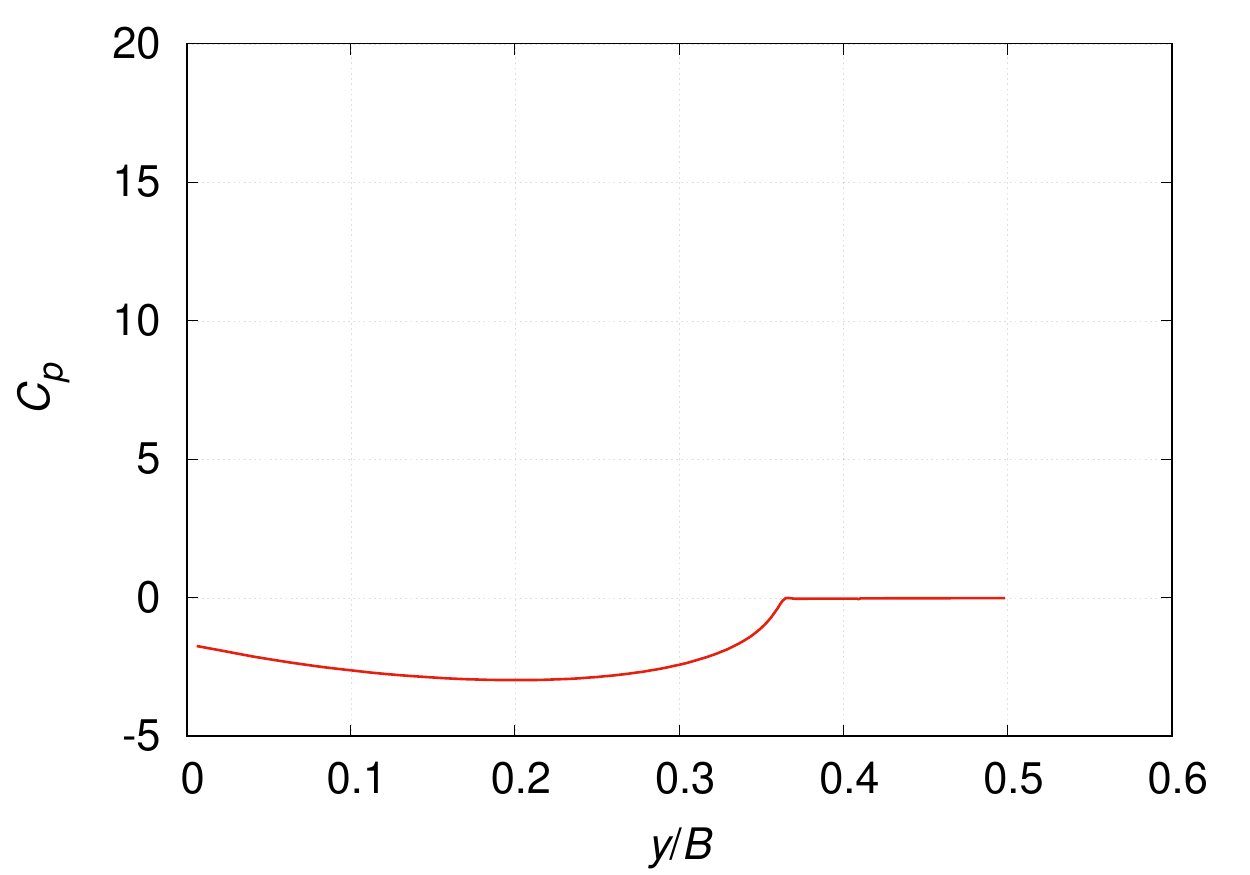}}\\
	\subfigure[$V/V_0=0$]{%
		\includegraphics[scale=0.5]{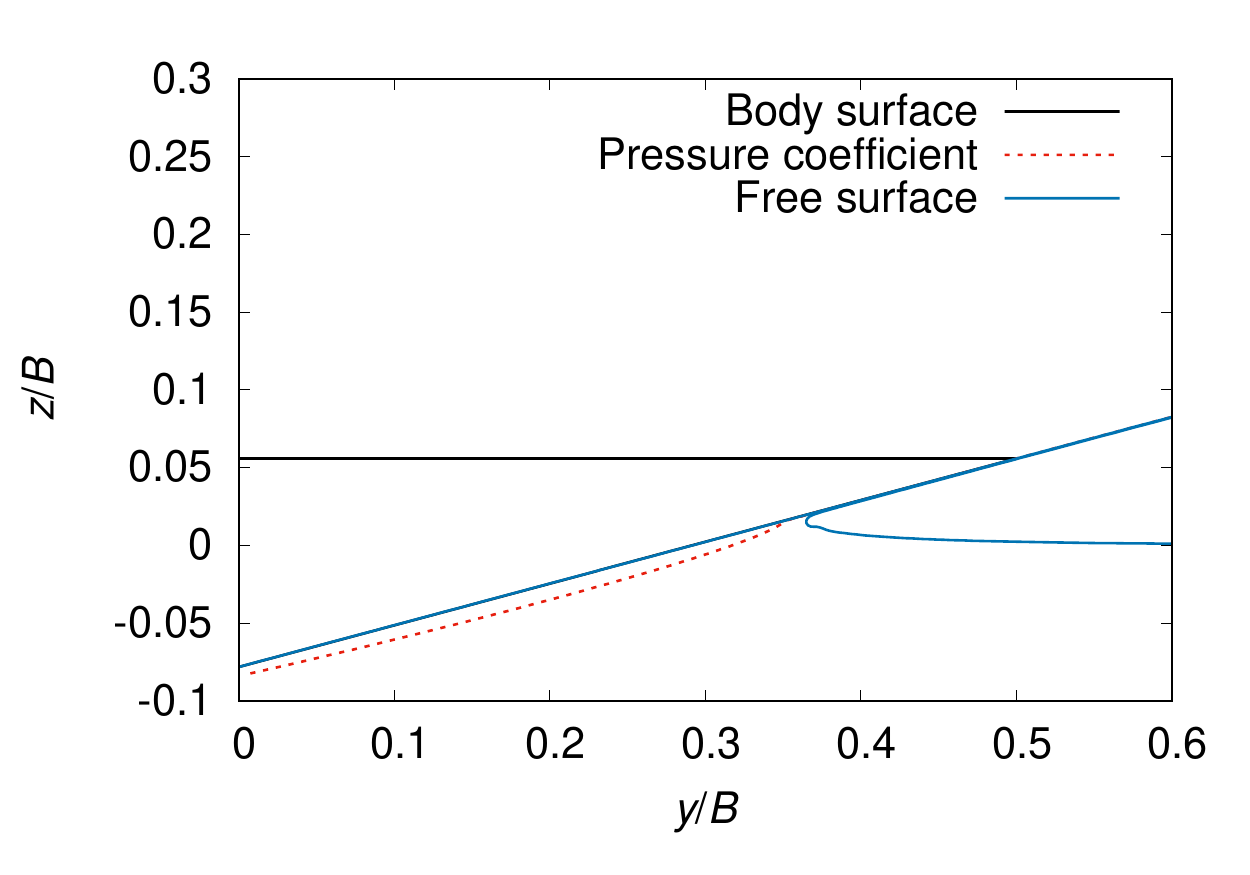} \hspace{2cm}
		\includegraphics[scale=0.5]{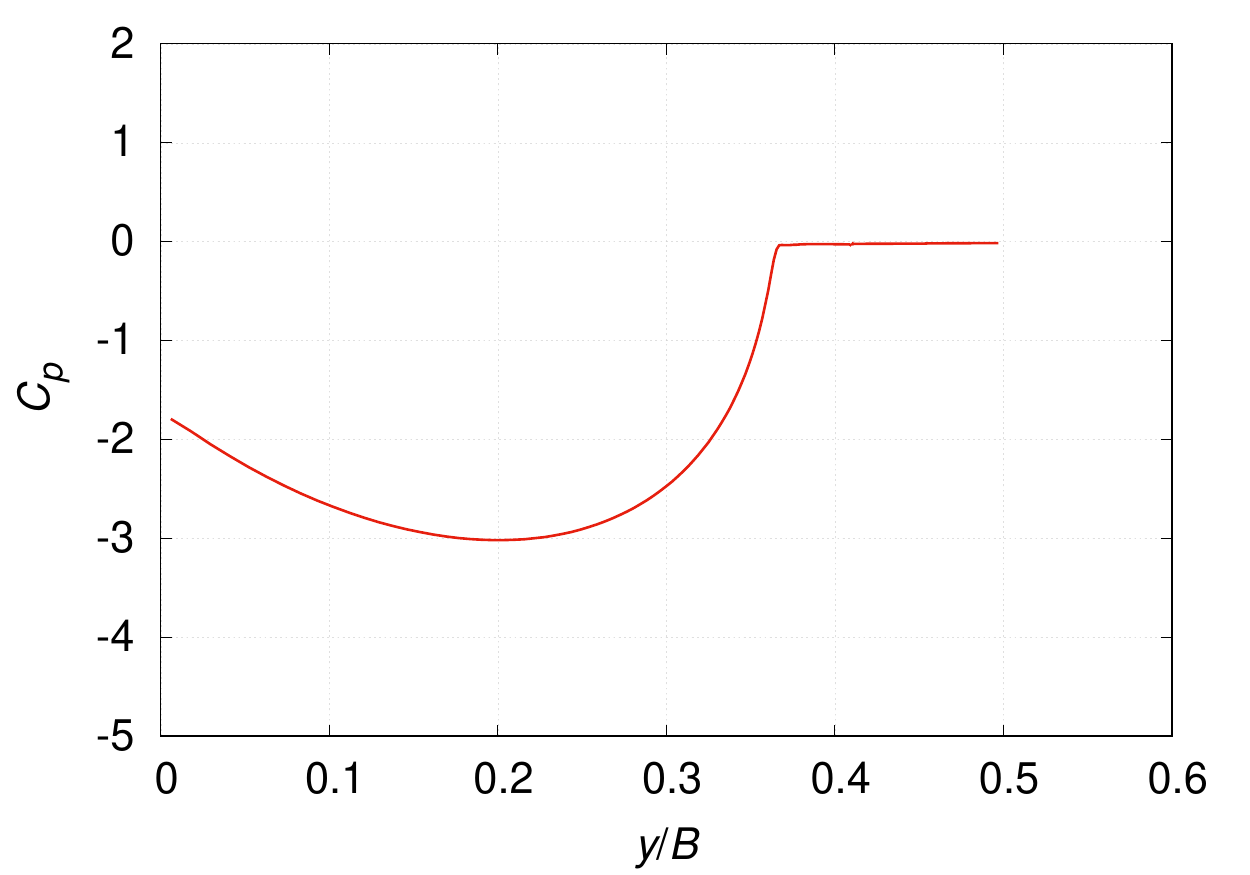}}\\
\end{figure}
\\
Figure \ref{C_force} shows the time histories of the hydrodynamic force and of the 
contact line position, evaluated by following the $y$-coordinate of the jet root, as 
indicated in Fig.\ref{clip}. The force, computed by pressure integration along the 
wetted area, is positive in the first part of the entry phase and turns negative 
afterwards, approaching zero during the exit phase. The negative force peak, occuring 
at the transition between the entry and exit stages, is higher in amplitude than the 
maximum force recorded in the entry phase, despite of the action of the gravity. The 
contact line is maximum at the instant when the body attains its maximum depth, i.e. at 
the transition between the entry and exit phases, but it is smaller than the maximum 
expected value $c_{max}$. This is an effect of the presence of the meniscus during the 
experiments \cite{Breton} which is considered in the numerical simulation by correcting 
the body displacement with a parameter $\delta_z=3mm$, which practically changes the 
initial reference vertical position. The asymmetry of the curve with respect to the 
peak indicates that in the exit phase the wetted surface decreases 
\textcolor{black}{slower} than it increases during the entry phase. The initial step on 
the curves of the contact line time history obtained with the proposed model, is due to 
the initial body depth necessary to start the simulation \cite{Iafrati2010}. The 
results obtained with the proposed fully non-linear model (FNL) are in very good 
agreement with the experimental data. It is worth noting that in the FNL simulations 
gravity effects are included. In order to evaluate the role of the gravity, especially 
during the exit phase, the numerical simulations were conducted with and without 
gravity. As shown in Fig. \ref{C_force}, the gravity affects both the hydrodynamic 
force and wetted surface, especially during the exit phase. In Fig. \ref{C_force}b the 
results obtained with water exit semi-analytical models in \cite{Breton} are also 
provided. The first one is the modified von K\'{a}rm\'{a}n (MVK), which is an extension 
to the axisymmetric bodies of the model proposed in \cite{Tassin}. The second one is 
the extension to axisymmetric bodies and non-constant exit velocity proposed in 
\cite{Breton} of the Korobkin model \cite{Kor2017a}. The semi-analytical models perform 
much less satisfactorily than the fully non-linear model, presumably because of 
neglecting the gravity effects. Based on the above considerations, the flexibility of 
the fully non-linear model seems very helpful for an accurate description of the exit 
phase. 
\section{Conclusions}
A fully non-linear potential flow model based on the hybrid BEM-FEM approach for the 
description of combined water entry and exit problems of 2D and axisymmetric bodies has 
been proposed. The model is an extension of that proposed and validated in 
\cite{Battistin2003} and \cite{Battistin2004} for the water entry with constant 
velocity applications. In particular, a water exit model that includes two different 
strategies aimed at improving the stability of the free-surface shape, has been 
proposed and validated against data available in the literature. Both strategies have 
been found able to improve the numerical stability of the solution providing similar 
results in the wedge case, where a very accurate prediction of the free-surface, 
pressure distribution and vertical hydrodynamic force time histories has been obtained. 
\textcolor{black}{The combined use of the two strategies both in the cone and the 
wedge cases has demonstrated further improvement of the stability of the free-surface 
dynamics. Such combined approach seems the best in terms of computational effort and 
robustness of the solver.} 

\noindent
The detailed information provided by the model also allow an accurate analysis 
of the pressure distribution along the wetted body portion. In particular, it has been 
shown that, similarly to what happens in the self-similar solution, the pressure, 
close to its peak, matches the non-linear term of the Bernoulli's equation. 
The results obtained for the cone, for which comparisons with accurate experimental 
data have been established in terms of hydrodynamic force and wetted area, clearly 
highlight the strong potential of the proposed model. 
Furthermore, the role played by the gravity, which is particularly important during the 
exit phase, as shown in the experiments of \cite{Breton}, has been observed. 

\noindent
As a future work, the model will be extended into a multisectional 2D+t procedure for 
application to both aircraft ditching and high-speed planing vessels. 

\section*{Funding}
This project has been partly funded from the European Union's Horizon 2020 Research and 
Innovation Programme under Grant Agreement No. 724139 (H2020-SARAH: increased SAfety \& 
Robust certification for ditching of Aircrafts \& Helicopters).

\end{document}